\documentclass[longauth]{aa}

\usepackage[english]{babel}
\usepackage{natbib}
\bibpunct{(}{)}{;}{a}{}{,} 
\usepackage{amsmath}
\usepackage{graphicx}
\usepackage{txfonts}
\usepackage[colorlinks=true, allcolors=blue]{hyperref}
\usepackage{enumitem}
\usepackage{verbatim}
\usepackage[hang,small,bf]{caption}
\usepackage{rotating}
\usepackage{url}
\usepackage{times}
\usepackage{color,colortbl}
\usepackage{siunitx}
\usepackage{amsfonts}
\usepackage{amssymb}
\usepackage{epsfig}
\usepackage{mathtools}
\usepackage[normalem]{ulem}
\usepackage{multirow}
\usepackage{tabularx}
\usepackage[capitalise,nameinlink]{cleveref}
\crefname{section}{Sect.}{Sects}
\usepackage{euclid}
\usepackage{changepage}
\newcommand{\onesigma}{1\mbox{--}\sigma}
\usepackage{adjustbox}
\usepackage{dsfont}
\usepackage{braket}
\usepackage{subfigure}
\usepackage{float}
\usepackage{mathrsfs} 
\usepackage{comment}
\usepackage{booktabs}
\usepackage[toc,page]{appendix}

\usepackage{cleveref}
\usepackage{orcidlink}
\usepackage{graphicx}
\usepackage{natbib}
\usepackage{scalerel}
\usepackage{enumitem}
\usepackage{multirow}
\usepackage{url}
\usepackage{glossaries}
\usepackage{cancel}
\usepackage{float}

\defcitealias{2020A&A...642A.191E}{EP-VII}
\defcitealias{Collaboration2016}{Planck Collaboration 2016}

\hypersetup{
    colorlinks=true,
    linkcolor=blue,
    citecolor=blue,
}
\binoppenalty=\maxdimen
\relpenalty=\maxdimen

\newcommand{\Omb}{\Omega_\mathrm{b}}
\newcommand{\Omm}{\Omega_\mathrm{m}}
\newcommand{\mnu}{M_\nu}
\newcommand{\GCph}{\mathrm{GC}_\mathrm{ph}}
\newcommand{\GCsp}{\mathrm{GC}_\mathrm{sp}}
\newcommand{\WL}{\mathrm{WL}}
\newcommand{\wz}{w_0}
\newcommand{\wa}{w_a}
\newcommand{\sige}{\sigma_{8}}
\newcommand{\FoM}{\mathrm{FoM}}

\newcommand{\wzwaCDM}{\wz\wa\mathrm{CDM}}

\newcommand{\cl}[3]{C^{#1}_{#2}({#3})}
\newcommand{\cnol}[2]{C^{#1}_{#2}}
\newcommand{\Halpha}{\mathrm{H}\alpha}
\newcommand{\de}{\mathrm{d}}

\newcommand{\CAMB}{\texttt{CAMB}}

\newcommand{\ns}{n_\textrm{s}}

\newcommand{\aIA}{\mathcal{A}_{\mathrm{IA}}}
\newcommand{\betaIA}{\beta_{\mathrm{IA}}}
\newcommand{\etaIA}{\eta_{\mathrm{IA}}}
\newcommand{\CIA}{C_{\mathrm{IA}}}

\newcommand{\wlwl}{\mathrm{wlwl}}
\newcommand{\phph}{\mathrm{phph}}
\newcommand{\spsp}{\mathrm{spsp}}
\newcommand{\wlph}{\mathrm{wlph}}
\newcommand{\phwl}{\mathrm{phwl}}
\newcommand{\wlsp}{\mathrm{wlsp}}

\newcommand{\phsp}{\mathrm{phsp}}

\newcommand{\xc}[2]{\mathrm{XC}(#1,#2)}
\newcommand{\threetwoptShort}{\ensuremath{\textrm{3$\times$2pt}}}
\newcommand{\threetwoptLong}{\WL+\GCph+\xc{\WL}{\GCph}}
\newcommand{\sixtwoptShort}{\ensuremath{\textrm{6$\times$2pt}}}

\newcommand{\positive}[1]{\textcolor{ForestGreen}{#1}}
\definecolor{ForestGreen}{RGB}{34,139,34}
\newcommand{\negative}[1]{\textcolor{red}{#1}}

\newcommand{\bcl}[2]{\mathbf{C}^{#1}(#2)}

\newcommand{\sgl}[3]{\Sigma^{#1}_{#2}(#3)}

\newcommand{\cov}[2]{\operatorname{Cov}\left[#1, #2\right]}
\newcommand{\invcov}[2]{\operatorname{Cov}^{-1}\left[#1, #2\right]}
\newcommand{\pdv}[2]{\frac{\partial #1}{\partial #2}}

\newenvironment{resultstable}[1]
    {\centering
    \setlength{\tabcolsep}{3.0pt}
    \begin{adjustbox}{max width=\textwidth}
    \begin{tabular}{#1}
    \toprule
    }
    {\bottomrule
    \end{tabular}
    \end{adjustbox}
    }

\renewcommand\aap{A\&A}

\renewcommand\apjl{ApJL}

\usepackage{soul}
\usepackage[switch]{lineno}

\begin{document}
\title{\Euclid preparation}
\subtitle{6$\times$2pt analysis of \Euclid's spectroscopic and photometric data sets}


\newcommand{\orcid}[1]{} 
\author{Euclid Collaboration: L.~Paganin\inst{\ref{aff1}}
\and M.~Bonici\orcid{0000-0002-8430-126X}\thanks{\email{marco.bonici@inaf.it}}\inst{\ref{aff2}}
\and C.~Carbone\orcid{0000-0003-0125-3563}\inst{\ref{aff2}}
\and S.~Camera\orcid{0000-0003-3399-3574}\inst{\ref{aff3},\ref{aff4},\ref{aff5}}
\and I.~Tutusaus\orcid{0000-0002-3199-0399}\inst{\ref{aff6},\ref{aff7},\ref{aff8},\ref{aff9}}
\and S.~Davini\inst{\ref{aff10}}
\and J.~Bel\inst{\ref{aff11}}
\and S.~Tosi\inst{\ref{aff12},\ref{aff10},\ref{aff13}}
\and D.~Sciotti\orcid{0009-0008-4519-2620}\inst{\ref{aff14},\ref{aff15},\ref{aff16}}
\and S.~Di~Domizio\orcid{0000-0003-2863-5895}\inst{\ref{aff12},\ref{aff10}}
\and I.~Risso\inst{\ref{aff1}}
\and G.~Testera\inst{\ref{aff10}}
\and D.~Sapone\orcid{0000-0001-7089-4503}\inst{\ref{aff17}}
\and Z.~Sakr\inst{\ref{aff18},\ref{aff6},\ref{aff19}}
\and A.~Amara\inst{\ref{aff20}}
\and S.~Andreon\orcid{0000-0002-2041-8784}\inst{\ref{aff13}}
\and N.~Auricchio\orcid{0000-0003-4444-8651}\inst{\ref{aff21}}
\and C.~Baccigalupi\orcid{0000-0002-8211-1630}\inst{\ref{aff22},\ref{aff23},\ref{aff24},\ref{aff25}}
\and M.~Baldi\orcid{0000-0003-4145-1943}\inst{\ref{aff26},\ref{aff21},\ref{aff27}}
\and S.~Bardelli\orcid{0000-0002-8900-0298}\inst{\ref{aff21}}
\and P.~Battaglia\orcid{0000-0002-7337-5909}\inst{\ref{aff21}}
\and R.~Bender\orcid{0000-0001-7179-0626}\inst{\ref{aff28},\ref{aff29}}
\and F.~Bernardeau\inst{\ref{aff30},\ref{aff31}}
\and C.~Bodendorf\inst{\ref{aff28}}
\and D.~Bonino\orcid{0000-0002-3336-9977}\inst{\ref{aff5}}
\and E.~Branchini\orcid{0000-0002-0808-6908}\inst{\ref{aff12},\ref{aff10},\ref{aff13}}
\and M.~Brescia\orcid{0000-0001-9506-5680}\inst{\ref{aff32},\ref{aff33},\ref{aff34}}
\and J.~Brinchmann\orcid{0000-0003-4359-8797}\inst{\ref{aff35}}
\and V.~Capobianco\orcid{0000-0002-3309-7692}\inst{\ref{aff5}}
\and V.~F.~Cardone\inst{\ref{aff15},\ref{aff16}}
\and J.~Carretero\orcid{0000-0002-3130-0204}\inst{\ref{aff36},\ref{aff37}}
\and S.~Casas\orcid{0000-0002-4751-5138}\inst{\ref{aff38}}
\and M.~Castellano\orcid{0000-0001-9875-8263}\inst{\ref{aff15}}
\and G.~Castignani\orcid{0000-0001-6831-0687}\inst{\ref{aff39},\ref{aff21}}
\and S.~Cavuoti\orcid{0000-0002-3787-4196}\inst{\ref{aff33},\ref{aff34}}
\and A.~Cimatti\inst{\ref{aff40}}
\and C.~Colodro-Conde\inst{\ref{aff41}}
\and G.~Congedo\orcid{0000-0003-2508-0046}\inst{\ref{aff42}}
\and C.~J.~Conselice\inst{\ref{aff43}}
\and L.~Conversi\orcid{0000-0002-6710-8476}\inst{\ref{aff44},\ref{aff45}}
\and Y.~Copin\orcid{0000-0002-5317-7518}\inst{\ref{aff46}}
\and L.~Corcione\orcid{0000-0002-6497-5881}\inst{\ref{aff5}}
\and A.~Costille\inst{\ref{aff47}}
\and F.~Courbin\orcid{0000-0003-0758-6510}\inst{\ref{aff48}}
\and H.~M.~Courtois\orcid{0000-0003-0509-1776}\inst{\ref{aff49}}
\and M.~Crocce\orcid{0000-0002-9745-6228}\inst{\ref{aff8},\ref{aff50}}
\and M.~Cropper\orcid{0000-0003-4571-9468}\inst{\ref{aff51}}
\and A.~Da~Silva\orcid{0000-0002-6385-1609}\inst{\ref{aff52},\ref{aff53}}
\and H.~Degaudenzi\orcid{0000-0002-5887-6799}\inst{\ref{aff54}}
\and G.~De~Lucia\orcid{0000-0002-6220-9104}\inst{\ref{aff23}}
\and A.~M.~Di~Giorgio\orcid{0000-0002-4767-2360}\inst{\ref{aff55}}
\and J.~Dinis\inst{\ref{aff53},\ref{aff52}}
\and F.~Dubath\orcid{0000-0002-6533-2810}\inst{\ref{aff54}}
\and C.~A.~J.~Duncan\inst{\ref{aff43},\ref{aff56}}
\and X.~Dupac\inst{\ref{aff45}}
\and S.~Dusini\orcid{0000-0002-1128-0664}\inst{\ref{aff57}}
\and A.~Ealet\inst{\ref{aff46}}
\and M.~Farina\orcid{0000-0002-3089-7846}\inst{\ref{aff55}}
\and S.~Farrens\orcid{0000-0002-9594-9387}\inst{\ref{aff58}}
\and S.~Ferriol\inst{\ref{aff46}}
\and M.~Frailis\orcid{0000-0002-7400-2135}\inst{\ref{aff23}}
\and E.~Franceschi\orcid{0000-0002-0585-6591}\inst{\ref{aff21}}
\and S.~Galeotta\orcid{0000-0002-3748-5115}\inst{\ref{aff23}}
\and B.~Garilli\orcid{0000-0001-7455-8750}\inst{\ref{aff2}}
\and K.~George\orcid{0000-0002-1734-8455}\inst{\ref{aff29}}
\and W.~Gillard\orcid{0000-0003-4744-9748}\inst{\ref{aff59}}
\and B.~Gillis\orcid{0000-0002-4478-1270}\inst{\ref{aff42}}
\and C.~Giocoli\orcid{0000-0002-9590-7961}\inst{\ref{aff21},\ref{aff60}}
\and A.~Grazian\orcid{0000-0002-5688-0663}\inst{\ref{aff61}}
\and F.~Grupp\inst{\ref{aff28},\ref{aff29}}
\and L.~Guzzo\orcid{0000-0001-8264-5192}\inst{\ref{aff62},\ref{aff13},\ref{aff63}}
\and S.~V.~H.~Haugan\orcid{0000-0001-9648-7260}\inst{\ref{aff64}}
\and W.~Holmes\inst{\ref{aff65}}
\and I.~Hook\orcid{0000-0002-2960-978X}\inst{\ref{aff66}}
\and F.~Hormuth\inst{\ref{aff67}}
\and A.~Hornstrup\orcid{0000-0002-3363-0936}\inst{\ref{aff68},\ref{aff69}}
\and S.~Ili\'c\orcid{0000-0003-4285-9086}\inst{\ref{aff70},\ref{aff71},\ref{aff6}}
\and K.~Jahnke\orcid{0000-0003-3804-2137}\inst{\ref{aff72}}
\and B.~Joachimi\orcid{0000-0001-7494-1303}\inst{\ref{aff73}}
\and E.~Keih\"anen\orcid{0000-0003-1804-7715}\inst{\ref{aff74}}
\and S.~Kermiche\orcid{0000-0002-0302-5735}\inst{\ref{aff59}}
\and A.~Kiessling\orcid{0000-0002-2590-1273}\inst{\ref{aff65}}
\and M.~Kilbinger\orcid{0000-0001-9513-7138}\inst{\ref{aff58}}
\and T.~Kitching\orcid{0000-0002-4061-4598}\inst{\ref{aff51}}
\and B.~Kubik\inst{\ref{aff46}}
\and M.~K\"ummel\orcid{0000-0003-2791-2117}\inst{\ref{aff29}}
\and M.~Kunz\orcid{0000-0002-3052-7394}\inst{\ref{aff7}}
\and H.~Kurki-Suonio\orcid{0000-0002-4618-3063}\inst{\ref{aff75},\ref{aff76}}
\and S.~Ligori\orcid{0000-0003-4172-4606}\inst{\ref{aff5}}
\and P.~B.~Lilje\orcid{0000-0003-4324-7794}\inst{\ref{aff64}}
\and V.~Lindholm\orcid{0000-0003-2317-5471}\inst{\ref{aff75},\ref{aff76}}
\and I.~Lloro\inst{\ref{aff77}}
\and G.~Mainetti\inst{\ref{aff78}}
\and D.~Maino\inst{\ref{aff62},\ref{aff2},\ref{aff63}}
\and E.~Maiorano\orcid{0000-0003-2593-4355}\inst{\ref{aff21}}
\and O.~Mansutti\orcid{0000-0001-5758-4658}\inst{\ref{aff23}}
\and O.~Marggraf\orcid{0000-0001-7242-3852}\inst{\ref{aff79}}
\and K.~Markovic\orcid{0000-0001-6764-073X}\inst{\ref{aff65}}
\and M.~Martinelli\orcid{0000-0002-6943-7732}\inst{\ref{aff15},\ref{aff16}}
\and N.~Martinet\orcid{0000-0003-2786-7790}\inst{\ref{aff47}}
\and F.~Marulli\orcid{0000-0002-8850-0303}\inst{\ref{aff39},\ref{aff21},\ref{aff27}}
\and R.~Massey\orcid{0000-0002-6085-3780}\inst{\ref{aff80}}
\and H.~J.~McCracken\orcid{0000-0002-9489-7765}\inst{\ref{aff31}}
\and E.~Medinaceli\orcid{0000-0002-4040-7783}\inst{\ref{aff21}}
\and S.~Mei\orcid{0000-0002-2849-559X}\inst{\ref{aff81}}
\and Y.~Mellier\inst{\ref{aff82},\ref{aff31}}
\and M.~Meneghetti\orcid{0000-0003-1225-7084}\inst{\ref{aff21},\ref{aff27}}
\and E.~Merlin\orcid{0000-0001-6870-8900}\inst{\ref{aff15}}
\and G.~Meylan\inst{\ref{aff48}}
\and M.~Moresco\orcid{0000-0002-7616-7136}\inst{\ref{aff39},\ref{aff21}}
\and L.~Moscardini\orcid{0000-0002-3473-6716}\inst{\ref{aff39},\ref{aff21},\ref{aff27}}
\and E.~Munari\orcid{0000-0002-1751-5946}\inst{\ref{aff23}}
\and S.-M.~Niemi\inst{\ref{aff83}}
\and J.~W.~Nightingale\orcid{0000-0002-8987-7401}\inst{\ref{aff84},\ref{aff80}}
\and C.~Padilla\orcid{0000-0001-7951-0166}\inst{\ref{aff36}}
\and S.~Paltani\inst{\ref{aff54}}
\and F.~Pasian\inst{\ref{aff23}}
\and K.~Pedersen\inst{\ref{aff85}}
\and W.~J.~Percival\orcid{0000-0002-0644-5727}\inst{\ref{aff86},\ref{aff87},\ref{aff88}}
\and V.~Pettorino\inst{\ref{aff58}}
\and S.~Pires\orcid{0000-0002-0249-2104}\inst{\ref{aff58}}
\and G.~Polenta\orcid{0000-0003-4067-9196}\inst{\ref{aff89}}
\and M.~Poncet\inst{\ref{aff71}}
\and L.~A.~Popa\inst{\ref{aff90}}
\and L.~Pozzetti\orcid{0000-0001-7085-0412}\inst{\ref{aff21}}
\and F.~Raison\orcid{0000-0002-7819-6918}\inst{\ref{aff28}}
\and R.~Rebolo\inst{\ref{aff41},\ref{aff91}}
\and A.~Renzi\orcid{0000-0001-9856-1970}\inst{\ref{aff92},\ref{aff57}}
\and J.~Rhodes\inst{\ref{aff65}}
\and G.~Riccio\inst{\ref{aff33}}
\and E.~Romelli\orcid{0000-0003-3069-9222}\inst{\ref{aff23}}
\and M.~Roncarelli\orcid{0000-0001-9587-7822}\inst{\ref{aff21}}
\and E.~Rossetti\inst{\ref{aff26}}
\and R.~Saglia\orcid{0000-0003-0378-7032}\inst{\ref{aff29},\ref{aff28}}
\and B.~Sartoris\inst{\ref{aff29},\ref{aff23}}
\and P.~Schneider\orcid{0000-0001-8561-2679}\inst{\ref{aff79}}
\and T.~Schrabback\orcid{0000-0002-6987-7834}\inst{\ref{aff93}}
\and M.~Scodeggio\inst{\ref{aff2}}
\and A.~Secroun\orcid{0000-0003-0505-3710}\inst{\ref{aff59}}
\and G.~Seidel\orcid{0000-0003-2907-353X}\inst{\ref{aff72}}
\and S.~Serrano\orcid{0000-0002-0211-2861}\inst{\ref{aff9},\ref{aff8},\ref{aff94}}
\and C.~Sirignano\orcid{0000-0002-0995-7146}\inst{\ref{aff92},\ref{aff57}}
\and G.~Sirri\orcid{0000-0003-2626-2853}\inst{\ref{aff27}}
\and L.~Stanco\orcid{0000-0002-9706-5104}\inst{\ref{aff57}}
\and J.-L.~Starck\orcid{0000-0003-2177-7794}\inst{\ref{aff58}}
\and J.~Steinwagner\inst{\ref{aff28}}
\and C.~Surace\inst{\ref{aff47}}
\and P.~Tallada-Cresp\'{i}\orcid{0000-0002-1336-8328}\inst{\ref{aff95},\ref{aff37}}
\and D.~Tavagnacco\orcid{0000-0001-7475-9894}\inst{\ref{aff23}}
\and A.~N.~Taylor\inst{\ref{aff42}}
\and I.~Tereno\inst{\ref{aff52},\ref{aff96}}
\and R.~Toledo-Moreo\orcid{0000-0002-2997-4859}\inst{\ref{aff97}}
\and F.~Torradeflot\inst{\ref{aff37},\ref{aff95}}
\and E.~A.~Valentijn\inst{\ref{aff98}}
\and L.~Valenziano\orcid{0000-0002-1170-0104}\inst{\ref{aff21},\ref{aff99}}
\and T.~Vassallo\orcid{0000-0001-6512-6358}\inst{\ref{aff29},\ref{aff23}}
\and A.~Veropalumbo\orcid{0000-0003-2387-1194}\inst{\ref{aff13},\ref{aff10}}
\and Y.~Wang\orcid{0000-0002-4749-2984}\inst{\ref{aff100}}
\and J.~Weller\orcid{0000-0002-8282-2010}\inst{\ref{aff29},\ref{aff28}}
\and A.~Zacchei\orcid{0000-0003-0396-1192}\inst{\ref{aff23},\ref{aff25}}
\and G.~Zamorani\orcid{0000-0002-2318-301X}\inst{\ref{aff21}}
\and J.~Zoubian\inst{\ref{aff59}}
\and E.~Zucca\orcid{0000-0002-5845-8132}\inst{\ref{aff21}}
\and A.~Biviano\inst{\ref{aff23},\ref{aff25}}
\and A.~Boucaud\orcid{0000-0001-7387-2633}\inst{\ref{aff81}}
\and E.~Bozzo\orcid{0000-0002-8201-1525}\inst{\ref{aff54}}
\and C.~Burigana\orcid{0000-0002-3005-5796}\inst{\ref{aff101},\ref{aff99}}
\and M.~Calabrese\orcid{0000-0002-2637-2422}\inst{\ref{aff102},\ref{aff2}}
\and D.~Di~Ferdinando\inst{\ref{aff27}}
\and G.~Fabbian\orcid{0000-0002-3255-4695}\inst{\ref{aff103},\ref{aff104}}
\and R.~Farinelli\inst{\ref{aff21}}
\and J.~Graci\'{a}-Carpio\inst{\ref{aff28}}
\and N.~Mauri\orcid{0000-0001-8196-1548}\inst{\ref{aff40},\ref{aff27}}
\and V.~Scottez\inst{\ref{aff82},\ref{aff105}}
\and M.~Tenti\orcid{0000-0002-4254-5901}\inst{\ref{aff27}}
\and M.~Viel\orcid{0000-0002-2642-5707}\inst{\ref{aff25},\ref{aff23},\ref{aff22},\ref{aff24},\ref{aff106}}
\and M.~Wiesmann\orcid{0009-0000-8199-5860}\inst{\ref{aff64}}
\and Y.~Akrami\orcid{0000-0002-2407-7956}\inst{\ref{aff107},\ref{aff108}}
\and V.~Allevato\inst{\ref{aff33}}
\and S.~Anselmi\orcid{0000-0002-3579-9583}\inst{\ref{aff57},\ref{aff92},\ref{aff109}}
\and M.~Ballardini\orcid{0000-0003-4481-3559}\inst{\ref{aff110},\ref{aff111},\ref{aff21}}
\and A.~Blanchard\orcid{0000-0001-8555-9003}\inst{\ref{aff6}}
\and S.~Borgani\orcid{0000-0001-6151-6439}\inst{\ref{aff112},\ref{aff25},\ref{aff23},\ref{aff24}}
\and S.~Bruton\orcid{0000-0002-6503-5218}\inst{\ref{aff113}}
\and R.~Cabanac\orcid{0000-0001-6679-2600}\inst{\ref{aff6}}
\and A.~Calabro\orcid{0000-0003-2536-1614}\inst{\ref{aff15}}
\and A.~Cappi\inst{\ref{aff21},\ref{aff114}}
\and C.~S.~Carvalho\inst{\ref{aff96}}
\and T.~Castro\orcid{0000-0002-6292-3228}\inst{\ref{aff23},\ref{aff24},\ref{aff25},\ref{aff106}}
\and G.~Ca\~{n}as-Herrera\orcid{0000-0003-2796-2149}\inst{\ref{aff83},\ref{aff115}}
\and K.~C.~Chambers\orcid{0000-0001-6965-7789}\inst{\ref{aff116}}
\and S.~Contarini\inst{\ref{aff39},\ref{aff27},\ref{aff21}}
\and A.~R.~Cooray\orcid{0000-0002-3892-0190}\inst{\ref{aff117}}
\and J.~Coupon\inst{\ref{aff54}}
\and G.~Desprez\inst{\ref{aff118}}
\and H.~Dole\orcid{0000-0002-9767-3839}\inst{\ref{aff119}}
\and A.~D\'{i}az-S\'{a}nchez\orcid{0000-0003-0748-4768}\inst{\ref{aff120}}
\and J.~A.~Escartin~Vigo\inst{\ref{aff28}}
\and S.~Escoffier\orcid{0000-0002-2847-7498}\inst{\ref{aff59}}
\and P.~G.~Ferreira\inst{\ref{aff56}}
\and I.~Ferrero\orcid{0000-0002-1295-1132}\inst{\ref{aff64}}
\and F.~Finelli\inst{\ref{aff21},\ref{aff99}}
\and F.~Fornari\orcid{0000-0003-2979-6738}\inst{\ref{aff99}}
\and L.~Gabarra\inst{\ref{aff56}}
\and K.~Ganga\orcid{0000-0001-8159-8208}\inst{\ref{aff81}}
\and J.~Garc\'ia-Bellido\orcid{0000-0002-9370-8360}\inst{\ref{aff107}}
\and E.~Gaztanaga\orcid{0000-0001-9632-0815}\inst{\ref{aff8},\ref{aff9},\ref{aff121}}
\and F.~Giacomini\orcid{0000-0002-3129-2814}\inst{\ref{aff27}}
\and G.~Gozaliasl\orcid{0000-0002-0236-919X}\inst{\ref{aff122},\ref{aff75}}
\and A.~Gregorio\orcid{0000-0003-4028-8785}\inst{\ref{aff112},\ref{aff23},\ref{aff24}}
\and A.~Hall\orcid{0000-0002-3139-8651}\inst{\ref{aff42}}
\and H.~Hildebrandt\orcid{0000-0002-9814-3338}\inst{\ref{aff123}}
\and J.~Hjorth\orcid{0000-0002-4571-2306}\inst{\ref{aff124}}
\and J.~J.~E.~Kajava\orcid{0000-0002-3010-8333}\inst{\ref{aff125},\ref{aff126}}
\and V.~Kansal\inst{\ref{aff127},\ref{aff128},\ref{aff129}}
\and D.~Karagiannis\orcid{0000-0002-4927-0816}\inst{\ref{aff130},\ref{aff131}}
\and C.~C.~Kirkpatrick\inst{\ref{aff74}}
\and L.~Legrand\orcid{0000-0003-0610-5252}\inst{\ref{aff7},\ref{aff132}}
\and A.~Loureiro\orcid{0000-0002-4371-0876}\inst{\ref{aff133},\ref{aff134}}
\and J.~Macias-Perez\orcid{0000-0002-5385-2763}\inst{\ref{aff135}}
\and G.~Maggio\orcid{0000-0003-4020-4836}\inst{\ref{aff23}}
\and M.~Magliocchetti\orcid{0000-0001-9158-4838}\inst{\ref{aff55}}
\and F.~Mannucci\orcid{0000-0002-4803-2381}\inst{\ref{aff136}}
\and R.~Maoli\orcid{0000-0002-6065-3025}\inst{\ref{aff14},\ref{aff15}}
\and C.~J.~A.~P.~Martins\orcid{0000-0002-4886-9261}\inst{\ref{aff137},\ref{aff35}}
\and S.~Matthew\inst{\ref{aff42}}
\and L.~Maurin\orcid{0000-0002-8406-0857}\inst{\ref{aff119}}
\and R.~B.~Metcalf\orcid{0000-0003-3167-2574}\inst{\ref{aff39},\ref{aff21}}
\and M.~Migliaccio\inst{\ref{aff138},\ref{aff139}}
\and P.~Monaco\orcid{0000-0003-2083-7564}\inst{\ref{aff112},\ref{aff23},\ref{aff24},\ref{aff25}}
\and G.~Morgante\inst{\ref{aff21}}
\and S.~Nadathur\orcid{0000-0001-9070-3102}\inst{\ref{aff121}}
\and L.~Patrizii\inst{\ref{aff27}}
\and A.~Pezzotta\orcid{0000-0003-0726-2268}\inst{\ref{aff28}}
\and V.~Popa\inst{\ref{aff90}}
\and C.~Porciani\orcid{0000-0002-7797-2508}\inst{\ref{aff79}}
\and D.~Potter\orcid{0000-0002-0757-5195}\inst{\ref{aff140}}
\and M.~P\"{o}ntinen\orcid{0000-0001-5442-2530}\inst{\ref{aff75}}
\and P.-F.~Rocci\inst{\ref{aff119}}
\and M.~Sahl\'en\orcid{0000-0003-0973-4804}\inst{\ref{aff141}}
\and A.~Schneider\orcid{0000-0001-7055-8104}\inst{\ref{aff140}}
\and M.~Schultheis\inst{\ref{aff114}}
\and M.~Sereno\orcid{0000-0003-0302-0325}\inst{\ref{aff21},\ref{aff27}}
\and C.~Tao\orcid{0000-0001-7961-8177}\inst{\ref{aff59}}
\and N.~Tessore\orcid{0000-0002-9696-7931}\inst{\ref{aff73}}
\and R.~Teyssier\orcid{0000-0001-7689-0933}\inst{\ref{aff142}}
\and S.~Toft\inst{\ref{aff69},\ref{aff143},\ref{aff144}}
\and A.~Troja\orcid{0000-0003-0239-4595}\inst{\ref{aff92},\ref{aff57}}
\and M.~Tucci\inst{\ref{aff54}}
\and C.~Valieri\inst{\ref{aff27}}
\and J.~Valiviita\orcid{0000-0001-6225-3693}\inst{\ref{aff75},\ref{aff76}}
\and D.~Vergani\orcid{0000-0003-0898-2216}\inst{\ref{aff21}}
\and G.~Verza\orcid{0000-0002-1886-8348}\inst{\ref{aff145},\ref{aff103}}
\and P.~Vielzeuf\orcid{0000-0003-2035-9339}\inst{\ref{aff59}}}
										   
\institute{Dipartimento di Fisica, Universit\`a degli studi di Genova, and INFN-Sezione di Genova, via Dodecaneso 33, 16146, Genova, Italy\label{aff1}
\and
INAF-IASF Milano, Via Alfonso Corti 12, 20133 Milano, Italy\label{aff2}
\and
Dipartimento di Fisica, Universit\`a degli Studi di Torino, Via P. Giuria 1, 10125 Torino, Italy\label{aff3}
\and
INFN-Sezione di Torino, Via P. Giuria 1, 10125 Torino, Italy\label{aff4}
\and
INAF-Osservatorio Astrofisico di Torino, Via Osservatorio 20, 10025 Pino Torinese (TO), Italy\label{aff5}
\and
Institut de Recherche en Astrophysique et Plan\'etologie (IRAP), Universit\'e de Toulouse, CNRS, UPS, CNES, 14 Av. Edouard Belin, 31400 Toulouse, France\label{aff6}
\and
Universit\'e de Gen\`eve, D\'epartement de Physique Th\'eorique and Centre for Astroparticle Physics, 24 quai Ernest-Ansermet, CH-1211 Gen\`eve 4, Switzerland\label{aff7}
\and
Institute of Space Sciences (ICE, CSIC), Campus UAB, Carrer de Can Magrans, s/n, 08193 Barcelona, Spain\label{aff8}
\and
Institut d'Estudis Espacials de Catalunya (IEEC),  Edifici RDIT, Campus UPC, 08860 Castelldefels, Barcelona, Spain\label{aff9}
\and
INFN-Sezione di Genova, Via Dodecaneso 33, 16146, Genova, Italy\label{aff10}
\and
Aix-Marseille Universit\'e, Universit\'e de Toulon, CNRS, CPT, Marseille, France\label{aff11}
\and
Dipartimento di Fisica, Universit\`a di Genova, Via Dodecaneso 33, 16146, Genova, Italy\label{aff12}
\and
INAF-Osservatorio Astronomico di Brera, Via Brera 28, 20122 Milano, Italy\label{aff13}
\and
Dipartimento di Fisica, Sapienza Universit\`a di Roma, Piazzale Aldo Moro 2, 00185 Roma, Italy\label{aff14}
\and
INAF-Osservatorio Astronomico di Roma, Via Frascati 33, 00078 Monteporzio Catone, Italy\label{aff15}
\and
INFN-Sezione di Roma, Piazzale Aldo Moro, 2 - c/o Dipartimento di Fisica, Edificio G. Marconi, 00185 Roma, Italy\label{aff16}
\and
Departamento de F\'isica, FCFM, Universidad de Chile, Blanco Encalada 2008, Santiago, Chile\label{aff17}
\and
Institut f\"ur Theoretische Physik, University of Heidelberg, Philosophenweg 16, 69120 Heidelberg, Germany\label{aff18}
\and
Universit\'e St Joseph; Faculty of Sciences, Beirut, Lebanon\label{aff19}
\and
School of Mathematics and Physics, University of Surrey, Guildford, Surrey, GU2 7XH, UK\label{aff20}
\and
INAF-Osservatorio di Astrofisica e Scienza dello Spazio di Bologna, Via Piero Gobetti 93/3, 40129 Bologna, Italy\label{aff21}
\and
SISSA, International School for Advanced Studies, Via Bonomea 265, 34136 Trieste TS, Italy\label{aff22}
\and
INAF-Osservatorio Astronomico di Trieste, Via G. B. Tiepolo 11, 34143 Trieste, Italy\label{aff23}
\and
INFN, Sezione di Trieste, Via Valerio 2, 34127 Trieste TS, Italy\label{aff24}
\and
IFPU, Institute for Fundamental Physics of the Universe, via Beirut 2, 34151 Trieste, Italy\label{aff25}
\and
Dipartimento di Fisica e Astronomia, Universit\`a di Bologna, Via Gobetti 93/2, 40129 Bologna, Italy\label{aff26}
\and
INFN-Sezione di Bologna, Viale Berti Pichat 6/2, 40127 Bologna, Italy\label{aff27}
\and
Max Planck Institute for Extraterrestrial Physics, Giessenbachstr. 1, 85748 Garching, Germany\label{aff28}
\and
Universit\"ats-Sternwarte M\"unchen, Fakult\"at f\"ur Physik, Ludwig-Maximilians-Universit\"at M\"unchen, Scheinerstrasse 1, 81679 M\"unchen, Germany\label{aff29}
\and
Institut de Physique Th\'eorique, CEA, CNRS, Universit\'e Paris-Saclay 91191 Gif-sur-Yvette Cedex, France\label{aff30}
\and
Institut d'Astrophysique de Paris, UMR 7095, CNRS, and Sorbonne Universit\'e, 98 bis boulevard Arago, 75014 Paris, France\label{aff31}
\and
Department of Physics "E. Pancini", University Federico II, Via Cinthia 6, 80126, Napoli, Italy\label{aff32}
\and
INAF-Osservatorio Astronomico di Capodimonte, Via Moiariello 16, 80131 Napoli, Italy\label{aff33}
\and
INFN section of Naples, Via Cinthia 6, 80126, Napoli, Italy\label{aff34}
\and
Instituto de Astrof\'isica e Ci\^encias do Espa\c{c}o, Universidade do Porto, CAUP, Rua das Estrelas, PT4150-762 Porto, Portugal\label{aff35}
\and
Institut de F\'{i}sica d'Altes Energies (IFAE), The Barcelona Institute of Science and Technology, Campus UAB, 08193 Bellaterra (Barcelona), Spain\label{aff36}
\and
Port d'Informaci\'{o} Cient\'{i}fica, Campus UAB, C. Albareda s/n, 08193 Bellaterra (Barcelona), Spain\label{aff37}
\and
Institute for Theoretical Particle Physics and Cosmology (TTK), RWTH Aachen University, 52056 Aachen, Germany\label{aff38}
\and
Dipartimento di Fisica e Astronomia "Augusto Righi" - Alma Mater Studiorum Universit\`a di Bologna, via Piero Gobetti 93/2, 40129 Bologna, Italy\label{aff39}
\and
Dipartimento di Fisica e Astronomia "Augusto Righi" - Alma Mater Studiorum Universit\`a di Bologna, Viale Berti Pichat 6/2, 40127 Bologna, Italy\label{aff40}
\and
Instituto de Astrof\'isica de Canarias, Calle V\'ia L\'actea s/n, 38204, San Crist\'obal de La Laguna, Tenerife, Spain\label{aff41}
\and
Institute for Astronomy, University of Edinburgh, Royal Observatory, Blackford Hill, Edinburgh EH9 3HJ, UK\label{aff42}
\and
Jodrell Bank Centre for Astrophysics, Department of Physics and Astronomy, University of Manchester, Oxford Road, Manchester M13 9PL, UK\label{aff43}
\and
European Space Agency/ESRIN, Largo Galileo Galilei 1, 00044 Frascati, Roma, Italy\label{aff44}
\and
ESAC/ESA, Camino Bajo del Castillo, s/n., Urb. Villafranca del Castillo, 28692 Villanueva de la Ca\~nada, Madrid, Spain\label{aff45}
\and
Universit\'e Claude Bernard Lyon 1, CNRS/IN2P3, IP2I Lyon, UMR 5822, Villeurbanne, F-69100, France\label{aff46}
\and
Aix-Marseille Universit\'e, CNRS, CNES, LAM, Marseille, France\label{aff47}
\and
Institute of Physics, Laboratory of Astrophysics, Ecole Polytechnique F\'ed\'erale de Lausanne (EPFL), Observatoire de Sauverny, 1290 Versoix, Switzerland\label{aff48}
\and
UCB Lyon 1, CNRS/IN2P3, IUF, IP2I Lyon, 4 rue Enrico Fermi, 69622 Villeurbanne, France\label{aff49}
\and
Institut de Ciencies de l'Espai (IEEC-CSIC), Campus UAB, Carrer de Can Magrans, s/n Cerdanyola del Vall\'es, 08193 Barcelona, Spain\label{aff50}
\and
Mullard Space Science Laboratory, University College London, Holmbury St Mary, Dorking, Surrey RH5 6NT, UK\label{aff51}
\and
Departamento de F\'isica, Faculdade de Ci\^encias, Universidade de Lisboa, Edif\'icio C8, Campo Grande, PT1749-016 Lisboa, Portugal\label{aff52}
\and
Instituto de Astrof\'isica e Ci\^encias do Espa\c{c}o, Faculdade de Ci\^encias, Universidade de Lisboa, Campo Grande, 1749-016 Lisboa, Portugal\label{aff53}
\and
Department of Astronomy, University of Geneva, ch. d'Ecogia 16, 1290 Versoix, Switzerland\label{aff54}
\and
INAF-Istituto di Astrofisica e Planetologia Spaziali, via del Fosso del Cavaliere, 100, 00100 Roma, Italy\label{aff55}
\and
Department of Physics, Oxford University, Keble Road, Oxford OX1 3RH, UK\label{aff56}
\and
INFN-Padova, Via Marzolo 8, 35131 Padova, Italy\label{aff57}
\and
Universit\'e Paris-Saclay, Universit\'e Paris Cit\'e, CEA, CNRS, AIM, 91191, Gif-sur-Yvette, France\label{aff58}
\and
Aix-Marseille Universit\'e, CNRS/IN2P3, CPPM, Marseille, France\label{aff59}
\and
Istituto Nazionale di Fisica Nucleare, Sezione di Bologna, Via Irnerio 46, 40126 Bologna, Italy\label{aff60}
\and
INAF-Osservatorio Astronomico di Padova, Via dell'Osservatorio 5, 35122 Padova, Italy\label{aff61}
\and
Dipartimento di Fisica "Aldo Pontremoli", Universit\`a degli Studi di Milano, Via Celoria 16, 20133 Milano, Italy\label{aff62}
\and
INFN-Sezione di Milano, Via Celoria 16, 20133 Milano, Italy\label{aff63}
\and
Institute of Theoretical Astrophysics, University of Oslo, P.O. Box 1029 Blindern, 0315 Oslo, Norway\label{aff64}
\and
Jet Propulsion Laboratory, California Institute of Technology, 4800 Oak Grove Drive, Pasadena, CA, 91109, USA\label{aff65}
\and
Department of Physics, Lancaster University, Lancaster, LA1 4YB, UK\label{aff66}
\and
von Hoerner \& Sulger GmbH, Schlossplatz 8, 68723 Schwetzingen, Germany\label{aff67}
\and
Technical University of Denmark, Elektrovej 327, 2800 Kgs. Lyngby, Denmark\label{aff68}
\and
Cosmic Dawn Center (DAWN), Denmark\label{aff69}
\and
Universit\'e Paris-Saclay, CNRS/IN2P3, IJCLab, 91405 Orsay, France\label{aff70}
\and
Centre National d'Etudes Spatiales -- Centre spatial de Toulouse, 18 avenue Edouard Belin, 31401 Toulouse Cedex 9, France\label{aff71}
\and
Max-Planck-Institut f\"ur Astronomie, K\"onigstuhl 17, 69117 Heidelberg, Germany\label{aff72}
\and
Department of Physics and Astronomy, University College London, Gower Street, London WC1E 6BT, UK\label{aff73}
\and
Department of Physics and Helsinki Institute of Physics, Gustaf H\"allstr\"omin katu 2, 00014 University of Helsinki, Finland\label{aff74}
\and
Department of Physics, P.O. Box 64, 00014 University of Helsinki, Finland\label{aff75}
\and
Helsinki Institute of Physics, Gustaf H{\"a}llstr{\"o}min katu 2, University of Helsinki, Helsinki, Finland\label{aff76}
\and
NOVA optical infrared instrumentation group at ASTRON, Oude Hoogeveensedijk 4, 7991PD, Dwingeloo, The Netherlands\label{aff77}
\and
Centre de Calcul de l'IN2P3/CNRS, 21 avenue Pierre de Coubertin 69627 Villeurbanne Cedex, France\label{aff78}
\and
Universit\"at Bonn, Argelander-Institut f\"ur Astronomie, Auf dem H\"ugel 71, 53121 Bonn, Germany\label{aff79}
\and
Department of Physics, Institute for Computational Cosmology, Durham University, South Road, DH1 3LE, UK\label{aff80}
\and
Universit\'e Paris Cit\'e, CNRS, Astroparticule et Cosmologie, 75013 Paris, France\label{aff81}
\and
Institut d'Astrophysique de Paris, 98bis Boulevard Arago, 75014, Paris, France\label{aff82}
\and
European Space Agency/ESTEC, Keplerlaan 1, 2201 AZ Noordwijk, The Netherlands\label{aff83}
\and
School of Mathematics, Statistics and Physics, Newcastle University, Herschel Building, Newcastle-upon-Tyne, NE1 7RU, UK\label{aff84}
\and
Department of Physics and Astronomy, University of Aarhus, Ny Munkegade 120, DK-8000 Aarhus C, Denmark\label{aff85}
\and
Waterloo Centre for Astrophysics, University of Waterloo, Waterloo, Ontario N2L 3G1, Canada\label{aff86}
\and
Department of Physics and Astronomy, University of Waterloo, Waterloo, Ontario N2L 3G1, Canada\label{aff87}
\and
Perimeter Institute for Theoretical Physics, Waterloo, Ontario N2L 2Y5, Canada\label{aff88}
\and
Space Science Data Center, Italian Space Agency, via del Politecnico snc, 00133 Roma, Italy\label{aff89}
\and
Institute of Space Science, Str. Atomistilor, nr. 409 M\u{a}gurele, Ilfov, 077125, Romania\label{aff90}
\and
Departamento de Astrof\'isica, Universidad de La Laguna, 38206, La Laguna, Tenerife, Spain\label{aff91}
\and
Dipartimento di Fisica e Astronomia "G. Galilei", Universit\`a di Padova, Via Marzolo 8, 35131 Padova, Italy\label{aff92}
\and
Universit\"at Innsbruck, Institut f\"ur Astro- und Teilchenphysik, Technikerstr. 25/8, 6020 Innsbruck, Austria\label{aff93}
\and
Satlantis, University Science Park, Sede Bld 48940, Leioa-Bilbao, Spain\label{aff94}
\and
Centro de Investigaciones Energ\'eticas, Medioambientales y Tecnol\'ogicas (CIEMAT), Avenida Complutense 40, 28040 Madrid, Spain\label{aff95}
\and
Instituto de Astrof\'isica e Ci\^encias do Espa\c{c}o, Faculdade de Ci\^encias, Universidade de Lisboa, Tapada da Ajuda, 1349-018 Lisboa, Portugal\label{aff96}
\and
Universidad Polit\'ecnica de Cartagena, Departamento de Electr\'onica y Tecnolog\'ia de Computadoras,  Plaza del Hospital 1, 30202 Cartagena, Spain\label{aff97}
\and
Kapteyn Astronomical Institute, University of Groningen, PO Box 800, 9700 AV Groningen, The Netherlands\label{aff98}
\and
INFN-Bologna, Via Irnerio 46, 40126 Bologna, Italy\label{aff99}
\and
Infrared Processing and Analysis Center, California Institute of Technology, Pasadena, CA 91125, USA\label{aff100}
\and
INAF, Istituto di Radioastronomia, Via Piero Gobetti 101, 40129 Bologna, Italy\label{aff101}
\and
Astronomical Observatory of the Autonomous Region of the Aosta Valley (OAVdA), Loc. Lignan 39, I-11020, Nus (Aosta Valley), Italy\label{aff102}
\and
Center for Computational Astrophysics, Flatiron Institute, 162 5th Avenue, 10010, New York, NY, USA\label{aff103}
\and
School of Physics and Astronomy, Cardiff University, The Parade, Cardiff, CF24 3AA, UK\label{aff104}
\and
Junia, EPA department, 41 Bd Vauban, 59800 Lille, France\label{aff105}
\and
ICSC - Centro Nazionale di Ricerca in High Performance Computing, Big Data e Quantum Computing, Via Magnanelli 2, Bologna, Italy\label{aff106}
\and
Instituto de F\'isica Te\'orica UAM-CSIC, Campus de Cantoblanco, 28049 Madrid, Spain\label{aff107}
\and
CERCA/ISO, Department of Physics, Case Western Reserve University, 10900 Euclid Avenue, Cleveland, OH 44106, USA\label{aff108}
\and
Laboratoire Univers et Th\'eorie, Observatoire de Paris, Universit\'e PSL, Universit\'e Paris Cit\'e, CNRS, 92190 Meudon, France\label{aff109}
\and
Dipartimento di Fisica e Scienze della Terra, Universit\`a degli Studi di Ferrara, Via Giuseppe Saragat 1, 44122 Ferrara, Italy\label{aff110}
\and
Istituto Nazionale di Fisica Nucleare, Sezione di Ferrara, Via Giuseppe Saragat 1, 44122 Ferrara, Italy\label{aff111}
\and
Dipartimento di Fisica - Sezione di Astronomia, Universit\`a di Trieste, Via Tiepolo 11, 34131 Trieste, Italy\label{aff112}
\and
Minnesota Institute for Astrophysics, University of Minnesota, 116 Church St SE, Minneapolis, MN 55455, USA\label{aff113}
\and
Universit\'e C\^{o}te d'Azur, Observatoire de la C\^{o}te d'Azur, CNRS, Laboratoire Lagrange, Bd de l'Observatoire, CS 34229, 06304 Nice cedex 4, France\label{aff114}
\and
Institute Lorentz, Leiden University, Niels Bohrweg 2, 2333 CA Leiden, The Netherlands\label{aff115}
\and
Institute for Astronomy, University of Hawaii, 2680 Woodlawn Drive, Honolulu, HI 96822, USA\label{aff116}
\and
Department of Physics \& Astronomy, University of California Irvine, Irvine CA 92697, USA\label{aff117}
\and
Department of Astronomy \& Physics and Institute for Computational Astrophysics, Saint Mary's University, 923 Robie Street, Halifax, Nova Scotia, B3H 3C3, Canada\label{aff118}
\and
Universit\'e Paris-Saclay, CNRS, Institut d'astrophysique spatiale, 91405, Orsay, France\label{aff119}
\and
Departamento F\'isica Aplicada, Universidad Polit\'ecnica de Cartagena, Campus Muralla del Mar, 30202 Cartagena, Murcia, Spain\label{aff120}
\and
Institute of Cosmology and Gravitation, University of Portsmouth, Portsmouth PO1 3FX, UK\label{aff121}
\and
Department of Computer Science, Aalto University, PO Box 15400, Espoo, FI-00 076, Finland\label{aff122}
\and
Ruhr University Bochum, Faculty of Physics and Astronomy, Astronomical Institute (AIRUB), German Centre for Cosmological Lensing (GCCL), 44780 Bochum, Germany\label{aff123}
\and
DARK, Niels Bohr Institute, University of Copenhagen, Jagtvej 155, 2200 Copenhagen, Denmark\label{aff124}
\and
Department of Physics and Astronomy, Vesilinnantie 5, 20014 University of Turku, Finland\label{aff125}
\and
Serco for European Space Agency (ESA), Camino bajo del Castillo, s/n, Urbanizacion Villafranca del Castillo, Villanueva de la Ca\~nada, 28692 Madrid, Spain\label{aff126}
\and
ARC Centre of Excellence for Dark Matter Particle Physics, Melbourne, Australia\label{aff127}
\and
Centre for Astrophysics \& Supercomputing, Swinburne University of Technology,  Hawthorn, Victoria 3122, Australia\label{aff128}
\and
W.M. Keck Observatory, 65-1120 Mamalahoa Hwy, Kamuela, HI, USA\label{aff129}
\and
School of Physics and Astronomy, Queen Mary University of London, Mile End Road, London E1 4NS, UK\label{aff130}
\and
Department of Physics and Astronomy, University of the Western Cape, Bellville, Cape Town, 7535, South Africa\label{aff131}
\and
ICTP South American Institute for Fundamental Research, Instituto de F\'{\i}sica Te\'orica, Universidade Estadual Paulista, S\~ao Paulo, Brazil\label{aff132}
\and
Oskar Klein Centre for Cosmoparticle Physics, Department of Physics, Stockholm University, Stockholm, SE-106 91, Sweden\label{aff133}
\and
Astrophysics Group, Blackett Laboratory, Imperial College London, London SW7 2AZ, UK\label{aff134}
\and
Univ. Grenoble Alpes, CNRS, Grenoble INP, LPSC-IN2P3, 53, Avenue des Martyrs, 38000, Grenoble, France\label{aff135}
\and
INAF-Osservatorio Astrofisico di Arcetri, Largo E. Fermi 5, 50125, Firenze, Italy\label{aff136}
\and
Centro de Astrof\'{\i}sica da Universidade do Porto, Rua das Estrelas, 4150-762 Porto, Portugal\label{aff137}
\and
Dipartimento di Fisica, Universit\`a di Roma Tor Vergata, Via della Ricerca Scientifica 1, Roma, Italy\label{aff138}
\and
INFN, Sezione di Roma 2, Via della Ricerca Scientifica 1, Roma, Italy\label{aff139}
\and
Department of Astrophysics, University of Zurich, Winterthurerstrasse 190, 8057 Zurich, Switzerland\label{aff140}
\and
Theoretical astrophysics, Department of Physics and Astronomy, Uppsala University, Box 515, 751 20 Uppsala, Sweden\label{aff141}
\and
Department of Astrophysical Sciences, Peyton Hall, Princeton University, Princeton, NJ 08544, USA\label{aff142}
\and
Niels Bohr Institute, University of Copenhagen, Jagtvej 128, 2200 Copenhagen, Denmark\label{aff143}
\and
Cosmic Dawn Center (DAWN)\label{aff144}
\and
Center for Cosmology and Particle Physics, Department of Physics, New York University, New York, NY 10003, USA\label{aff145}}

\date{\today}

\authorrunning{Euclid Collaboration: L.\ Paganin et al.}

\titlerunning{Euclid preparation: \sixtwoptShort\ analysis of \Euclid's spectro and photo data sets}

 
  \abstract
   {}
   {In this paper we present cosmological parameter forecasts for the so-called \Euclid \sixtwoptShort\ statistics, which include the galaxy clustering and weak lensing main probes \emph{together with} previously neglected cross-covariance and cross-correlation signals between imaging/photometric and spectroscopic data. The aim is understanding the impact of such cross-terms on the expected \Euclid\ performance.}
   {We adopt the Fisher information matrix approach to produce \sixtwoptShort\ cosmological forecasts from \Euclid, considering two different techniques: the so-called \textit{harmonic} and \textit{hybrid} approaches, respectively. In the first, we treat all the different \Euclid\ probes in the same way, i.e. we consider only angular 2pt-statistics for spectroscopic and photometric galaxy distributions, as well as for weak lensing, fully analysing all their possible cross-covariances and cross-correlations in the spherical harmonic domain. In the second, thanks to lessons learnt from the harmonic approach, we do not account for negligible cross-covariances between the 3D spectroscopic galaxy distribution and the 2D photometric/imaging data, but consider the combination of their cross-correlation with the auto-correlation signals.}
   {We find that both cross-covariances and cross-correlation signals between the two \Euclid\ main probes, i.e. the spectroscopic galaxy sample and the photometric/imaging data, have a negligible impact on the cosmological parameter constraints and, therefore, on the \Euclid\ performance. In the case of the hybrid approach, we attribute this result to the effect of the cross-correlation between weak lensing and photometric data, which is dominant with respect to other cross-correlation signals, and to the better performance of the full anisotropic 3D spectroscopic galaxy clustering with respect to the projected one. In the case of the 2D harmonic approach, we attribute this result to two main theoretical limitations of the 2D projected statistics implemented in this work according to the analysis of \cite{2020A&A...642A.191E}: the high shot noise and the limited redshift range of the spectroscopic sample, with respect to the photometric one, together with the suppression of radial information from subdominant contributions such as redshift-space distortions and lensing magnification. Therefore, our analysis suggests that 2D and 3D \Euclid\ data can be safely treated as independent, with a great saving in computational resources.} 
   {}


   \maketitle
   
\section{Introduction}
\label{sec:introduction}

\Euclid is a medium-sized ESA mission devoted to the investigation of the nature of dark matter (DM) and dark energy (DE) and the study of the galaxy formation and evolution~\citep{laureijs2011euclid, Euclid:2024yrr}. The \Euclid satellite was launched on July 1 2023 and will observe about one third of the sky, performing one of the largest galaxy surveys ever made.  It will probe the last 10 billion years of the Universe expansion history via its main cosmological probes which are weak lensing ($\WL$) and galaxy clustering ($\mathrm{GC}$). Through $\WL$ measurements it is possible to probe the matter distribution of the Universe, as $\WL$ represents the slight deformation of galaxy images induced by the gravitational potential produced by such a distribution.
$\mathrm{GC}$ consists in the determination of the statistical properties of the distribution of galaxies, the so-called \textit{dark matter tracers}. In particular, it is characterised by the so-called baryon acoustic oscillations (BAO), whose scale can be adopted as a \textit{standard ruler} and used to constrain the expansion rate of the Universe in different redshift bins.
\Euclid will study these probes with two instruments: the Visible Imager (VIS)~\citep{cropper2016, Euclid:2024yvv} and the Near-Infrared Spectro-Photometer (NISP)~\citep{Maciaszek, Euclid:2024sqd}.  VIS  will provide high resolution images of around $1.5$ billion galaxies for weak lensing measurements. NISP, used in the photometric mode, will allow measurements of the photometric redshifts of the same galaxies observed with VIS, when combined with ground-based photometry. When used in the spectroscopic mode, NISP will measure the spectroscopic redshifts of around $20$ million $\Halpha$-emitting galaxies, with a precision better than a factor of $50$ with respect to the photometric redshift determination. \Euclid will therefore produce two galaxy samples, a photometric and a spectroscopic one. In this sense, the $\mathrm{GC}$ probe can be split into the photometric galaxy clustering ($\GCph$) and the spectroscopic galaxy clustering ($\GCsp$). $\GCsp$ and the so-called \Euclid \threetwoptShort\ statistics (composed by WL, $\GCph$ and their cross-correlation) represent the two main probes of \Euclid.

In this work we present \Euclid cosmological parameter forecasts which include the cross-correlations between $\GCsp$ and \threetwoptShort\ statistics. The data analysis of the survey needs in fact to be accurately planned, and to this aim, pre-data forecasts of the expected scientific performance are needed. In a previous official \Euclid\ forecast \citep[][hereafter `EP-VII']{2020A&A...642A.191E} it has been shown that the cross-correlation (XC) between $\WL$ and $\GCph$ significantly improves the \Euclid\ constraints on cosmological parameters. 
The aim of this work is to extend previous analyses which neglected the cross-correlations between the imaging/photometric probes and the spectroscopic probe, in order to understand their impact on the expected constraints from \Euclid. In particular, we include the cross-correlations between $\GCsp$ and $\GCph$ and between $\GCsp$ and $\WL$, which defines the so-called \Euclid\ \sixtwoptShort\ statistics, forecasting the impact of these cross-correlations on the \Euclid\ performance. \textcolor{black}{Several works in the literature have investigated how to analyze and combine photometric and spectroscopic surveys, studying different approaches to minimising the information loss~\citep{Asorey2012, Eriksen:2014wda, Joudaki:2017zdt,  Camera:2018jys, Loureiro:2018qva, GrasshornGebhardt:2020wsw, Taylor:2021bhg}. Recently, \cite{Taylor:2022rgy} also studied the theoretical modelling of the cross-covariance between photometric and spectroscopic probes and their impact on the forecast of the measurements of cosmological parameters.}


This paper is organised as follows. In \cref{sec:model} we describe in detail the adopted modelling of the $\GCsp$, $\GCph$ and $\WL$ observables. In \cref{sec:fisher} we present the Fisher information matrix approach implemented to produce our \sixtwoptShort\ cosmological parameter forecasts from \Euclid; we adopt two different techniques: the so-called \textit{harmonic} and \textit{hybrid} approaches which we describe in the following. In \cref{sec:results} we present our the results, and finally we draw our conclusions.

\section{Cosmological model and observables}
\label{sec:model}

In this forecast the cosmological model investigated is a flat $\wzwaCDM$ cosmology. The DE equation of state is described by the CPL parametrisation \citep{Linder2002,chevallier2001accelerating}
\begin{equation}
w_{\rm DE}(z) = \wz + \wa\, \frac{z}{1+z}\;,
\label{eq:CPL}
\end{equation}
and the Hubble factor hence takes the form
\begin{equation}
    \frac{H^2(z)}{H_0^2} = \Omega_{\mathrm{m}}\,(1+z)^3+\left(1-\Omega_{\mathrm{m}}\right)\,(1+z)^{3\,\left(1+w_0+w_a\right)}\, \mathrm{e}^{-3\,w_a\,z/(1+z)}\;.
\end{equation}
%
%
%
%
The cosmological parameters involved in the analysis are summarised in \cref{tab:cosmo_pars_fiducials}, where the values in the reference cosmology are reported.
\begin{table*}
\centering
\caption{Values of the cosmological parameters considered of the reference cosmology; $\Omb$ and $\Omm$ refer to present value of the baryon and matter density, respectively. All the parameters are let free to vary except for the sum of the neutrino masses, which has been fixed to its reference value $\mnu = \SI{0.06}{eV}$.}
\begin{tabular}{ccccccccc}
\toprule
\text{Parameter} & $\Omb$ & $\Omm$ & $\wz$ & $\wa$ & $h$ & $\ns$ & $\sige$ & $\mnu[\mathrm{eV}]$ \\
\midrule
\text{Fiducial value} & 0.05 & 0.32 & $-1$ & 0 & 0.67 & 0.96 & 0.816 & 0.06 \\
\bottomrule
\end{tabular}
\label{tab:cosmo_pars_fiducials}
\end{table*}
The probes considered in this forecast are: the weak lensing ($\WL$), the photometric galaxy clustering ($\GCph$), and the spectroscopic galaxy clustering ($\GCsp$). For the $\WL$ and $\GCph$ probes, the observable employed is the tomographic angular power spectrum $C_{ij}(\ell)$. For the $\GCsp$ probe, both the usual power spectrum in Fourier space and the harmonic power spectrum have been considered as observables. The Fourier power spectrum approach is the same followed in \citetalias{2020A&A...642A.191E} and briefly summarised in \cref{sec:gc_sp}. 

An angular power spectrum is essentially the harmonic transform of a two-point angular correlation function.  Each cosmological probe $A$ (in a given redshift bin) can be associated with a field $f^{A}(\hat{n})$ projected on the sky, which can be expanded in the spherical harmonics orthonormal basis $Y_{\ell m}$\footnote{\textcolor{black}{For an all sky analysis, the spherical harmonics are replaced by the spin-spherical harmonics~\citep{Stebbins:1996wx}.}},
\begin{equation}\label{eq:spherical_harmonics_expansion}
f^A(\hat{n}) = 
\sum_{\ell = 0}^{\infty} \sum_{m = -\ell}^{\ell}
a_{\ell m}^{A}\, Y_{\ell m} (\hat{n})\;.
\end{equation}
The $a^{A}_{\ell m}$ are the coefficients of the $f^A$ spherical harmonics transform,
\begin{equation}\label{eq:alm_def}
a_{\ell m}^{A} = 
\int \de\Omega \, Y_{\ell m}^{*}(\hat{n})\, f^A(\hat{n})\;.
\end{equation}
This forecast employs 10 tomographic redshift bins for the $\WL$ and $\GCph$ probes, and 4 bins for the $\GCsp$ probe.
The tomographic angular correlation between the probe $A$, at the redshift bin $i$ ($A_i$), and the probe $B$, at the redshift bin $j$ ($B_j$), is then defined as the implicit relation
\begin{equation}\label{eq:cl_implicit_def}
\left \langle 
a_{\ell m}^{Ai} \,
\left( a_{\ell' m'}^{Bj} \right)^\ast
\right \rangle = \cl{AB}{ij}{\ell} \,\delta_{\ell\ell^{\prime}} \,\delta_{m m^{\prime}}\;,
\end{equation}
where the angular brackets denote the theoretical expectation value.

In this forecast, we compute the angular power spectra following the Limber approximation \citep{1992ApJ...388..272K},
\begin{equation}\label{eq:limber_cls}
\cl{AB}{ij}{\ell} \simeq \int_{z_\mathrm{min}}^{z_\mathrm{max}} \frac{c\,\de z}{H(z)} \, \frac{W_i^{A}(z)\,W_{j}^{B}(z)}{r^{2}(z)}\,P_{\delta\delta}\left[k=\frac{\ell+1/2}{r(z)},z\right]\;,
\end{equation}
where $r(z)$ is the radial comoving distance, $W_i^{A}$ is the weight (or window) function for the probe $A$ in the  $i$-th bin, and $P_{\delta\delta}$ is the total matter power spectrum. The power spectra are the same used in \citetalias{2020A&A...642A.191E}, obtained from the \CAMB\ Boltzmann code. The nonlinear correction model employed in $P_{\delta\delta}$ is a revised version of the \texttt{halofit} recipe \citep{Takahashi:2012em}, modified to correctly consider the massive neutrinos contributions~\citep{bird2012}.

\begin{figure*}
\centering
\includegraphics{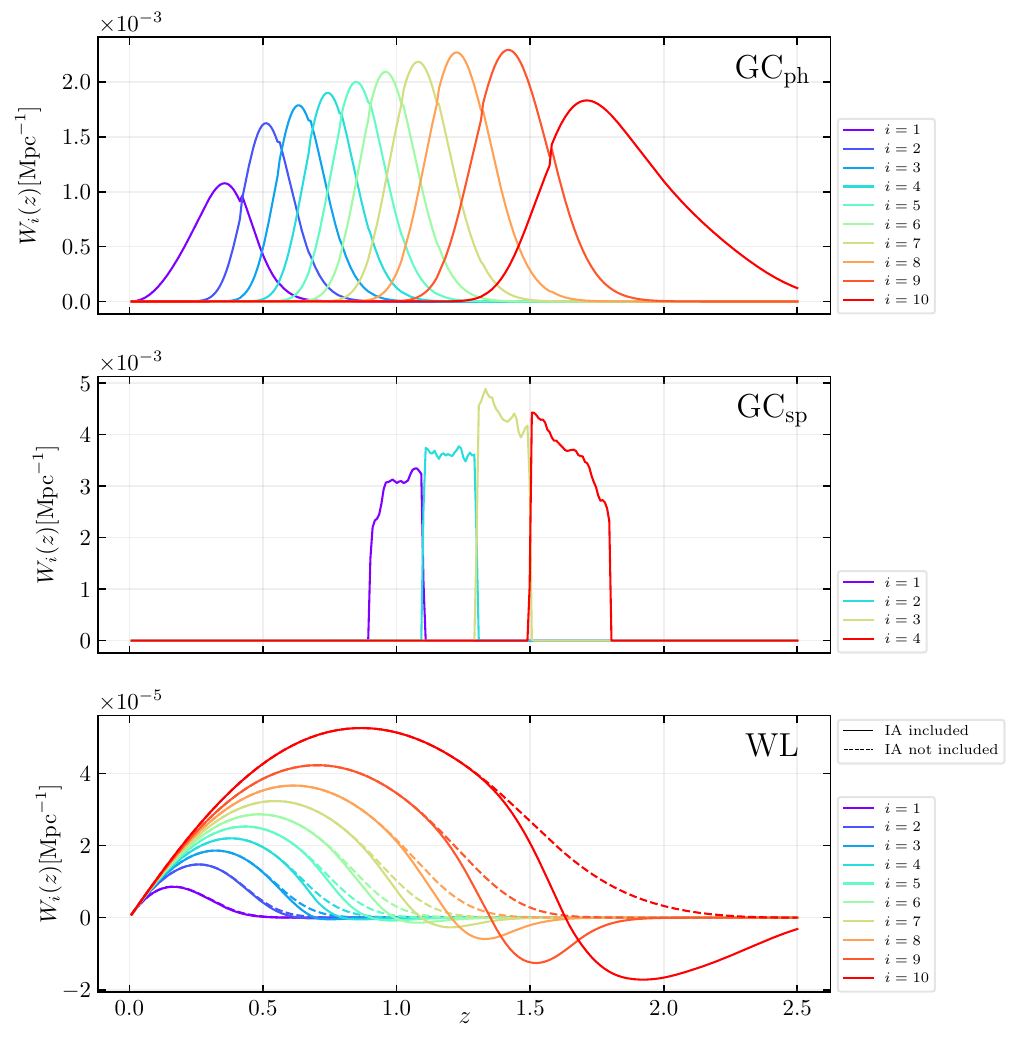}
\caption{Weight functions for the three probes considered in the forecast. For $\WL$ at high redshift bins, the weight function \cref{eq:weak_lensing_full_weight} (solid line) becomes negative, due to the contribution of intrinsic alignment (IA). The shear weight function \cref{eq:shear_weight} (dashed line) remains instead always positive as it should be.}
\label{fig:weight_functions_plot}
\end{figure*}

The functional form of the weight function $W_i^{A}$ depends on the probe $A$.
The weight functions for the probes considered in the forecast are shown in \cref{fig:weight_functions_plot}.

The fundamental ingredient for the computation of $W_i^A$ is the redshift distribution per unit solid angle, $\de N^A(z)/\de z/\de\Omega$. The redshift distribution has been modelled analytically for the $\WL$ and $\GCph$ probes, and obtained via simulations for the $\GCsp$ probe.

The normalised redshift density of the probe $A$ in the $i$-th bin can be computed from the redshift distribution as
\begin{align}
\tilde n^A_i(z)&=\int_{z_i^{-}}^{z_i^{+}} \de z_{\rm p} \, \frac{\de N^A}{\de z \,\de\Omega}(z)\,p_A\left(z_{\rm p} | z\right)\;,\\
n^A_i(z)&=\tilde n^A_i(z)\,\left[\int_{z_{\min}}^{z_{\max}} \de z\,\tilde n^A_i(z)\right]^{-1}\;,\label{eq:norm_density_def}
\end{align}
where the function $p_A(z_{\rm p}|z)$ is the probability that a galaxy with true redshift $z$ will be measured with a redshift $z_{\rm p}$.

Formally, the redshift integration range extends from $z_{\min} = 0$ to $z_{\max} = + \infty$; however, given the shape of the integrand functions, the integration range has been truncated at $z_{\min} = 0.001$ and $z_{\max} = 3$ in this work.
The tomographic bin edges $z_i^{-}$ and $z_i^{+}$ for $\WL$, $\GCph$, and $\GCsp$ are reported in \cref{tab:tomo_bins}.

\begin{table*}
\centering
\caption{Tomographic bin edges for weak lensing ($\WL$), photometric galaxy clustering ($\GCph$), and spectroscopic galaxy clustering ($\GCsp$). \textcolor{black}{The left table shows the WL and $\GCph$ tomograbic bin edges while the right table shows the $\GCsp$ ones.} The corresponding galaxy biases for photometric ($b^{\rm ph}_i$) and spectroscopic ($b^{\rm ph}_i$) galaxy clustering are also reported.}
\begin{tabular}{ccccccccccc}
\toprule
$z_i^-$ & 0.001 & 0.418 & 0.560 & 0.678 & 0.789 & 0.900 & 1.019 & 1.155 & 1.324 & 1.576 \\
$z_i^+$ & 0.418 & 0.560 & 0.678 & 0.789 & 0.900 & 1.019 & 1.155 & 1.324 & 1.576 & 2.500 \\
\midrule
$b^{\rm ph}_i$ & 1.100 & 1.220 & 1.272 & 1.317 & 1.358 & 1.400 & 1.445 & 1.497 & 1.565 & 1.743 \\
\bottomrule
\end{tabular}
\begin{tabular}{ccccc}
\toprule
$z_i^{-}$ & 0.90 & 1.10 & 1.30 & 1.50 \\
$z_i^{+}$ & 1.10 & 1.30 & 1.50 & 1.80 \\
\midrule
$b^{\rm sp}_i$ & 1.46 & 1.61 & 1.75 & 1.90 \\
\bottomrule
\end{tabular}
\label{tab:tomo_bins}
\end{table*}

This probability distribution $p_A(z_{\rm p}|z)$ models the redshift measurement errors for the probe $A$.
The model chosen is the same of \citetalias{2020A&A...642A.191E}:
\begin{multline}\label{eq:prob_z_distr}
p_{A}(z_{\rm p} | z) = 
\frac{1-f_{\rm out}}{\sqrt{2\,\pi}\,(1+z)\,\sigma_{\rm{b}}} 
\exp \left\{-\frac{1}{2}\left[\frac{z-c_{\rm{b}}\,z_{\rm p}-z_{\rm{b}}}{(1+z)\,\sigma_{\rm{b}}}\right]^{2}\right\} \\
+ \frac{f_{\rm out}}{\sqrt{2\,\pi}\,(1+z)\,\sigma_{\rm{o}}}
\exp \left\{-\frac{1}{2}\left[\frac{z-c_{\rm{o}}\,z_{\rm p}-z_{0}}{(1+z)\,\sigma_{\rm{o}}}\right]^{2}\right\}\;.
\end{multline}

This model includes multiplicative and additive biases in the redshift determination, both for a fraction ($1-f_{\rm out}$) of sources with well measured redshifts, and for a fraction ($f_{\rm out}$) of catastrophic outliers, i.e. galaxies with severely incorrect estimate of the redshift.
Different parameters of $f_{\rm out}$ and the biases have been used, depending whether the redshift measurement is photometric or spectroscopic; in particular, $f_{\rm out}$ is taken to be zero for $\GCsp$, $\sigma_{\rm b} = 0.05$ for $\GCph$, and $\sigma_{\rm b} = 0.001$ for $\GCsp$.
The values of the parameters for $p_A(z_{\rm p}|z)$, which are kept fixed in our analysis, are summarised in \cref{tab:prob_convolve_params}.

\begin{table*}
\centering
\caption{Values of the parameters adopted for the probability distributions $p_A(z_{\rm p}|z)$ defined in \cref{eq:prob_z_distr}.
The uncertainty $\sigma_{\rm b}$ on the correctly measured (not catastrophic) redshifts has been chosen from the \Euclid design requirements \citep{laureijs2011euclid}.
For $\GCsp$ the outlier parameters are not reported since they are irrelevant, being the outliers fraction $f_{\rm out} = 0$.}
\begin{tabular}{c|ccccccc}
\toprule
\text{probe} & $c_{\mathrm{b}}$ & $z_{\mathrm{b}}$ & $\sigma_{\mathrm{b}}$ & $c_{\mathrm{o}}$ & $z_{\mathrm{o}}$ & $\sigma_{\mathrm{o}}$ & $f_{\text {out }}$ \\
\midrule
$\WL$, $\GCph$ & 1.0 & 0.0 & 0.050 & 1.0 & 0.1 & 0.05 & 0.1 \\
$\GCsp$ & 1.0 & 0.0 & 0.001 & \text{--} & \text{--} & \text{--} & 0 \\
\bottomrule
\end{tabular}
\label{tab:prob_convolve_params}
\end{table*}

The (true) redshift distribution of the photometric samples is modelled as in the \Euclid redbook \citep{laureijs2011euclid}
\begin{equation}\label{eq:photo_true_nz}
\frac{\de N^{\mathrm{ph}}}{\de z \,\de\Omega}(z) = N^{\mathrm{ph}}_0\,\left(\frac{z}{z_{0}}\right)^{2}\, \exp \left[-\left(\frac{z}{z_{0}}\right)^{3 / 2}\right]\;,
\end{equation}
where $z_0 = 0.9/\sqrt{2}$ and the normalisation factor $N^{\mathrm{ph}}_0$ is chosen such that the surface density of galaxies is equal to $30$ galaxies per square arcminute, corresponding to an expected total number of galaxies of about $\num{1.6e9}$.

In the next subsections we give a short description of the probes involved in this analysis and their theoretical modelling.
\subsection{Weak lensing}
\label{sec:wl}
The gravitational field of large-scale cosmic structure deflects the path of light rays emitted by distant galaxies, distorting the images of the galaxies detected by the observers \citep{Kaiser:2000if, Bacon:2000sy,Kitching:2016zkn,Lemos:2017arq}. At the linear level these distortions can be decomposed locally into convergence $\kappa$ and a (complex) shear $\gamma$, which are respectively related to the size and shape distortion of the image.

In this work we only consider the shear signal $\gamma$, following \citetalias{2020A&A...642A.191E}. The corresponding weight function for cosmic shear in \cref{eq:limber_cls} is
\begin{equation}\label{eq:shear_weight}
W_i^\gamma(z) = 
\frac{3}{2}\,\left(\frac{H_{0}}{c}\right)^{2}\,\Omm\,(1+z)\, r(z)\,\int_{z}^{z_{\rm max}} \de z'\, n^{\rm ph}_i(z')\,\frac{r(z')-r(z)}{r(z')}\;.
\end{equation}
The integral makes it clear that weak lensing is a cumulative effect: the detected shapes of galaxies are influenced by all the matter along the line of sight.

The forecast also includes the \emph{intrinsic alignment} (IA), which is one of the main observable effects altering weak lensing measurements \citep{Joachimi:2015mma, Kiessling:2015sma, Kirk:2015nma}. IA refers to the alignments of nearby galaxies caused by tidal interactions which occur during galaxy formation and evolution. This produces spurious correlations over the ones due to cosmic shear. 
The IA effect can be included in the lensing angular power spectrum. A possible way is by using the extended nonlinear alignment model \citep{Bridle:2007ft} which consists in the following addition to the cosmic shear weight function
\begin{equation}\label{eq:weak_lensing_full_weight}
W_i^{\rm wl}(z) = W_i^{\gamma}(z) - \aIA \,\CIA\, \Omm \frac{H(z)\,\mathcal{F}_{\rm IA}(z)}{D(z)\,c}\, n^{\rm ph}_i(z)\;.
\end{equation}
In the above equation, $D(z)$ is the linear scale-independent growth factor. The function $\mathcal{F}_{\rm IA}$ is defined as 
\begin{equation}\label{eq:FIA_def}
\mathcal{F}_{\rm IA}(z) = (1+z)^{\etaIA}\,
\left[\frac{\langle L\rangle(z)}{L_{\star}(z)}\right]^{\betaIA}\;,
\end{equation}
with $\langle L\rangle(z)$ and $L_{\star}(z)$ are the mean and characteristic luminosity functions respectively. The intrinsic alignment parameters $\aIA$, $\etaIA$, $\betaIA$ are treated as nuisance parameters in the analysis. The reference values for the IA parameters are the same adopted in \citetalias{2020A&A...642A.191E}, namely $\{\aIA = 1.72, \, \etaIA = -0.41, \, \betaIA = 2.17\}$. The parameter $\CIA = 0.0134$ is fixed, since it is completely degenerate with $\aIA$.\\
\subsection{Photometric galaxy clustering}
\label{sec:gc_ph}
The Galaxy Clustering probes exploit the statistical properties of the galaxy distribution.
For the $\GCph$ probe, $W^{\mathrm{ph}}_i(z)$ in \cref{eq:limber_cls} is the galaxy clustering weight function
\begin{equation}\label{eq:galaxy_weight}
W^{\mathrm{ph}}_i(z) = b^{\mathrm{ph}}(z)\, \frac{H(z)}{c}\, n^{\mathrm{ph}}_i(z)\;,
\end{equation}
where $b$ is the galaxy bias and $n^{\mathrm{ph}}_i(z)$ is defined through \cref{eq:norm_density_def,eq:photo_true_nz}. 

Galaxies are biased tracers of the dark matter distribution; the former are related to latter through the galaxy bias $b$:
\begin{equation}\label{eq:bias_definition}
\delta^{\rm g}(k, z) = b(z)\, \delta(k, z)\;,
\end{equation}
with $k$ being the wavenumber. In general the galaxy bias is a function of $z$ and $k$, but in this work the $k$-dependence has been neglected, following the treatment of \citetalias{2020A&A...642A.191E}. Moreover, as in  \citetalias{2020A&A...642A.191E}, we neglect subdominant contributions to clustering such as {\it redshift-space distortions} (RSD) and lensing magnification.

The redshift evolution of the galaxy bias is modelled as in \citetalias{2020A&A...642A.191E}: for $\GCph$ a piecewise function is employed such that 
\begin{equation}
b^{\mathrm{ph}}(z) = b^{\mathrm{ph}}_i = \sqrt{1 + \bar{z}_i}\;, \qquad z_i^- < z < z_i^+\;,
\end{equation}
where $\bar{z}_i = (z_i^- + z_i^+)/2$ is the mean redshift of the $i$th bin.
Such bin edges, $z_i^-$ and $z_i^+$, are defined in \cref{tab:tomo_bins}, where we also report the values of $b^{\mathrm{ph}}_i$. \textcolor{black}{Following the approach of \citetalias{2020A&A...642A.191E}, we marginalise over their values.}
\subsection{Spectroscopic galaxy clustering}
\label{sec:gc_sp}

The $\Euclid$ mission will employ both its photometric and  spectroscopic samples to study galaxy clustering.
Spectroscopic redshift determination in $\Euclid$ is based on $\Halpha$-emitting galaxies in the redshift range $z \in [0.9, 1.8]$. 
The total number of galaxies in the spectroscopic sample is about $\num{2e7}$, which is smaller by a factor $\sim 80$ than the total number photometric galaxies.
Despite the lack of counting statitics, a sample based on
precise spectroscopic redshifts allows us to go beyond projected 2D statistics, exploiting the information in the full 3D galaxy distribution.

The spectroscopic galaxy clustering is usually treated by using the 3D Fourier galaxy power spectrum.
In this work, we also treated the $\GCsp$ probe in the harmonic domain.


The 3D Fourier approach can model several physical effects, such as RSD, the Alcock-Paczynski (AP) projection effects, the nonlinear damping of the Baryon Acoustic Oscillations (BAO), and the residual shot noise. The full nonlinear model for the 3D power spectrum of the $\Halpha$ galaxies employed in $\Euclid$ is described in Sec.~3.2 of \citetalias{2020A&A...642A.191E}.

The harmonic approach is based on \cref{eq:limber_cls}, with the $\GCsp$ weight function given by 
\begin{equation}
\label{eq:sp_weight}
W^{\mathrm{sp}}_i(z) = b^{\mathrm{sp}}(z) \,\frac{H(z)}{c}\, n^{\mathrm{sp}}_i(z)\;.
\end{equation}
The plot of these functions for the reference cosmology is shown in the middle panel of \cref{fig:weight_functions_plot}.

Both for the 3D Fourier power spectrum and angular power spectra, the underlying redshift distribution $\de N^{sp}(z)/\de z/\de\Omega$ is obtained from private communication with the GC-E2E work package group; the galaxy density has been obtained according to Model 3 from~\cite{Pozzetti:2016cch}. 
However, differently from the 3D case, for the 2D projected $\GCsp$ we neglect contributions from AP and RSD.

The spectroscopic redshift distribution is convolved with the probability $p_{\mathrm{sp}}(z_{\rm p}|z)$ in \cref{eq:prob_z_distr}, with redshift uncertainty $\sigma_{\rm b}$ set to $\num{0.001}$ and the fraction of outliers $f_{\rm out}$ set to zero, as specified in the \Euclid scientific requirements \citep{laureijs2011euclid}. The other parameters in \cref{eq:prob_z_distr} are summarised in \cref{tab:prob_convolve_params}.
The convolution mitigates the sharpness of the boundaries of the spectroscopic bins, and prevents potential numerical instabilities in the computation of the redshift integrals.

The spectroscopic galaxy bias $b^{\mathrm{sp}}(z)$ is modelled as a piecewise constant function, and the values $b^{\mathrm{sp}}_i$ in the bins are summarised in \cref{tab:tomo_bins} along with the bin edges. Both the bias values and the bins are the same that have been used in \citetalias{2020A&A...642A.191E}. In this work also a finer binning has been considered, in particular 12, 24 and 40 equally spaced bins in the range $0.9 < z < 1.8$. The values of the bias for finer binning are obtained by linear interpolation of the values listed in \cref{tab:tomo_bins}.

\section{Fisher information matrix}
\label{sec:fisher}
The Fisher information matrix is defined as the expectation value of the Hessian of the log-likelihood:
\begin{equation}
F_{\alpha \beta} = -\left\langle\frac{\partial^{2} \ln L}{\partial \theta_\alpha \partial \theta_\beta}\right\rangle\;,
\end{equation}
where $\alpha$ and $\beta$ are the model parameter indices, including both cosmological and nuisance parameters. The expected parameter covariance matrix is the inverse of the Fisher matrix:
\begin{equation}
\mathcal{C}_{\alpha \beta}=\left(F^{-1}\right)_{\alpha \beta}\;.
\end{equation}
%
The marginalised $1\,\sigma$ uncertainties, where $\sigma$ is the Gaussian standard deviation, on the model parameters are the square roots of the diagonal elements of the parameter covariance matrix:
\begin{equation}
\sigma_{\alpha}=\sqrt{\mathcal{C}_{\alpha \alpha}}\;.
\end{equation}
One of the metrics used to assess the scientific performance  of $\Euclid$ is the $\wz\text{-}\wa$ Figure-of-Merit ($\FoM$) which has been defined as in \citetalias{2020A&A...642A.191E} 
\begin{equation}\label{eq:fom_def}
\FoM \equiv 
\sqrt{\operatorname{det}\left( \tilde{F}_{\wz\wa} \right)}
= \left[\sigma_{\wz}^2\, \sigma_{\wa}^2 - \mathcal{C}^2_{\wz\wa}\right]^{-1/2}\;.
\end{equation}
The symbol $\tilde{F}_{\wz\wa}$ denotes  the Fisher information matrix relative to the dark energy equation of state parameters $\wz$ and $\wa$,see \cref{eq:CPL}, marginalised over all the other free parameters. 
The $\FoM$ is inversely proportional to the area of the $1\,\sigma$ marginalised contour ellipse in the $\wz\text{-}\wa$ plane. Tighter constraints on $\wz$ and $\wa$ lead to smaller ellipses, which in turn means higher FoMs.

In this work we assume that the data vector $\mathcal{D}$, which contains the values of the $C(\ell)$'s or the $P(k)$'s considered, is distributed according to a multivariate Gaussian. Under this assumption, the Fisher matrix element $F_{\alpha\beta}$ can be calculated as
\begin{equation}\label{eq:fisher_most_compact_form}
F_{\alpha\beta} = 
\pdv{\vec{\mathcal{D}}}{\theta_\alpha}\,
\invcov{\vec{\mathcal{D}}}{\vec{\mathcal{D}}}\,
\pdv{\vec{\mathcal{D}}}{\theta_\beta} \;,
\end{equation}
where $\cov{\vec{\mathcal{D}}}{\vec{\mathcal{D}}}$ represents the covariance matrix.

The scenarios and settings considered in the Fisher computations of this work are summarised in \cref{tab:scenarios_settings}.
\begin{table*}[]
\centering
\caption{Summary of the setting scenarios considered in the forecast. For $\WL$ the general settings are not reported since no distinction other than the multipole range has been made. The scenario in which the $\GCsp$ shot noise is reduced \emph{is not intended to be realistic}, and has been studied only to probe the theoretical limitations of the harmonic approach.}
\begin{tabular}{llll}
\toprule
\multicolumn{4}{c}{Forecast settings in the harmonic approach} \\
\midrule
\multirow{6}{*}{Multipoles settings} & \multirow{3}{*}{Optimistic} & $\GCph$ & $10 \leq \ell \leq 3000$ \\
 & & $\GCsp$ & $10 \leq \ell \leq 3000$ \\
 & & $\WL$ & $10 \leq \ell \leq 5000$ \\[1ex]
 & \multirow{3}{*}{Pessimistic} & $\GCph$ & $10 \leq \ell \leq 750$ \\
 & & $\GCsp$ & $10 \leq \ell \leq 750$ \\
 & & $\WL$ & $10 \leq \ell \leq 1500$ \\
 \midrule
\multirow{4}{*}{General settings} & \multirow{2}{*}{Baseline} & $\GCph$ & all the $10$ bins of \cref{tab:tomo_bins} \\
 & & $\GCsp$ & reduced shot noise of \cref{eq:reduced_gcsp_noise} \\[1ex]
 & \multirow{2}{*}{Alternative} & $\GCph$ & only $4$ bins of \cref{tab:tomo_bins} with $z_i^-, z_i^+ \in [0.9, 1.8]$\\
 & & $\GCsp$ & reduced shot noise of \cref{eq:reduced_gcsp_noise}\\
\bottomrule
\end{tabular}
\label{tab:scenarios_settings}
\end{table*}

The following subsections describe the data vectors and the covariance matrices used in the approaches considered in this work, as well as the shot noise implementation.\\
\subsection{Harmonic space approach}
\label{sec:fisher_fullcl}
The assumption that the $a^{A}_{\ell m}$ coefficients of the observed fields follow a Gaussian distribution leads to the analytical expression for the Fisher matrix presented in \cref{eq:fisher_most_compact_form}. In this case the Gaussian covariance matrix for the harmonic power spectra $\cl{AB}{ij}{\ell}$ is
\begin{multline}\label{eq:4thorder_cov}
\operatorname{Cov} \left [ 
\cl{AB}{ij}{\ell}, \cl{CD}{km}{\ell'}
\right ] = \frac{\delta_{\ell\ell'}}{2\,\ell+1}\\\times
\left  [
\sgl{AC}{ik}{\ell}\,\sgl{BD}{jm}{\ell'} +
\sgl{AD}{im}{\ell}\,\sgl{BC}{jk}{\ell'}
\right ]\;,
\end{multline}
with $\delta_{\ell\ell'}$ the Kronecker delta symbol. 
\textcolor{black}{Under the assumption that we account for the partial sky coverage only through the observed fraction of the sky $f_{\rm sky}$}, the masking effects are negligible, and the galaxy counts are affected by Poisson shot noise only, the matrices $\sgl{AB}{ij}{\ell}$ can be expressed by 
\begin{equation}
\sgl{AB}{ij}{\ell} = 
\frac{1}{\sqrt{f_{\rm sky}\,\Delta \ell}}\, \left ( \cl{AB}{ij}{\ell} + N_{ij}^{AB}(\ell)\right )\;,
\end{equation}
where $N^{AB}_{ij}(\ell)$ is the Poisson shot noise matrix described later in \cref{sec:shot_noise} and $\Delta \ell$ is the spacing between the multipoles in which the $C(\ell)$ are sampled.

The data-vector in the harmonic space approach contains the independent values of the $C(\ell)$'s to be included in the computation.
Since the tomographic angular power spectra are matrices, it is convenient to \emph{vectorise} them. In this context the term ``vectorise'' refers to matrix vectorisation. In this paragraph we are providing some examples of matrix vectorization for our probes; a formal description can be found in \cref{sec:appendix_cross}.

Let us consider the $\GCph$ auto-correlation $\cl{\phph}{ij}{\ell}$ at fixed  multipole $\ell$; this is usually represented as a $10\times 10$ symmetric matrix with $10(10+1)/2=55$ independent components. 
This data vector can be also represented as a row vector of the independent components:
\begin{equation}\label{eq:datavector_generic}
\mathbf{C}(\ell) = \left \lbrace
\mathbf{C}_1(\ell), \mathbf{C}_2(\ell), \dots, \mathbf{C}_{55}(\ell)\right \rbrace\;,
\end{equation}
where the index of the vector enumerates the $55$ independent components of $\mathbf{C}(\ell)$ considered in the analysis.

When considering 3 probes -- $\WL$, $\GCph$, and $\GCsp$ -- there are $6$ angular power spectra matrices:
\begin{itemize}
\item 3 auto-correlations: $\bcl{\wlwl}{\ell}$, $\bcl{\spsp}{\ell}$, $\bcl{\phph}{\ell}$;
\item 3 cross-correlations: $\bcl{\wlph}{\ell}$, $\bcl{\wlsp}{\ell}$, $\bcl{\phsp}{\ell}$.
\end{itemize}
%


The angular power spectra are evaluated in a grid of $N_{\ell}$ multipoles; the full data-vector $\vec{\mathcal{D}}$ also includes all the values of the power spectra evaluated over all multipoles, viz.\
\begin{equation}
\vec{\mathcal{D}} = 
\left \lbrace 
\bcl{}{\ell_1}, \bcl{}{\ell_2}, \dots,
\bcl{}{\ell_{N_\ell}} 
\right \rbrace\;.
\end{equation}
The covariance matrix associated with this data-vector is block diagonal, since the Kronecker delta of \cref{eq:4thorder_cov} ensures that different multipoles are uncorrelated:
\begin{equation}\label{eq:block_diag_multipole_covariance}
\cov{\vec{\mathcal{D}}}{\vec{\mathcal{D}}} =
\begin{pmatrix}
\cov{\bcl{}{\ell_1}}{\bcl{}{\ell_1}} & 0 & 0 \\
0 & \ddots & 0 \\
0 & 0 & \cov{\bcl{}{\ell_{{N_\ell}}}}{\bcl{}{\ell_{N_\ell}}}
\end{pmatrix}\;.
\end{equation}
The diagonal blocks, $\cov{\bcl{}{\ell}}{\bcl{}{\ell}}$, account for all the correlations between the $\bcl{}{\ell}$'s:
\begin{multline}\label{eq:cov_matrix_generic_form}
\cov{\mathbf{C}(\ell)}{\mathbf{C}(\ell)}
= \\
\begin{pmatrix}
\cov{\mathbf{C}_1(\ell)}{\mathbf{C}_1(\ell)} &
\cdots & \cov{\mathbf{C}_1(\ell)}{\mathbf{C}_\mathcal{N}(\ell)} \\
\vdots & \ddots & \vdots \\
\cov{\mathbf{C}_\mathcal{N}(\ell)}{\mathbf{C}_1(\ell)} &
\cdots & \cov{\mathbf{C}_\mathcal{N}(\ell)}{\mathbf{C}_\mathcal{N}(\ell)}
\end{pmatrix}\;,
\end{multline}
where the single blocks $\cov{\mathbf{C}_i(\ell)}{\mathbf{C}_j(\ell)}$ are computed according to \cref{eq:4thorder_cov}. 
With these definitions, the Fisher matrix element $F_{\alpha\beta}$ can be calculated from \cref{eq:fisher_most_compact_form} as
\begin{equation}\label{eq:fisher_vec_4th_order_generic_fixed_ell}
F_{\alpha\beta} = 
\sum_{\ell = \ell_{1}}^{\ell_{N_{\ell}}} F_{\alpha\beta}(\ell)
= \sum_{\ell = \ell_{1}}^{\ell_{N_{\ell}}}
\pdv{\mathbf{C}(\ell)}{\theta_\alpha}^T
\,\cov{\mathbf{C}(\ell)}{\mathbf{C}(\ell)}^{-1}\,
\pdv{\mathbf{C}(\ell)}{\theta_\beta}
 \;,
\end{equation}
where the first equality follows from the block diagonal form of the covariance.\\
%
\subsection{Naming conventions}
\label{sec:naming}
For later convenience, here are reported the adopted conventions for naming the Fisher matrices that have been computed in this work:
\begin{enumerate}
\item The name of a Fisher matrix is representative of the data vector, and it is composed by different labels, which identify the $C(\ell)$'s contained in the data-vector itself.
\item Within the name of a Fisher matrix, the auto-correlation $C^{\rm AA}(\ell)$ of the probe $\rm A$ is labelled simply as $\rm A$, while the cross-correlation $C^{\rm AB}(\ell)$ between the two probes $\rm A$ and $\rm B$ is denoted as $\xc{\mathrm{A}}{\mathrm{B}}$.
\item Square brackets are used to delimit the data-vector extent.
\item The pairwise cross-covariances between the $C(\ell)$'s included in a given data-vector are always considered in the computation of the corresponding Fisher matrix.
\item The sum of two Fisher matrices $[F_1]$ and $[F_2]$ is simply denoted by $[F_1] + [F_2]$. This simple sum corresponds to combine $[F_1]$ and $[F_2]$ without taking into account the cross-covariances between their data-vectors, i.e. it is an \emph{independent sum}.
\end{enumerate}
Let us return to the example where only the auto-correlation of $\GCph$ is considered.
In this case the data-vector at fixed $\ell$ is 
\begin{equation}
\mathbf{C}(\ell) = \left \lbrace \bcl{\phph}{\ell} \right \rbrace\;,
\end{equation}
and ranges over $55$ elements (assuming $10$ tomographic bins). Consequently, the covariance matrix is made by a single $55\times 55$ block:
\begin{equation}
\cov{\bcl{\phph}{\ell}}{\bcl{\phph}{\ell}}\;,
\end{equation}
which takes into account the auto-covariance of $\GCph$ only. 

When considering two or more probes, multiple combinations can be constructed, depending whether or not \emph{cross-covariances} and \emph{cross-correlations} are included in the computation. In this work we reserve the word ``cross-covariance'' to the off-diagonal blocks of the covariance matrix \cref{eq:cov_matrix_generic_form}, e.g. the blocks $\cov{\mathbf{C}^{ph}_i(\ell)}{\mathbf{C}^{wl}_j(\ell)}$. The term ``cross-correlation'' (signal) is instead used to denote the data-vector entry corresponding to the correlation between two probes, e.g. $\bcl{\wlph}{\ell}$.

The possible combinations that can be constructed using two probes, $\rm A$ and $B$, are described in detail in Appendix \ref{sec:appendix_cross}.\\
\subsection{Hybrid approach}
\label{sec:fisher_2dx3d}

The harmonic space approach has both advantages and disadvantages. One of the main advantages of the harmonic space approach is the straightforward way to compute cross-covariances. The main disadvantage of treating the clustering signal in the harmonic domain is that the projection on the celestial sphere results in a partial loss of information from the galaxy density distribution along the radial direction.

For $\GCph$, this loss is limited, since the redshift resolution is already hampered by large photometric errors.
In the spectroscopic sample, the redshifts are measured with much better precision, therefore the 2D projection results in a larger loss of constraining power from $\GCsp$. The tomographic technique can be employed in order to partially recover the radial information about the distances of the galaxies.

The natural approach to treat $\GCsp$ is the ``3D'' or $P_k$.
Directly using a 3D galaxy power spectrum allows us to naturally exploit the high redshift resolution of the spectroscopic sample. Nonetheless, in this case it is difficult to compute cross-correlations and cross-covariances between $\WL, \GCph$, and $\GCsp$.

In this work we therefore considered two approaches to combine $\GCsp$ with $\WL$ and $\GCph$. The first is to treat all probes with the angular power spectra, as described in the previous section. In the second approach, the Fisher matrix contribution for the $\GCsp$ auto-correlation is taken into account by adding \emph{a posteriori} the Fisher matrix computed in the 3D approach of \citetalias{2020A&A...642A.191E}, referred here as $\GCsp(P_k)$.


\textcolor{black}{The core idea of the latter approach is to add the $\GCsp$ auto-correlation using the 3D observable and its cross-correlations with the other probes, treating it as a 2D observable, in order to be able to compute the covariance matrix analytically: for this reason we dubbed it the hybrid approach.}

The starting point of the hybrid approach is to include the harmonic $\GCph$ auto-correlation and the harmonic $\GCph\times\GCsp$ cross-correlation in the data-vector
\begin{equation}
\mathbf{C}(\ell) = \left \lbrace
\bcl{\phph}{\ell}, \bcl{\phsp}{\ell}
\right \rbrace \;,
\end{equation}
with covariance matrix
\begin{equation}
\begin{pmatrix}
\cov{\phph}{\phph} & \cov{\phph}{\phsp}\\
\cov{\phsp}{\phph} & \cov{\phsp}{\phsp}\\
\end{pmatrix} \;,
\end{equation}
where we have dropped $\mathbf{C}$ and the exponent in order to have a lighter notation.

We refer to the resulting Fisher matrix with the notation $[\GCph+\xc{\GCph}{\GCsp}]$.

The Fisher matrix of $\GCsp(P_k)$ is independently added:
\begin{multline}
F_{\alpha\beta}[\GCph+\xc{\GCph}{\GCsp}\, + \, \GCsp(P_k)] \equiv \\
F_{\alpha\beta}[\GCph+\xc{\GCph}{\GCsp}] + F_{\alpha\beta}[\GCsp(P_k)]\;.
\end{multline}
This procedure is equivalent to neglecting the covariance terms between the $\GCsp$ auto-correlation and the other observables. More details are elaborated in Appendix~\ref{sec:appendix_hybrid_cross}.\\
\subsection{Poisson shot noise}
\label{sec:shot_noise}
The Poisson shot noise \citep{Baldauf:2013hka} has been implemented similarly to what has been done in \citetalias{2020A&A...642A.191E}. It is assumed that only the auto-correlation $C(\ell)$'s in the same tomographic bin are affected by a shot noise, which is independent of the multipole:
\begin{equation}\label{eq:shotnoise_kronecker}
N^{AB}_{ij}(\ell) = \delta_{AB}\,\delta_{ij}\,N^{A}_i\;.
\end{equation}
The quantity $N^{A}_i$ represents the shot noise associated with the probe $A$ at tomographic bin $i$. For photometric and spectroscopic galaxy clustering in the harmonic domain this is simply given by
\begin{align}
N^{\rm ph}_i &\equiv \left[\int_{z_i^-}^{z_i^+} \de z\, \frac{\de \mathcal{N}^{\rm ph}}{\de z \de \Omega}\right]^{-1}\;,\\
N^{\rm sp}_i &\equiv \left[\int_{z_i^-}^{z_i^+} \de z\, \frac{\de \mathcal{N}^{\rm sp}}{\de z \de \Omega}\right]^{-1}\;.\label{eq:gc_shot_noise}
\end{align}
For WL the definition is instead given by the $\GCph$ shot noise multiplied by the variance $\sigma^2_\epsilon$ of the intrinsic galaxy ellipticity
\begin{equation}\label{eq:wl_shot_noise}
N^{\rm wl}_i \equiv \sigma^2_\epsilon\,\left[\int_{z_i^-}^{z_i^+} \de z\, \frac{\de \mathcal{N}^{\rm ph}}{\de z \de \Omega}\right]^{-1}\;.
\end{equation}
The value assumed for $\sigma_\epsilon$ is $0.3$ as in \citetalias{2020A&A...642A.191E}.

As the Kronecker delta $\delta_{AB}$ of \cref{eq:shotnoise_kronecker} states, no shot noise has been considered for the cross-correlation $C(\ell)$'s. It is in fact commonly assumed for the noises of different tracers to be uncorrelated \citepalias{2020A&A...642A.191E}. This approximation is expected to work well for the cross-correlation between weak lensing and galaxy clustering, since these are different tracers of the same underlying dark matter distribution.

In the cross-correlation of $\GCph$ with $\GCsp$ the tracers are the galaxies for both the probes, so in principle a shot noise term in the cross power spectra (cross-noise) should be present. In this work we checked the $\delta_{\rm ph sp}$ approximation of \cref{eq:shotnoise_kronecker} with the following approach. Given that the Poisson shot noise affecting the two-point function comes from the count of galaxy \emph{self-pairs} -- see the introduction of \cite{Baldauf:2013hka} -- the cross-noise is due to those galaxies which are both in the photometric and in the spectroscopic sample. The scenario with the highest noise is therefore the one in which the spectroscopic sample is a \emph{proper subset} of the photometric one. So, assuming this worst case scenario, a \emph{subtraction} of the spectroscopic galaxy distribution from the photometric one has been performed. After the subtraction there is be no more overlap between the two samples, and their cross-noise becomes zero by construction. It has been checked that the constraints do not change appreciably after the subtraction, with the FoM and the marginalised uncertainties of the free parameters being negligibly affected. \textcolor{black}{However, we point out that the removal of the spectroscopic galaxies from the photometric samples might cause a bias in the clustering signal approach; the approach we followed was just a way to quantify the impact of the shot noise in the worst case scenario.}

The shot noise affects in a direct way the diagonal covariance blocks corresponding to the auto-correlation power spectra, as it enters in all the factors of \cref{eq:4thorder_cov}. This is what happens for example in the case of the auto-covariance of $\GCsp$
\begin{multline}\label{eq:spsp_diag_cov_block}
\cov{\cl{\spsp}{ij}{\ell}}{\cl{\spsp}{km}{\ell}} 
\propto\Big\{\left[\cnol{\spsp}{ik} + N^{\spsp}_{ik} \right] \,
 \left[\cnol{\spsp}{jm} + N^{\spsp}_{jm} \right]\\
+\left[\cnol{\spsp}{im} + N^{\spsp}_{im} \right]\,
 \left[\cnol{\spsp}{jk} + N^{\spsp}_{jk} \right]\Big\}
\;,
\end{multline}
where the $\ell$ dependence on the right-hand side (e.g. $N^{\spsp}_{ik}(\ell)$) has been omitted for compactness.
However, the fact that the auto-correlation power spectra are contaminated by the shot noise indirectly alters also the other blocks of the covariance matrix. For example the diagonal block corresponding to the auto-covariance of $\bcl{\phsp}{\ell}$ reads

\begin{multline}\label{eq:phsp_diag_cov_block}
\cov{\cl{\phsp}{ij}{\ell}}{\cl{\phsp}{km}{\ell}} 
\propto\Big\{\left[\cnol{\phph}{ik} + N^{\phph}_{ik} \right] \,
 \left[\cnol{\spsp}{jm} + N^{\spsp}_{jm} \right]\\
+\cnol{\phsp}{im}\, \cnol{\phsp}{jk}\Big\}
\;.
\end{multline}
and therefore also the terms coming from the inclusion of the $\xc{\GCph}{\GCsp}$ are affected by the shot noise of both $\GCph$ and $\GCsp$. Moreover, since the number of galaxies in the spectroscopic sample is $\sim 80$ times smaller than the galaxies in the photometric one, from \cref{eq:gc_shot_noise} it is clear that the $\GCsp$ shot noise is larger than the one associated with $\GCph$. In order to quantify the impact of the $\GCsp$ noise, the forecasts have been performed also in an unrealistic alternative scenario, where this noise is artificially reduced as if the number of spectroscopic galaxies were equal to the number of the photometric ones. This is done in practice by introducing an alternative reduced shot noise for $\GCsp$, defined as follows
\begin{equation}\label{eq:reduced_gcsp_noise}
\tilde{N}^{\rm sp}_i
\equiv \frac{N_{\rm sp}^{\rm tot}}{N_{\rm ph}^{\rm tot}}\, N^{\rm sp}_i\;,
\end{equation}
where $N_\mathrm{sp}^\mathrm{tot}$ ($N_\mathrm{ph}^\mathrm{tot}$) is the total number of spectroscopic (photometric) galaxies, computed by integrating the galaxy distribution over its redshift range. As it is discussed in \cref{sec:results} it turns out that the results in the harmonic approach drastically change using this reduced noise. \textcolor{black}{We emphasise that this test was performed only to check what is the origin of the different GCsp and GCph constraints in the harmonic approach and it is not representative of any realistic survey scenario.}

\section{Results}
\label{sec:results}

In this section we present the results of the forecast. The results are mainly reported as marginalised relative $\onesigma$ uncertainties on the FoM and the parameters in the reference cosmology in \cref{tab:cosmo_pars_fiducials}: 
\begin{equation}\label{perc_diff_definition}
\delta X\,[\%] \equiv 100\, \frac{X_{b} - X_{a}}{X_{a}}\;,
\end{equation}
where $X$ generically denotes the FoM or the relative uncertainty $\sigma_{\theta}/\theta_{\rm fid}$. The subscript $a$ denotes the Fisher matrix whose constraints are used as reference, while $b$ is the Fisher matrix under examination, also referred as the \emph{minuend} in the following.


The scenarios considered in this forecast are summarised in \cref{tab:scenarios_settings}. Concerning the multipole range, two scenarios have been studied: one \emph{optimistic} and one \emph{pessimistic}. In the optimistic scenario the multipole range for galaxy clustering is set to $\ell \in [10, 3000]$, while it is $\ell \in [10, 5000]$ for $\WL$. In the pessimistic scenario instead the multipole range is $\ell \in [10, 750]$ for $\GCph$, $\GCsp$, and $\ell \in [10, 1500]$ for $\WL$. 

In order to better understand the differences between $\GCph$ and $\GCsp$ when both are treated in the harmonic domain,
one more setting has been added, in which $\GCsp$ employs the alternative reduced shot noise described in \cref{sec:shot_noise}. We remark that this scenario \emph{is not meant to be realistic} and it is considered only to investigate the different constraining power of the $\GCsp$ and $\GCph$ probes. 

For the same reason, in the case of $\GCph$ the forecast has been performed with an alternative tomographic binning, where only the $4$ photometric bins in the redshift range $z \in [0.9, 1.8]$ are considered (see \cref{tab:tomo_bins}). This setting make both $\GCph$ and $\xc{\WL}{\GCph}$ comparable with $\GCsp$ and $\xc{\WL}{\GCsp}$, respectively. In fact, apart from the shape of the galaxy distribution and the values of the galaxy bias, in the harmonic domain the two main differences between $\GCph$ and $\GCsp$ are the shot noise and the redshift range of the galaxy catalogue.

This section is organised as follows.  
In \cref{sec:results_phxsp} the results from the combination of $\GCph$ and $\GCsp$ are presented, including also a comparison between $\GCph$ and $\GCsp$ when treated in the harmonic domain.
In \cref{sec:results_wlxsp} the results are reported for the combination of $\WL$ and $\GCsp$, also comparing the impact on the constraints from $\xc{\WL}{\GCsp}$ against $\xc{\WL}{\GCph}$.
In \cref{sec:results_6x2pt}, the results of the so-called \Euclid \sixtwoptShort\ statistics are reported; 
this term refers to the combination of the all possible two-point functions that can be constructed from $\WL, \GCph$, and $\GCsp$. The constraints coming from the combination of the full set of \Euclid main probes are shown, focusing on the importance of cross-covariances and cross-correlations between them. \textcolor{black}{We have checked the numerical stability of our results with respect to several hyperparameters, such as the $k$ and $z$ sampling of the matter power spectra; we find the variations of the marginalised errors and FoM to be negligible.}

In the following, the word \emph{harmonic} indicates that all the observables are treated using the angular power spectrum formalism. Instead, in the \emph{hybrid} \sixtwoptShort\ approach the data vector is entirely composed of $C(\ell)$'s except for the spectroscopic auto-correlation, whose contribution is included as described in \cref{sec:fisher_2dx3d} and in \cref{sec:appendix_hybrid_cross}.\\
\subsection{Combining photometric and spectroscopic clustering}
\label{sec:results_phxsp}

\begin{table*}[t]
\caption{Table reporting the $\FoM$ for $\GCph$, $\GCsp$ and their cross-correlation. The $\Delta\FoM$ column contains the variation of the $\FoM$ with respect to the independent sum $[\GCph] + [\GCsp]$ for the given number of bins. The $\Delta\FoM (\%)$ column is calculated by taking $\Delta\FoM$ as a percentage of the $\FoM$ of $[\GCph] + [\GCsp]$.}
\centering
\begin{resultstable}{cl*{3}{c}}
\multicolumn{5}{c}{$\GCph\times\GCsp$ FoM forecasts} \\
\midrule
$\GCsp$\, \text{bins} & \text{Fisher matrix} & $\FoM$ & $\Delta\FoM$ & $\Delta\FoM$(\%)\\
\midrule
\text{--} & [$\GCph$] & 63.12 & \text{--} & \text{--} \\
 \midrule
\multirow{4}{*}{4}
& [$\GCph$] + [$\GCsp(P_k)$]                   & 230.27 & \text{--} & \text{--} \\
& [$\GCph$+$\xc{\GCph}{\GCsp}$] + [$\GCsp(P_k)$] & 234.54 & \text{--} & \text{--} \\
& [$\GCph$] + [$\GCsp$]                        & 65.69  & \text{--} & \text{--} \\
& [$\GCph$+$\GCsp$]                            & 63.95  & $-1.75$ & $\negative{-2.66\%}$ \\
& [$\GCph$+$\GCsp$+$\xc{\GCph}{\GCsp}$]          & 69.63  & +3.94 & \positive{+6.00\%} \\
\midrule
\multirow{4}{*}{12}
& [$\GCph$] + [$\GCsp$] & 72.02 & \text{--} & \text{--}\\
& [$\GCph$+$\GCsp$] & 70.48 & $-1.55$ & $\negative{-2.15\%}$ \\
& [$\GCph$+$\GCsp$+$\xc{\GCph}{\GCsp}$] & 79.87 & +7.85 & \positive{+10.90\%} \\
\midrule
\multirow{4}{*}{24}
& [$\GCph$] + [$\GCsp$] & 85.02 & \text{--} & \text{--}\\
& [$\GCph$+$\GCsp$] & 83.88 & $-1.13$ & $\negative{-1.33\%}$ \\
& [$\GCph$+$\GCsp$+$\xc{\GCph}{\GCsp}$] & 108.35 & +23.34 & \positive{+27.45\%}\\
\midrule
\multirow{4}{*}{40}
& [$\GCph$] + [$\GCsp$] & 111.22 & \text{--} & \text{--} \\
& [$\GCph$+$\GCsp$] & 110.39 & $-0.83$ & $\negative{-0.75\%}$ \\
& [$\GCph$+$\GCsp$+$\xc{\GCph}{\GCsp}$] & 153.71 & +42.48 & \positive{+38.20\%} \\
\end{resultstable}
\label{tab:results_phsp_fom}
\end{table*}

In this section we report the results for the combination of $\GCph$ and $\GCsp$. We take as reference values for the results the constraints coming from the independent combination of $\GCph$ and $\GCsp$, that is [$\GCph$] + [$\GCsp$] in the harmonic approach and [$\GCph$ ] + [$\GCsp(P_k )$] in the hybrid one. We remind that the notation $[\GCph] + [\GCsp]$ represents  the simple sum of the Fisher matrices of the two probes, which does not account for their cross-covariance.
The forecasts for this combination are reported in the optimistic scenario only for brevity.

\begin{figure*}
\centering
\includegraphics[width=0.80\textwidth]{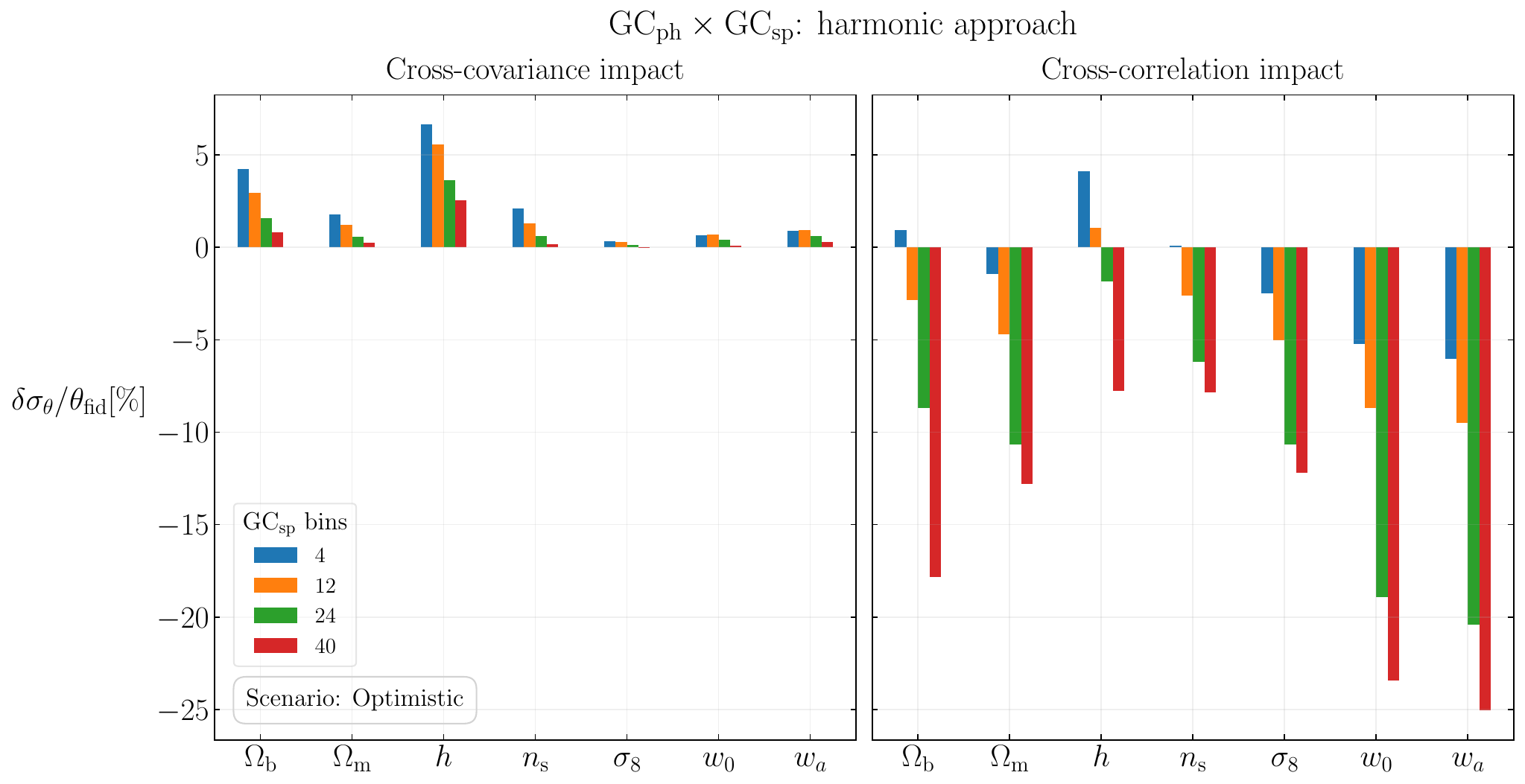}
\caption{Impact on $\onesigma$ parameter errors of cross-covariance and cross-correlation in the harmonic approach for the combination of $\GCph$ and $\GCsp$. The left panel shows the impact of the cross-covariance, quantified with the percentage differences between $[\GCph+\GCsp]$ and $[\GCph] + [\GCsp]$. In the right panel the impact of the cross-correlation is reported, quantified by percentage differences of $[\GCph+\GCsp+\xc{\GCph}{\GCsp}]$ with respect to $[\GCph]+[\GCsp]$. Note the opposite sign of the percentage differences for FoM and uncertainties.}
\label{fig:phsp_fullcl_percdiff_plot}
\end{figure*}

\subsubsection{Impact of the $\GCph$-$\GCsp$ cross-covariance on parameter constraints}

\cref{tab:results_phsp_fom} shows that, as expected, the cross-covariance slightly worsens the FoM. 
Anyway the contribution is always smaller than $3\%$, decreasing as the number $\GCsp$ bins increase. This trend can be understood as follows: the finer the tomographic binning of $\GCsp$ the smaller the support of the integrand of the off-diagonal terms $\bcl{\phsp}{\ell}$. Intuitively, the value of the integral over each of the $4$ thick bins is diluted into more thinner bins when a finer binning is used. Hence the off-diagonal block $\cov{\bcl{\phph}{\ell}}{\bcl{\spsp}{\ell}}$ and its transposed counterpart become larger and sparser as the number of bins is increased, and therefore the cross-covariance contribution becomes smaller. Physically this could be understood considering that for 40 bins the loss of information due to the projection transversely to the line of sight is less severe than for 4 bins only.

The same trend is observed also for the marginalised uncertainties on the cosmological parameters, as it can be seen by \cref{fig:phsp_fullcl_percdiff_plot}. The parameters mostly affected by the covariance are $\Omb$ and $h$, whose constraints in the 4 bin setting worsen by $4\%$ and $6\%$ respectively. However, as soon as the binning is refined, the effect gets smaller, becoming about $0.9\%$ for $\Omb$ and $2.5\%$ for $h$. The uncertainties on $\Omm$, $\ns$, and $\sige$ are instead affected by less than $2\%$ for all the binning settings. This outcome is confirmed also by \cref{fig:phsp_fullcl_vs_2dx3d_error_barplot}, which reports the relative marginalised uncertainties and the FoM as horizontal bars. The $[\GCph+\GCsp]$ bars (in blue) always have practically the same length as the $[\GCph] + [\GCsp]$ bars (in cyan), and they look more and more similar as the number of $\GCsp$ bins increases. 

In conclusion the cross-covariance between $\GCph$ and $\GCsp$ can be considered negligible, as it does not change the uncertainties on the cosmological parameters by more than $6\%$ and the FoM by more than $3\%$.

\begin{figure*}[t]
\centering
\includegraphics[width=0.60\textwidth]{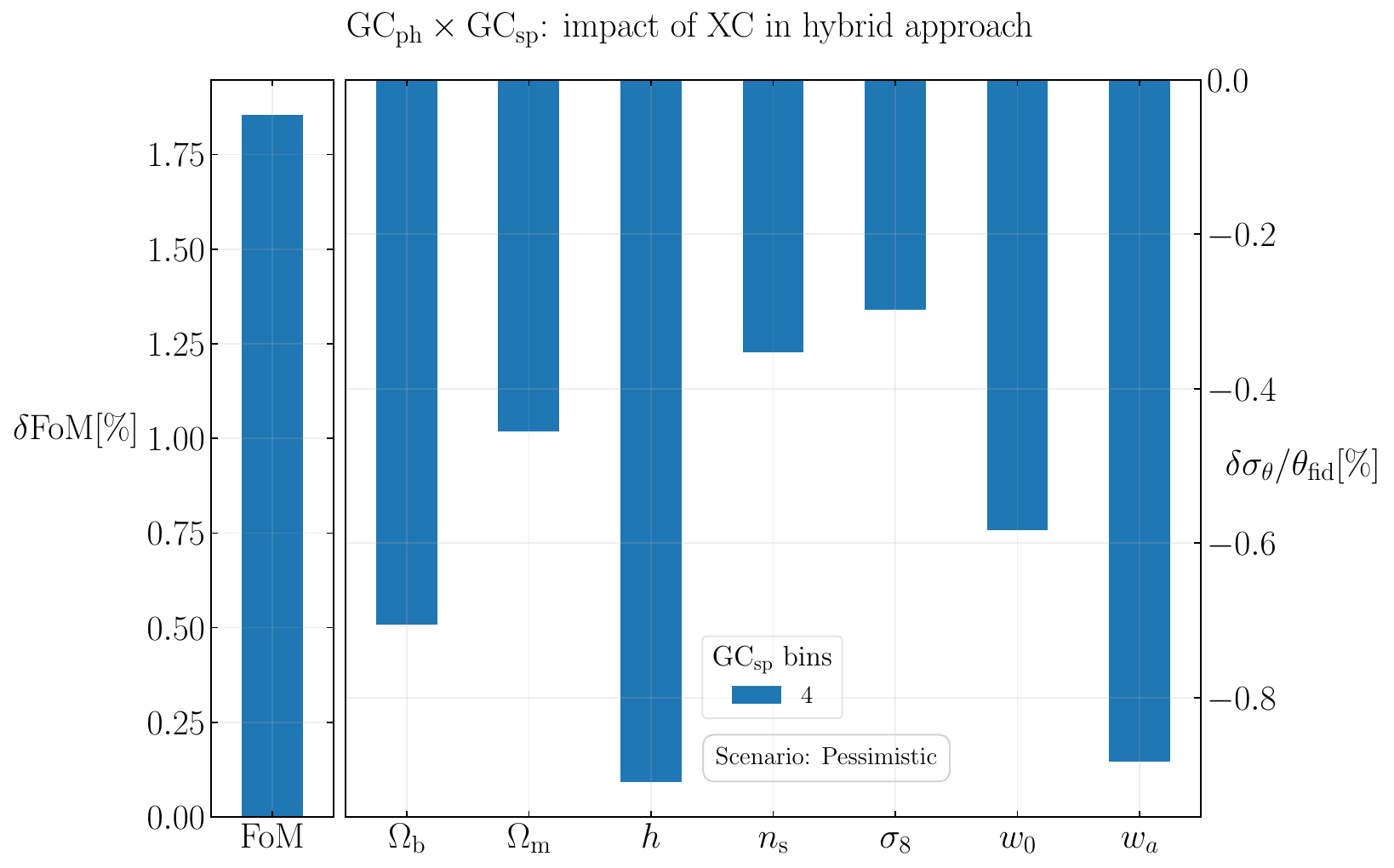}
\caption{Impact on FoM and marginalised $\onesigma$ errors, of the $\xc{\GCph}{\GCsp}$ in the hybrid approach, for the optimistic scenario. The reference here are the constraints coming from $[\GCph]+[\GCsp(P_k)]$, while the minuend for the percentage difference is $[\GCph+\xc{\GCph}{\GCsp}]+[\GCsp(P_k)]$. In the hybrid approach only $4$ tomographic bins were used to compute the $\xc{\GCph}{\GCsp}$ angular power spectra, for consistency with the Fourier power spectrum analysis. Note the opposite sign of the percentage differences for FoM and uncertainties.}
\label{fig:phsp_hybrid_xc_impact}
\end{figure*}

\subsubsection{\large{Impact of the $\xc{\GCph}{\GCsp}$ signal on parameter constraints}}

\paragraph{\textit{Harmonic approach}}

\begin{figure*}[h!]
\centering
\includegraphics[width=0.95\textwidth]{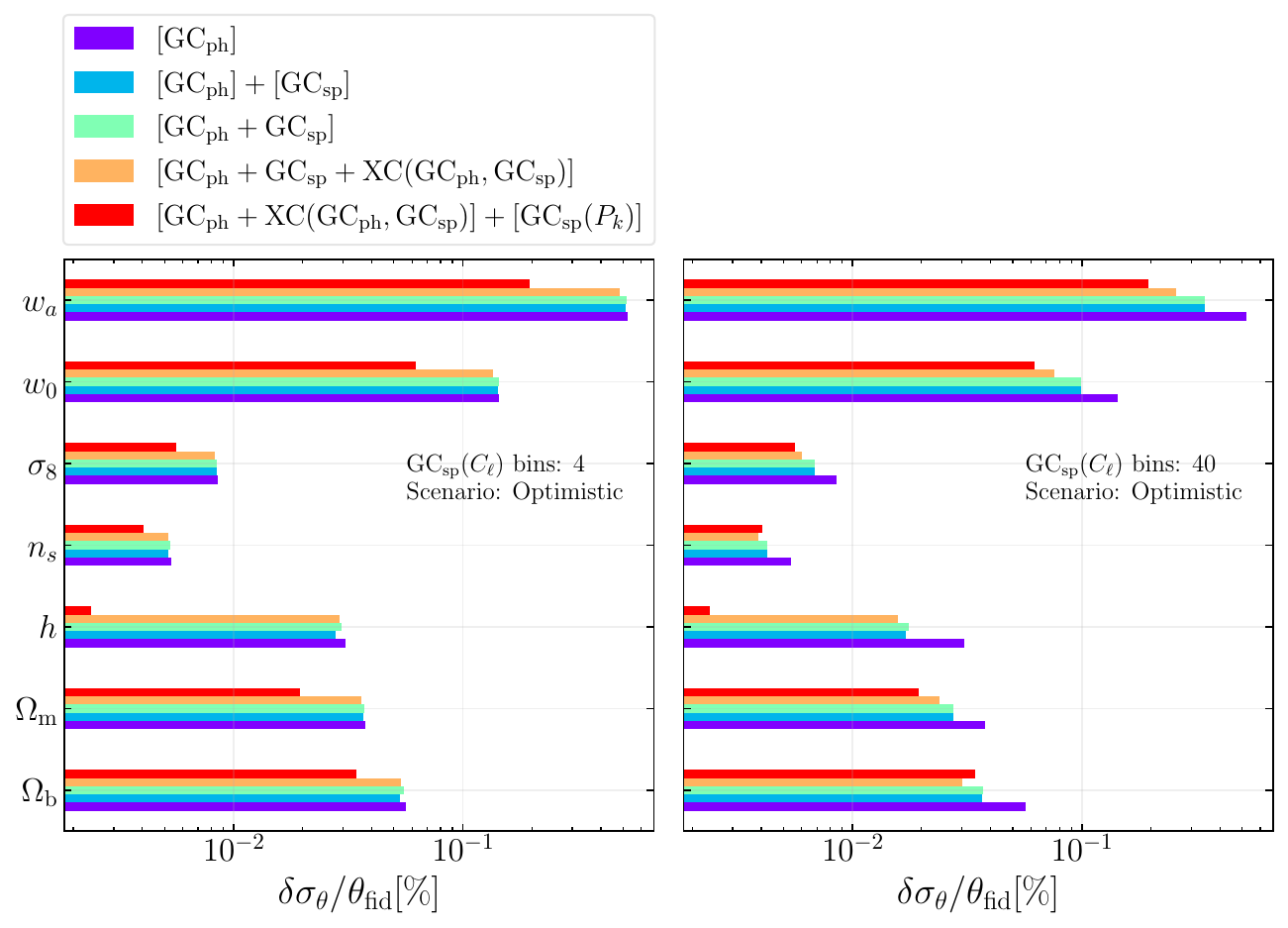}
    \caption{Comparison between harmonic and hybrid harmonic-Fourier approaches for the case of $\GCph\times\GCsp$.}
\label{fig:phsp_fullcl_vs_2dx3d_error_barplot}
\end{figure*}

In the harmonic approach the contribution to the FoM coming from the $\xc{\GCph}{\GCsp}$ cross-correlation is always positive, as reported in \cref{tab:results_phsp_fom}, and it is about $6\%$ $(38\%)$ with $4$ $(40)$ bins for $\GCsp$. This gain is visible also in \cref{fig:phsp_fullcl_vs_2dx3d_error_barplot}, where the FoM and the marginalised uncertainties from the  $[\GCph+\GCsp+\xc{\GCph}{\GCsp}]$ Fisher matrix are represented with green bars. The improvements on the errors for $\wz, \wa$, and consequently on the FoM, are particularly visible for finer binnings.

The marginalised uncertainties on the cosmological parameters also improve, but the gain is more modest than the one on the FoM, as it is shown in the right panel of \cref{fig:phsp_fullcl_percdiff_plot}. In fact, the biggest improvements are for $\wz$ and $\wa$, whose uncertainties get smaller at most by $23\%$ and $25\%$, respectively. Instead, the uncertainty on the baryon density parameter, $\Omb$, slightly increases by about $1\%$ with 4 bins and becomes smaller by $18\%$ with 40 bins. The small worsening on this parameter in the case of 4 bins has been attributed to the contribution of the cross-covariances between the $C(\ell)$'s included in the data-vector of the Fisher matrix $[\GCph+\GCsp+\xc{\GCph}{\GCsp}]$, which are not taken into account in the simple sum $[\GCph] + [\GCsp]$. It is possible that with 4 bins the gain coming from the inclusion of the $\rm XC$ signal is compensated by the covariance contribution, producing a net (small) worsening. However, as soon as the number of $\GCsp$ bins increases, the positive contribution of the cross-correlation signal starts to dominate, and the constraints on $\Omb$ to improve too.
The uncertainties on the other parameters also improve, in particular $\Omm$ and $\sige$ gain at most $12\%$ when $40$ bins are used, as well as $h$ and $\ns$ which improve by $7\%$ at maximum.

\begin{figure*}[t]
\centering
\includegraphics[width=0.75\textwidth]{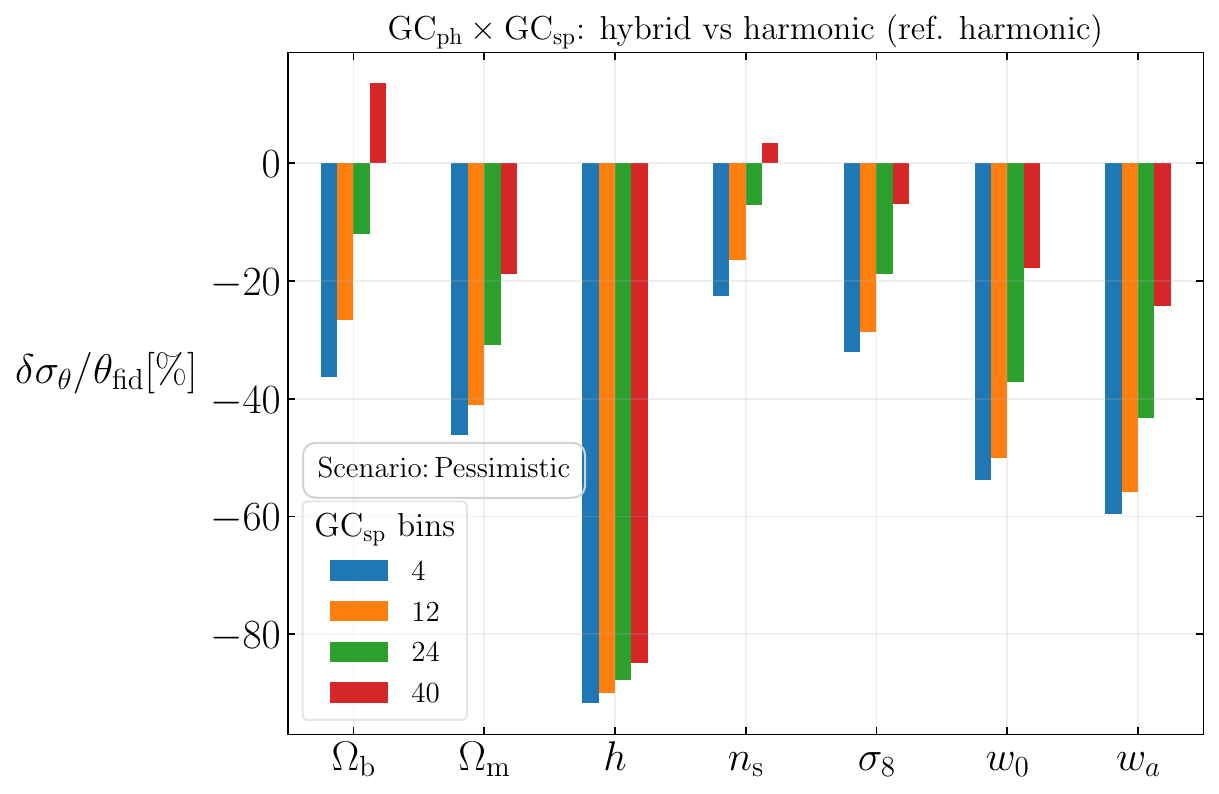}
\caption{Percentage difference between marginalised $\onesigma$ errors of the hybrid and harmonic approaches, for the combination of $\GCph$ and $\GCsp$ in the baseline optimistic scenario, \textcolor{black}{when the cross-covariance is included}. The percentage differences are normalised to the results of the harmonic approach.}
\label{fig:phsp_hybrid_vs_harmonic}
\end{figure*}

\paragraph{\textit{Hybrid approach}}
In the hybrid approach instead, the $\xc{\GCph}{\GCsp}$ has quite a small impact on the constraints: the marginalised uncertainties and FoM from the $[\GCph+\xc{\GCph}{\GCsp}] + [\GCsp(P_k)]$ Fisher matrix are very similar to the ones from $[\GCph] + [\GCsp(P_k)]$, as is qualitatively shown by the barplots of \cref{fig:phsp_fullcl_vs_2dx3d_error_barplot}. The percentage variations on the constraints resulting from adding the $\xc{\GCph}{\GCsp}$ are shown in \cref{fig:phsp_hybrid_xc_impact}. The variation on the FoM is contained between $+1.5\%$ and $+2\%$, so there is a very small improvement when including $\xc{\GCph}{\GCsp}$ in the hybrid approach. This is also the case for the marginalised uncertainties which improve less than $1\%$; \textcolor{black}{however, we would like to point out that this behavior could be due to numerical uncertainty.} 
Therefore, it can be concluded that, in the hybrid approach, the inclusion of the $\xc{\GCph}{\GCsp}$ cross-correlation in the combination of $\GCph$ and $\GCsp$ has a negligible impact on the results.

\paragraph{Hybrid approach vs harmonic approach}
As \cref{fig:phsp_hybrid_vs_harmonic} shows, the hybrid approach performs better than the harmonic one in constraining almost all the cosmological parameters, even if the constraining power of the harmonic approach significantly improves with increasing the number of $\GCsp$ tomographic bins. The harmonic approach has a FoM of $69$ ($153$) for $4$ ($40$) spectroscopic bins, while the hybrid one provides a FoM of $234$, which is $236\%$ ($52\%$) larger than in the former case. Thus the hybrid approach is better even when $40$ spectroscopic bins are used for the harmonic one. In particular, this is true for the reduced Hubble constant $h$. In this case the hybrid approach performs remarkably better, with a gain on the marginalised uncertainty which is always between $75\%$ and $100\%$, depending only slightly on the number of $\GCsp$ bins used for the harmonic approach. This is expected, since the hybrid approach takes advantage of the 3D power spectrum as an observable, in which radial BAO and RSD are included. The constraints on the other parameters appear instead quite sensitive to the $\GCsp$ binning. The differences on the $\wz$ and $\wa$ uncertainties significantly decrease from more than $50\%$ with $4$ bins to about $20\%$ with $40$ bins, as it could be expected from the FoM differences between the hybrid approach and the harmonic one. For $\Omm$ and $\sige$ the hybrid approach is still better than the harmonic one, even if the difference between the two approaches decreases significantly with the number of bins too.

The only exceptions are given by the baryon density, $\Omb$, and the spectral index, $\ns$. In particular, the $\Omb$ uncertainty from the hybrid approach is more than $25\%$ smaller than from the harmonic one with $4$ spectroscopic bins, but the situation gets reversed with $40$ bins. In the latter case the harmonic approach provides $\sim 10\%$ better constraints on $\Omb$ than the hybrid one. The $\ns$ uncertainty in the harmonic approach with $4$ bins is about $20\%$ larger than in the hybrid one. Instead, when $40$ bins are used for the harmonic approach, the $\ns$ uncertainty given by the hybrid approach is a few percents larger than the one given by the former.

To conclude, for the combination of $\GCph$ and $\GCsp$, the hybrid approach always provides better constraints than the harmonic one. Increasing the number of bins in the harmonic approach improves its performances. Using $40$ bins allows us to reach the performances of the hybrid approach for some parameters -- $\ns$ and $\Omb$ -- but a large gap still remains for the other constraints, especially for $h$ and the FoM.

\begin{figure*}[t]
\centering
\includegraphics[width=0.80\textwidth]{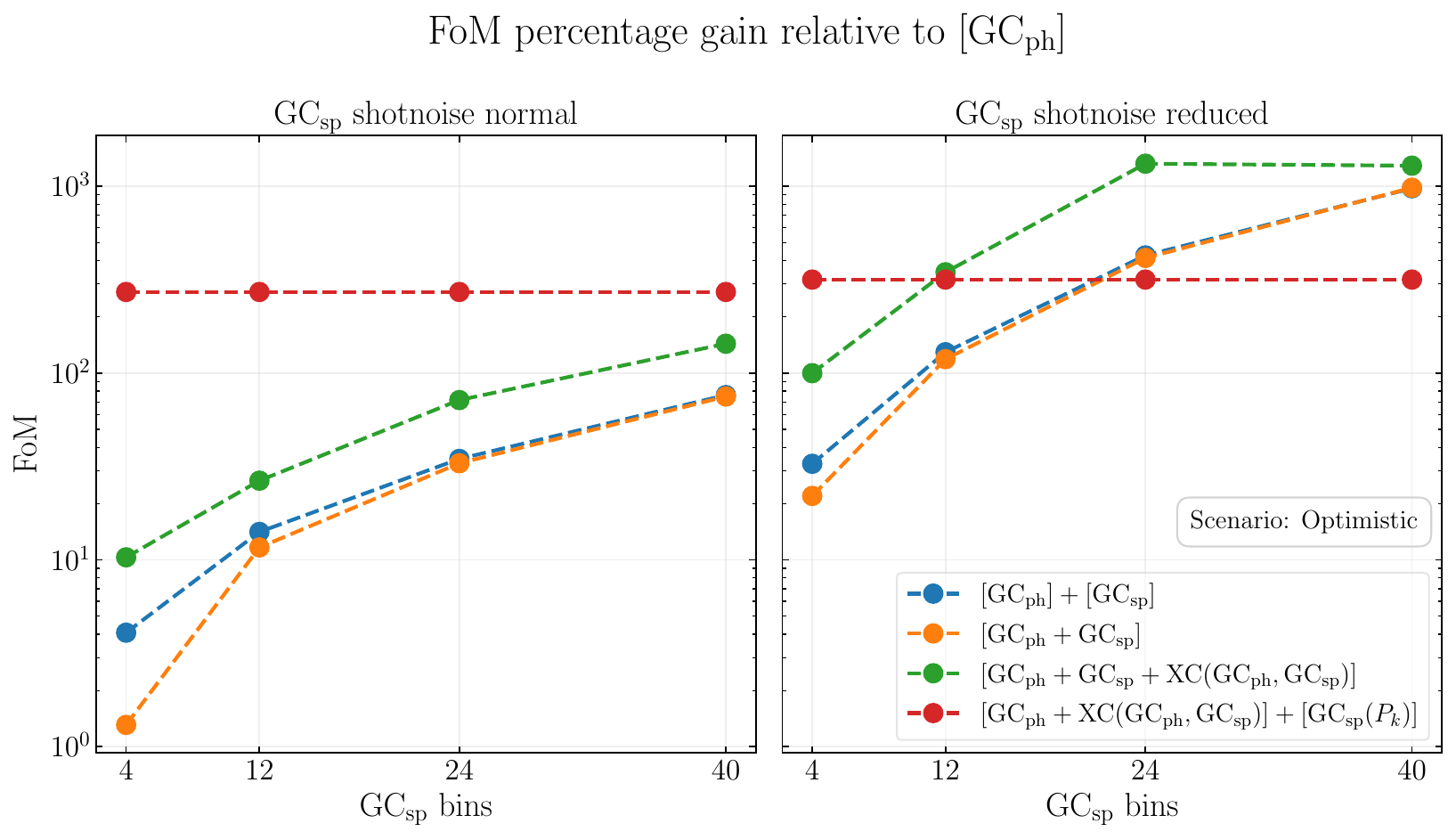}
\caption{Percentage gain on the FoM relative to photometric clustering alone ($[\GCph]$). In the left panel the computation has been done using the standard $\GCsp$ shot noise \cref{eq:gc_shot_noise}, while in the right panel the reduced version \cref{eq:reduced_gcsp_noise} has been used.}
\label{fig:phsp_fom_gain}
\end{figure*}
\subsubsection{Comparing photometric and spectroscopic clustering in the harmonic domain}
In the harmonic approach, two alternative configurations have been considered, both for $\GCph$ and $\GCsp$. The baseline configuration for $\GCph$ consists in using all the $10$ redshift bins reported in \cref{tab:tomo_bins}, while in the alternative configuration only the $4$ bins strictly contained in the range $[0.9, 1.8]$ are considered.

For spectroscopic clustering, the noise baseline settings correspond to using the shot noise as calculated from \cref{eq:gc_shot_noise}, while the alternative consists in using the artificially reduced version \cref{eq:reduced_gcsp_noise}. In fact, from further investigation it turned out that a great limitation of the harmonic approach is due to the shot noise associated with the $\GCsp$ auto-correlation $C(\ell)$'s. This is in fact much higher than the one associated with $\GCph$, since the expected number of $\Halpha$-emitting galaxies in the \Euclid catalogue ($\sim \num{2e7}$) is smaller than the expected number of galaxies in the photometric sample ($\sim \num{1.6e9}$). Therefore, in order to quantify the impact of the spectroscopic shot noise, the forecasts include also the artificially reduced shot noise setting, as explained in \cref{sec:shot_noise}.

The results are shown in \cref{fig:phsp_fom_gain}: in the left panel the standard $\GCsp$ shot noise of \cref{eq:gc_shot_noise} is employed, while in the right one the reduced noise of \cref{eq:reduced_gcsp_noise} is used. Lowering the shot noise systematically boosts the results of the harmonic approach by about one order of magnitude, making its performance comparable or even better than in the hybrid case. In particular, as can be expected, the observable which most improves its performance is the harmonic $\GCsp$ auto-correlation.

\begin{figure*}[t]
\centering
\includegraphics[width=0.70\textwidth]{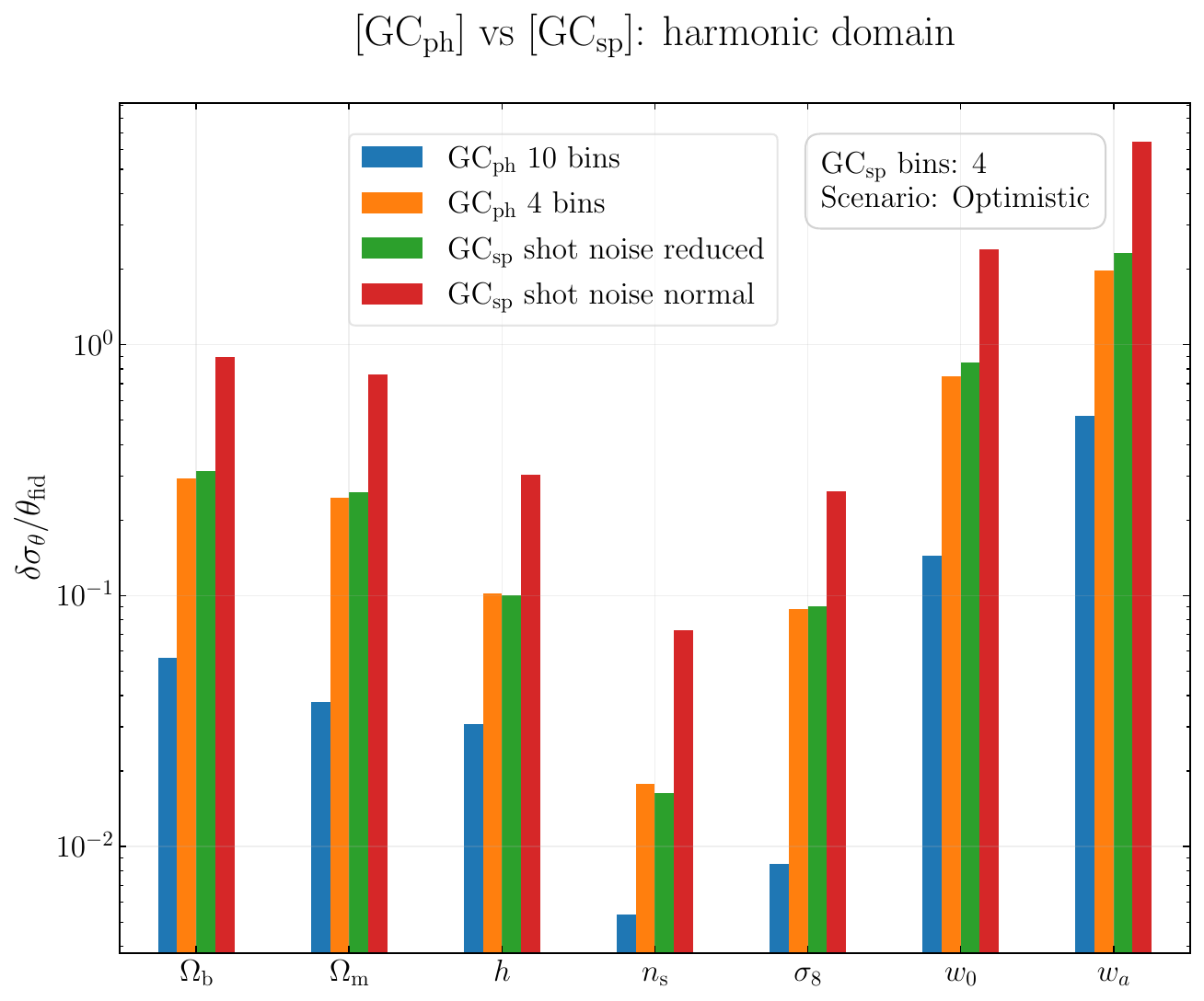}
\caption{Comparison between the constraints coming from $[\GCph]$ only and $[\GCsp]$ only in the harmonic approach, in the different configurations described in \cref{sec:results}.  The vertical bars represent the values of the FoM and the relative marginal uncertainties.} 
\label{fig:gcph_vs_gcsp_harmonic}
\end{figure*}

Instead, when considering the behaviour of the hybrid $[\GCph+\xc{\GCph}{\GCsp}] + [\GCsp(P_k)]$ Fisher matrix, as compared to the corresponding one in the harmonic domain, its performance remains stable against the change of the shot noise level, since the possible change of the latter would enter only $\GCsp(P_k)$, which is fixed. Therefore, while in the harmonic case the survey performance increases with the noise reduction and the increasing of the number of spectroscopic bins, in the hybrid case the performance remains unchanged since both the noise level and the bin number are fixed.


The results of these scenarios are reported in \cref{fig:gcph_vs_gcsp_harmonic}. The constraints from $\GCph$ only restricted to the $4$ bins in the spectroscopic range (blue bars) are very close to the ones of $\GCsp$ when the shot noise is reduced (yellow bars). Thus in these special conditions the two probes are essentially equivalent. This is expected, since the functional form of the weight for galaxy clustering is the same, as in \cref{eq:galaxy_weight,eq:sp_weight}, with the photometric function $W_i^{\rm ph}$ differing from the spectroscopic one $W_i^{\rm sp}$ only for the shape of the galaxy distribution and the values of the galaxy bias.

\Cref{fig:gcph_vs_gcsp_harmonic} shows also the two probes in the baseline configuration, in which $10$ bins are used for $\GCph$ and the more realistic shot noise of \cref{eq:gc_shot_noise}) is used for $\GCsp$. On the one hand, the usage of all bins for $\GCph$ reduces the uncertainties on cosmological parameters. On the other hand, the realistic shot noise of $\GCsp$ significantly affects its performances, making the uncertainties larger.

The conclusion is therefore that the shot noise and the redshift range of the galaxy sample are what make the difference between $\GCph$ and $\GCsp$ in terms of constraining power, when both are treated in the harmonic domain. When $\GCph$ is restricted to the same range of $\GCsp$ and the shot noise of the latter is reduced to the same level of the former, their constraints become comparable.
\\

\subsection{Combining weak lensing and spectroscopic clustering}
\label{sec:results_wlxsp}
In this section we report the results from the combination of $\WL$ and $\GCsp$. The Fisher matrix in the harmonic approach is $[\WL+\GCsp+\xc{\WL}{\GCsp}]$, while in the hybrid one it is $[\WL+\xc{\WL}{\GCsp}] + [\GCsp(P_k)]$. We quantify the impact on parameter constraints of the $\WL$-$\GCsp$ cross-covariance and the $\xc{\WL}{\GCsp}$ cross-correlation. We take as reference values the results coming from the Fisher matrix corresponding to independent combination of $\WL$ and $\GCsp$, that is $[\WL]+[\GCsp]$ in the harmonic approach and $[\WL]+[\GCsp(P_k)]$ in the hybrid one. For brevity, the forecasts of this combination are reported in the optimistic scenario only.

\begin{table*}[t]
\caption{Table reporting the $\FoM$ for $\WL$, $\GCsp$ and their cross-correlation. The $\Delta\FoM$ column contains the variation of the figure of merit with respect to the independent sum $[\WL] + [\GCsp]$ for the given number of bins. The $\Delta\FoM (\%)$ column is calculated by taking $\Delta\FoM$ as a percentage of the FoM of $[\WL] + [\GCsp]$.}
\centering
\begin{resultstable}{cl*{3}{c}}
\multicolumn{5}{c}{$\WL\times\GCsp$ FoM forecasts} \\
\midrule
$\GCsp$ \, \text{bins} & \text{Fisher matrix} & $\FoM$ & $\Delta\FoM$ & $\Delta\FoM$ (\%)\\
\midrule
\multirow{3}{*}{4} 
& [$\WL$] + [$\GCsp(P_k)$]                 & 158.13 & \text{--} & \text{--} \\
& [$\WL$+$\xc{\WL}{\GCsp}$] + [$\GCsp(P_k)$] & 182.74 & \text{--} & \text{--} \\
& [$\WL$] + [$\GCsp$]                      & 74.72  & \text{--} & \text{--} \\
& [$\WL+\GCsp$]                          & 74.37  & $-0.35$ & $\negative{-0.47\%}$\\
& [$\WL+\GCsp+\xc{\WL}{\GCsp}$]          & 103.56 & +28.83 & \positive{+38.59\%}\\
\midrule
\multirow{3}{*}{12} 
& [$\WL$] + [$\GCsp$] & 92.90 & \text{--} & \text{--}\\
& [$\WL+\GCsp$] & 92.77 & $-0.13$ & $\negative{-0.14\%}$\\
& [$\WL+\GCsp+\xc{\WL}{\GCsp}$] & 131.38 & +38.47 & \positive{+41.41\%}\\
\midrule
\multirow{3}{*}{24} 
& [$\WL$] + [$\GCsp$] & 111.41 & \text{--}& \text{--}\\
& [$\WL$+$\GCsp$] & 111.42 & +0.01 & \positive{+0.0021\%}\\
& [$\WL+\GCsp+\xc{\WL}{\GCsp}$] & 155.18 & +43.76 & \positive{+39.28\%}\\
\midrule
\multirow{3}{*}{40} 
& [$\WL$] + [$\GCsp$] & 141.12 & \text{--} & \text{--}\\
& [$\WL+\GCsp$] & 141.17 & +0.06 & \positive{+0.042\%}\\
& [$\WL+\GCsp+\xc{\WL}{\GCsp}$] & 188.45 & +47.34 & \positive{+33.55\%}\\
\end{resultstable}
\label{tab:results_wlsp_fom}
\end{table*}


\begin{figure*}[t]
\centering
\includegraphics[width=0.90\textwidth]{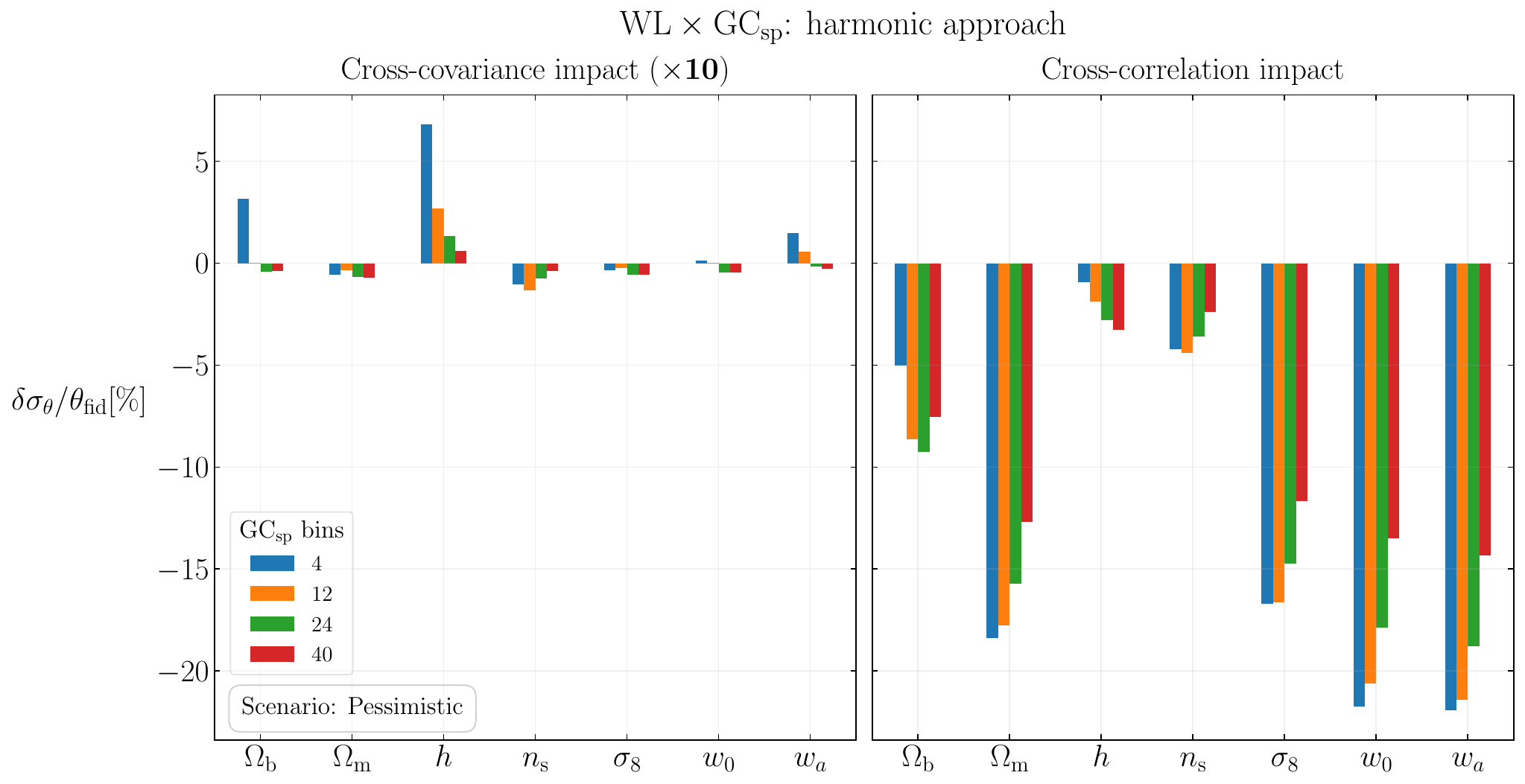}
\caption{Impact, on $\onesigma$ parameter errors, of cross-covariance and cross-correlation in the harmonic approach for the combination of $\WL$ and $\GCsp$. The left panel shows the impact of the cross-covariance, quantified with the percentage differences between $[\WL+\GCsp]$ and $[\WL] + [\GCsp]$. In the right panel the impact of the cross-correlation is reported, quantified by percentage differences of $[\WL+\GCsp+\xc{\WL}{\GCsp}]$ with respect to $[\WL]+[\GCsp]$. The percentages on the left panel have been multiplied by $10$ in order to make them visible with the same vertical scale of the right panel.}
\label{fig:wlsp_fullcl_percdiff_plot}
\end{figure*}

\subsubsection{Impact of the $\WL$-$\GCsp$ cross-covariance on parameter constraints}
\cref{tab:results_wlsp_fom} reports the FoM values resulting from the combinations of $\WL$ and $\GCsp$. In this case the cross-covariance is even more negligible than in the $\GCph$-$\GCsp$ case, always impacting by less than $0.5\%$ on the $\FoM$. This effect gets smaller as the number of bins increases, starting from $-0.47\%$ with $4$ spectroscopic bins up to $+0.042\%$ with $40$ spectroscopic bins respectively. This last is unexpectedly positive, but the variation is so small that can be considered as a numerical fluctuation around zero.

The effect of the cross-covariance is very small also on the marginalised $\onesigma$ uncertainties on the cosmological parameters, shown in the left panel of \cref{fig:wlsp_fullcl_percdiff_plot}. The marginalised uncertainties affected the most are the ones on $h$ and $\Omb$, with variations of $0.7\%$ and $0.3\%$ respectively. The variations on all the other parameters are always well below $0.2\%$, and in all cases the absolute value of these variations decreases as the number of $\GCsp$ bins increases, confirming the same trend observed for the FoM.

The conclusion here is that, when combining $\WL$ and $\GCsp$ in the harmonic domain, their cross-covariance can be safely neglected. \textcolor{black}{The limited impact of the WL-$\GCsp$ cross-covariance might have been expected also considering the limited redshift overlap of the two window functions, as can be seen in Fig.\ref{fig:weight_functions_plot}.} A similar result for another experiment other than \Euclid has been obtained in \cite{Joachimi:2020abi}. Here the authors performed a joint data analysis combining weak lensing measurements from the Kilo-Degree Survey (KiDS-1000) and spectroscopic clustering from the Baryon Acoustic Oscillations Survey (BOSS) and 2-degree Field Lensing Survey (2dFLenS). The $\WL$ was treated using the harmonic power spectrum as an observable, as has been done in this work. Moreover, the cross-covariance matrix was computed in the harmonic domain, considering only the correlation between $\WL$ and the transverse component of $\GCsp$. The authors estimated the covariance matrix for the data through an analysis of over $\num{20 000}$ fast full-sky mock galaxy catalogues, finding that the off-diagonal (cross-covariance) terms were negligible with respect to the diagonal (auto-covariance) ones.

\begin{figure*}[t]
\centering
\includegraphics[width=0.75\textwidth]{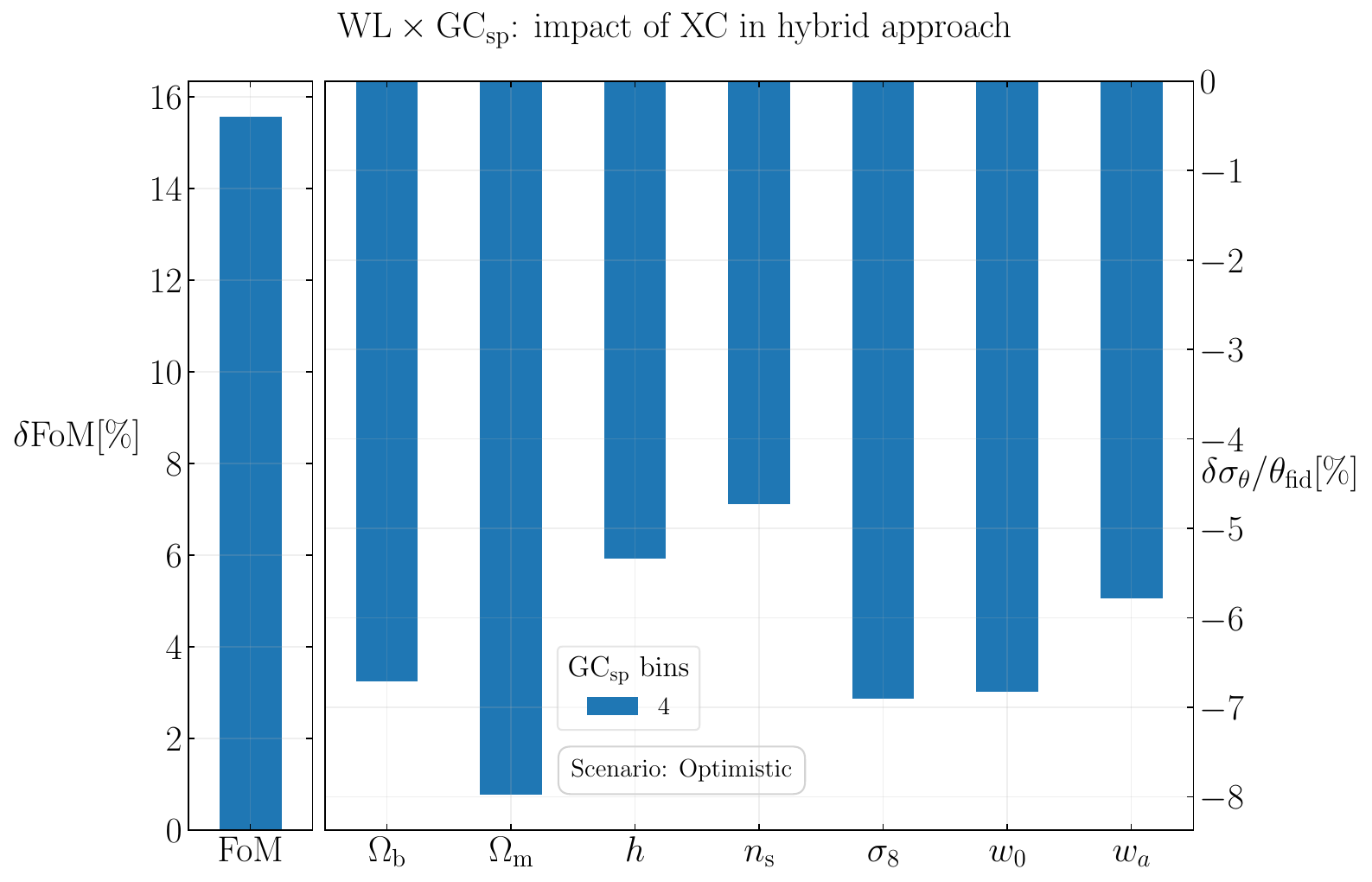}
\caption{Impact on FoM and marginalised $\onesigma$ errors, of the $\xc{\WL}{\GCsp}$ in the hybrid approach, for the optimistic scenario. In the hybrid approach only $4$ tomographic bins were used to compute the $\xc{\WL}{\GCsp}$ angular power spectra, for consistency with the Fourier power spectrum analysis. Note the opposite sign of the percentage differences for FoM and uncertainties.}
\label{fig:wlsp_hybrid_xc_impact}
\end{figure*}

\subsubsection{Impact of the $\xc{\WL}{\GCsp}$ signal on parameter constraints}
\paragraph{\textit{Harmonic approach}}
When combining $\WL$ and $\GCsp$, the $\xc{\WL}{\GCsp}$ has a quite significant impact on the results of the harmonic approach, as \cref{tab:results_wlsp_fom} shows. The percentage gain on the $\FoM$ is always larger than $+30\%$, and slightly depends on the number of spectroscopic bins used. The mild dependence on the number of $\GCsp$ bins can be explained by observing that the $\WL$ weight functions of \cref{eq:weak_lensing_full_weight} have a broad support, which becomes larger as the tomographic index increases, as \cref{fig:weight_functions_plot} shows. This suggests that increasing the radial resolution may not help in improving the constraints coming from the $\xc{\WL}{\GCsp}$.

It is also worth noting that the FoM percentage gain does not strictly increase with the number of spectroscopic bins.
This not intuitive behavior is due to the normalization of the FoM percentage difference, which is the FoM of the $[\WL] + [\GCsp]$ Fisher matrix. This quantity depends on the number of spectroscopic bins, and it increases slightly faster than the variation induced by the cross-correlation in the FoM of the $[\WL+\GCsp+\xc{\WL}{\GCsp}]$ Fisher matrix. This can be seen from \cref{tab:results_wlsp_fom}: the difference $\Delta\FoM$ between the FoMs of the $[\WL+\GCsp+\xc{\WL}{\GCsp}]$ and $[\WL] + [\GCsp]$ Fisher matrices grows more slowly with the number of bins than the FoM of $[\WL] + [\GCsp]$ alone.

The marginalised $\onesigma$ uncertainties decrease when the $\xc{\WL}{\GCsp}$ cross-correlation signal is included, especially the ones on $\wz, \wa, \sige$ and $\Omm$, as it can be seen from \cref{fig:wlsp_fullcl_percdiff_plot}. For these parameters the improvement is always larger than $12\%$ with $4$ $\GCsp$ bins, and it is about $20\%$ at most when $40$ bins are used. The gain on the $\Omb$ uncertainty is more modest, being contained between $5\%$ and $10\%$. The gain on the uncertainties of the reduced Hubble constant $h$ and the scalar spectral index $\ns$ is always less than $5\%$.

The improvement in the marginalised uncertainties $\xc{\WL}{\GCsp}$ cross-correlation seems to decrease when increasing the number of spectroscopic bins, as happens for the FoM. Again, the reason of this behavior is that the performances of the Fisher matrix taken as reference -- that is $[\WL] + [\GCsp]$ -- improve faster than the relative contribution of the cross-correlation. This is true for all parameters except for $h$, which is the parameter affected the least.

\paragraph{\textit{Hybrid approach}}
The impact on the constraints of the $\xc{\WL}{\GCsp}$ cross-correlation in the hybrid approach is less significant than in the harmonic one, as shown in \cref{fig:wlsp_hybrid_xc_impact}.
The FoM of the $[\WL+\xc{\WL}{\GCsp}] + [\GCsp(P_k)]$ Fisher matrix is $183$, which is $\sim 15\%$ higher than the one of the independent combination $[\WL] + [\GCsp(P_k)]$, which is $158$. The improvements in the marginalised uncertainties are all contained between $5\%$ and $7\%$, and there are no significant differences between the various parameters. The uncertainty on $\Omm$ is the most affected, gaining about $8\%$, while the least affected is the uncertainty on $\ns$, which is slightly lower than $5\%$.

\begin{figure*}[t]
\centering
\includegraphics[width=0.75\textwidth]{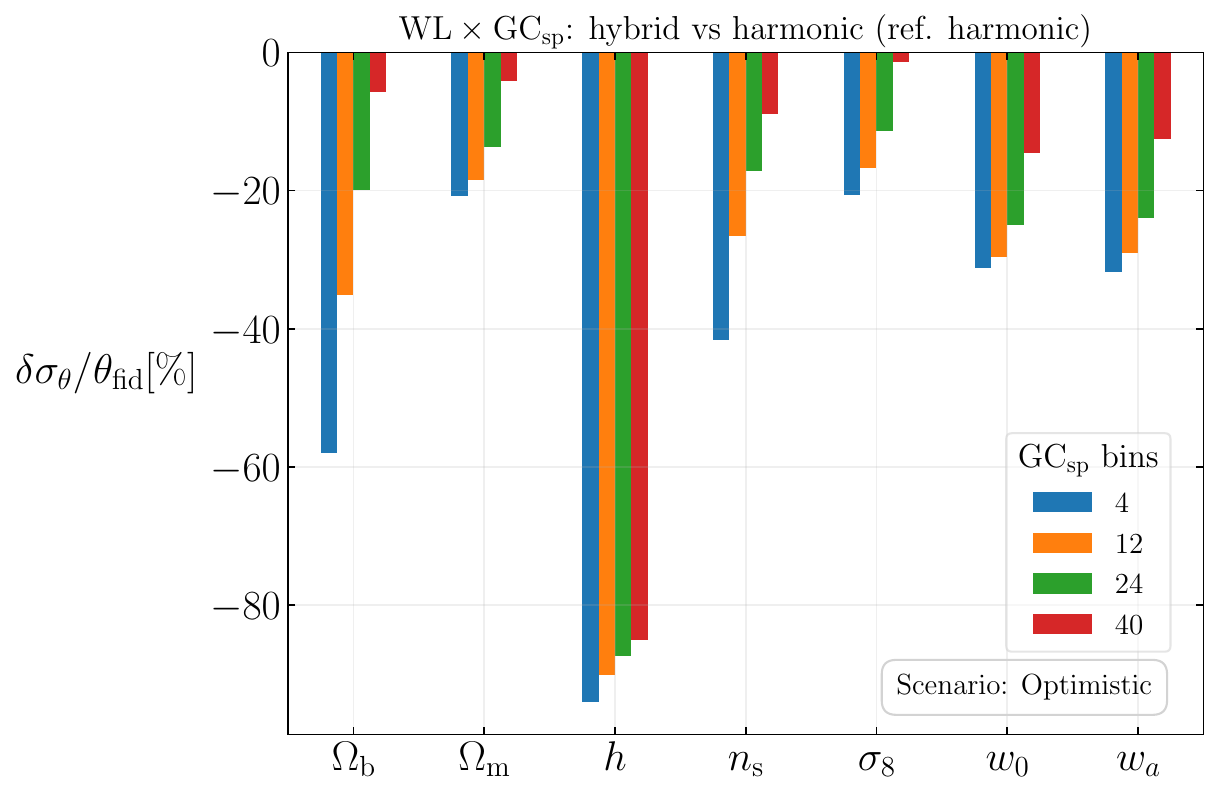}
\caption{Percentage difference between marginalised $\onesigma$ errors of the hybrid and harmonic approaches, for the combination of $\WL$ and $\GCsp$ in the baseline optimistic scenario. The percentage differences are normalised to the results of the harmonic approach.}
\label{fig:wlsp_hybrid_vs_harmonic}
\end{figure*}

\begin{figure*}[t]
\centering
\includegraphics[width=0.80\textwidth]{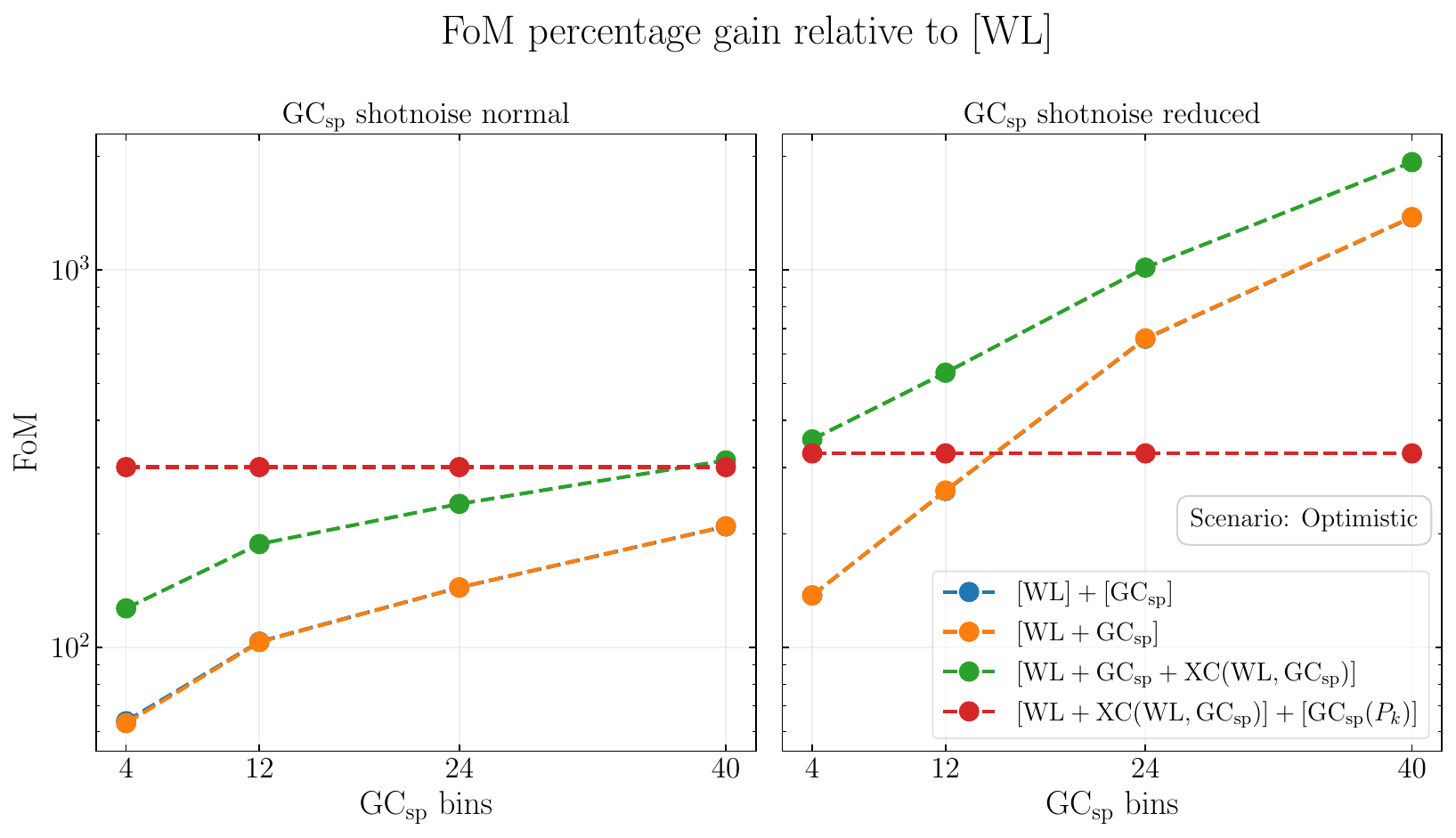}
\caption{Percentage gain on the FoM relative to weak lensing alone ($[\WL]$). In the left panel the computation has been done using the standard $\GCsp$ shot noise \cref{eq:gc_shot_noise}, while in the right panel the reduced version \cref{eq:reduced_gcsp_noise} has been used.}
\label{fig:wlsp_fom_gain}
\end{figure*}

\begin{figure*}[t]
\centering
\includegraphics[width=0.95\textwidth]{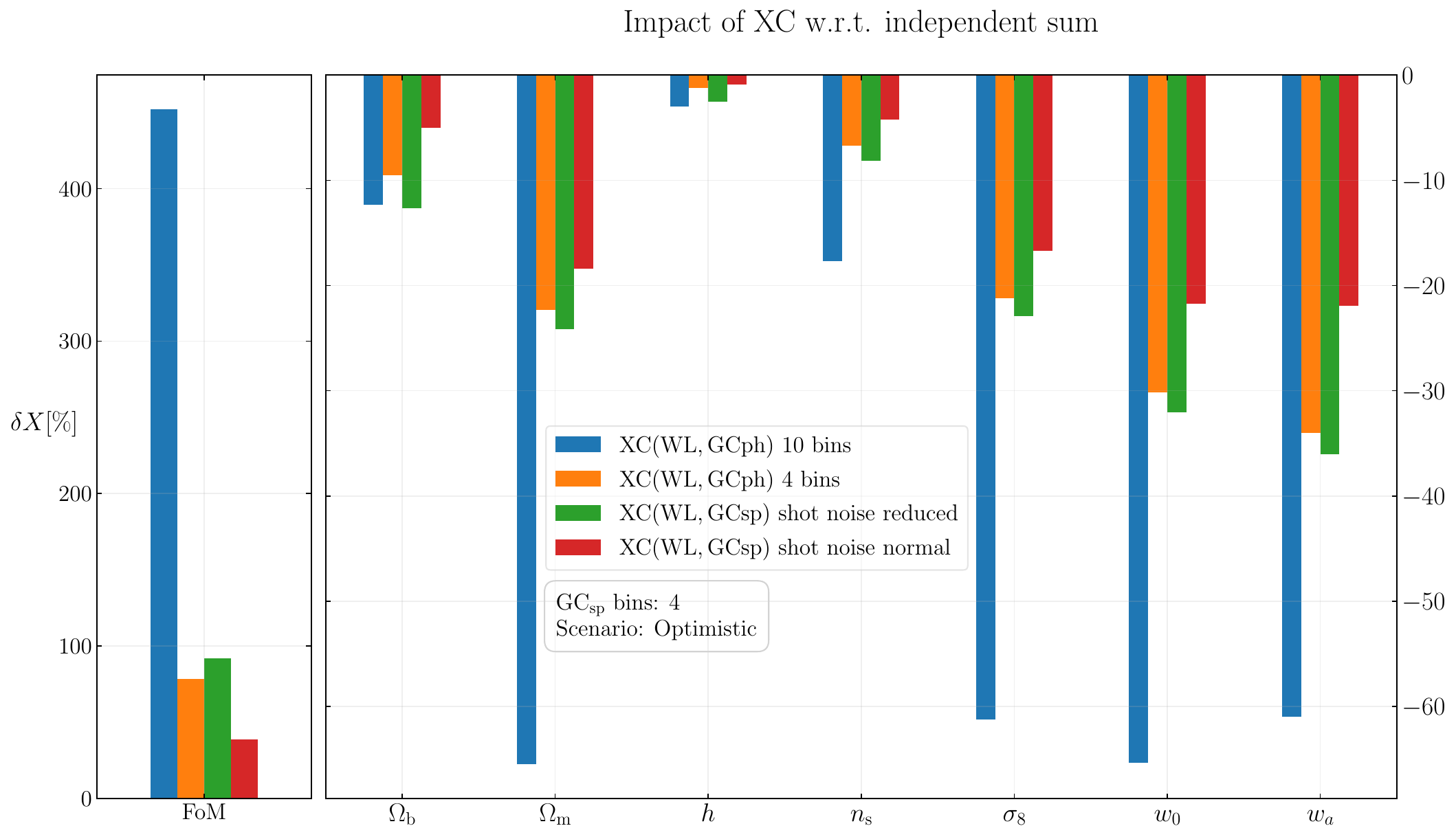}
\caption{Impact, on FoM and marginalised $\onesigma$ uncertainties, of the $\xc{\WL}{\GCph}$ and $\xc{\WL}{\GCsp}$ cross-correlations in different configurations. The vertical bars represent the percentage differences on the uncertainties normalised to the independent sum of the two probes. For $\xc{\WL}{\GCph}$ the reference is $[\WL] + [\GCph]$, while for $\xc{\WL}{\GCsp}$ is $[\WL] + [\GCsp]$. Note the opposite sign of the percentage differences for FoM and uncertainties.}
\label{fig:xcwlph_vs_xcwlsp}
\end{figure*}

\subsubsection{Hybrid approach vs harmonic approach}
When combining $\WL$ and $\GCsp$, the harmonic approach can reach the FoM of the hybrid one. This happens when $40$ $\GCsp$ bins are used to compute the harmonic Fisher matrix $[\WL+\GCsp+\xc{\WL}{\GCsp}]$. The value reached by the FoM is $188$, which is only $3\%$ higher than the on given by the hybrid approach. Moreover, \cref{fig:wlsp_hybrid_vs_harmonic} shows that the marginalised uncertainties in $\wz$-$\wa$ in the hybrid approach are smaller than their harmonic counterparts. Nonetheless, the FoM in the harmonic approach is slightly higher than the one in the hybrid approach. This is due to the correlation $\mathcal{C}_{\wz\wa}$ between the parameters, which enters the definition of the FoM in \cref{eq:fom_def}. The correlation is higher for the harmonic approach, and this compensates for the larger uncertainties, with a net result of a slightly higher FoM. 

Regarding the marginalised uncertainties, \cref{fig:wlsp_hybrid_vs_harmonic} shows that the hybrid approach always performs better than the harmonic one. The reduced Hubble constant $h$ is the parameter for which the difference is highest. In particular, the hybrid approach produces an uncertainty on $h$ which is $\sim 90\%$ ($75\%$) smaller than the one given by the harmonic approach with $4$ ($40$) spectroscopic bins. For the uncertainties on the other parameters the gap is smaller, and it reduces significantly as the number of spectroscopic bins increases. The most sensitive uncertainty is the one on $\Omb$, for which the gap between the two approaches decreases from more than $50\%$ to less than $10\%$ when $4$ and $40$ bins are used for the harmonic approach respectively. The uncertainty on $\ns$ is quite sensitive to the number of bins too, and the difference between the two approaches ranges from $40\%$ to $10\%$ when the number of bins of the harmonic approach increases from $4$ to $40$. Finally, the differences on the marginalised uncertainties on $\Omm$ and $\sige$ are smaller, ranging from $20\%$ to less than $5\%$.

\subsubsection{Comparing $\xc{\WL}{\GCsp}$ with $\xc{\WL}{\GCph}$}
In \citetalias{2020A&A...642A.191E} it has been shown that the $\xc{\WL}{\GCph}$ cross-correlation considerably improves the constraints on the cosmological parameters. We found the same result in this work: in the optimistic scenario the Fisher matrix $[\WL+\GCph+\xc{\WL}{\GCph}]$ yields a FoM which is a factor of $\sim 5$ higher than the one given by the $[\WL] + [\GCph]$ Fisher matrix. This means that the percentage gain induced by the cross-correlation signal is about $400\%$. The $\xc{\WL}{\GCsp}$ cross-correlation has a smaller impact on the constraints, as the $[\WL+\GCsp+\xc{\WL}{\GCsp}]$ Fisher matrix produces a FoM at most $\sim 40\%$ higher than the one of the $[\WL] + [\GCsp]$ combination. This is what happens when both $\GCph$ and $\GCsp$ are treated in the baseline settings, i.e. when the standard shot noise of \cref{eq:gc_shot_noise} is used for $\GCsp$ and $10$ tomographic bins are used for $\GCph$. The effect of the shot noise on $\GCsp$ can be seen from \cref{fig:wlsp_fom_gain}, which displays the percentage gain on the FoM of various combinations with respect to $\WL$ alone. The left panel refers to the standard $\GCsp$ shot noise, the right panel refers to the alternative reduced noise of \cref{eq:reduced_gcsp_noise}. When the noise is reduced, the Fisher matrix of the harmonic approach with $4$ bins is already competitive with the one of the hybrid approach, yielding a gain of $\sim 300\%$ relative to $\WL$ alone. The observable gaining the most from the noise reduction is the $\GCsp$ auto-correlation, as the two curves of $[\WL+\GCsp]$ and $[\WL+\GCsp+\xc{\WL}{\GCsp}]$ converges towards each other as the number of bins increases. 

A direct comparison between the gain coming from $\xc{\WL}{\GCph}$ and $\xc{\WL}{\GCsp}$ cross-correlations is shown in \cref{fig:xcwlph_vs_xcwlsp}. When $\GCsp$ and $\GCph$ have the same shot noise level and the same redshift range, the effect of the $\xc{\WL}{\GCsp}$ on the constraints becomes comparable with the one of $\xc{\WL}{\GCph}$.

Therefore, the conclusion is the same as that drawn for the direct comparison between the $\GCph$ and $\GCsp$ auto-correlations. The shot noise and the redshift range of the galaxy catalogue are what makes the differences between $\xc{\WL}{\GCph}$ and $\xc{\WL}{\GCsp}$ in terms of constraining power.
\begin{figure*}[h!]
\centering
\includegraphics[width=0.95\textwidth]{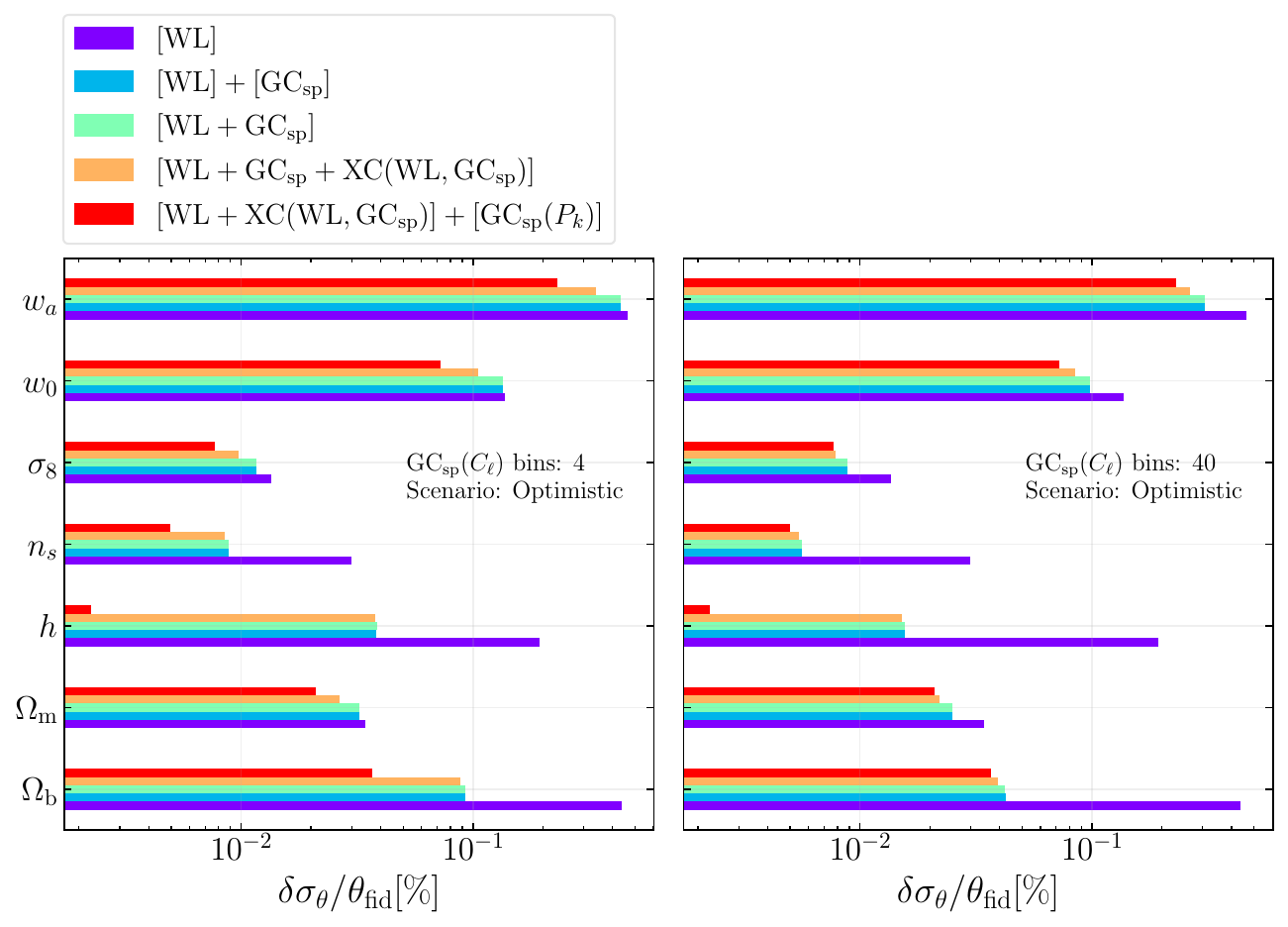}
\caption{Comparison between harmonic and hybrid harmonic-Fourier approaches for the case of $\WL\times\GCsp$.}
\label{fig:wlsp_fullcl_vs_2dx3d_error_barplot}
\end{figure*}
\\

\section{The \Euclid \sixtwoptShort\ statistics}
\label{sec:results_6x2pt}\
This section contains the main results of our paper, the Euclid full $\sixtwoptShort$ pt constraints and its comparison with the other approaches. Here we present the results of the combination of all the \Euclid main probes: $\WL, \GCph$, and $\GCsp$. The starting point is the (photometric) \threetwoptShort\ statistics, defined as
\begin{equation}\label{eq:3x2pt_def}
\threetwoptShort\ \equiv [\threetwoptLong]\;.
\end{equation}
In terms of this combination, the \sixtwoptShort\ statistics can be expressed as
\begin{multline}
\label{eq:6x2pt_harmonic_defs_from_3x2pt}
[\sixtwoptShort\ \, (\mathrm{harmonic})] = [\threetwoptShort\ + \GCsp \, \\ + \xc{\WL}{\GCsp} + \xc{\GCph}{\GCsp}] \;,
\end{multline}
for the full harmonic approach, and
\begin{multline}\label{eq:6x2pt_hybrid_defs_from_3x2pt}
[\sixtwoptShort\ \, (\mathrm{hybrid})] =\; [\threetwoptShort\ + \xc{\WL}{\GCsp} \\
+ \xc{\GCph}{\GCsp}] + [\GCsp(P_k)] \;.
\end{multline}
for the hybrid approach.

The discussion will be focused on two main points: 
\begin{itemize}
\item the importance of the cross-covariance between $\GCsp$ and the \threetwoptShort\ statistics;
\item the contribution to the constraints of the $\xc{\GCph}{\GCsp}$ and $\xc{\WL}{\GCsp}$ cross-correlations.
\end{itemize}
The cross-covariance between $\GCsp$ and \threetwoptShort\ statistics is studied only in the harmonic approach, since in the hybrid approach it is neglected. The effect of the cross-correlations is assessed for both the harmonic and the hybrid approaches, and in both the pessimistic and optimistic scenarios defined in \cref{tab:scenarios_settings}.

\begin{figure*}[t]
\centering
\includegraphics[width=0.95\textwidth]{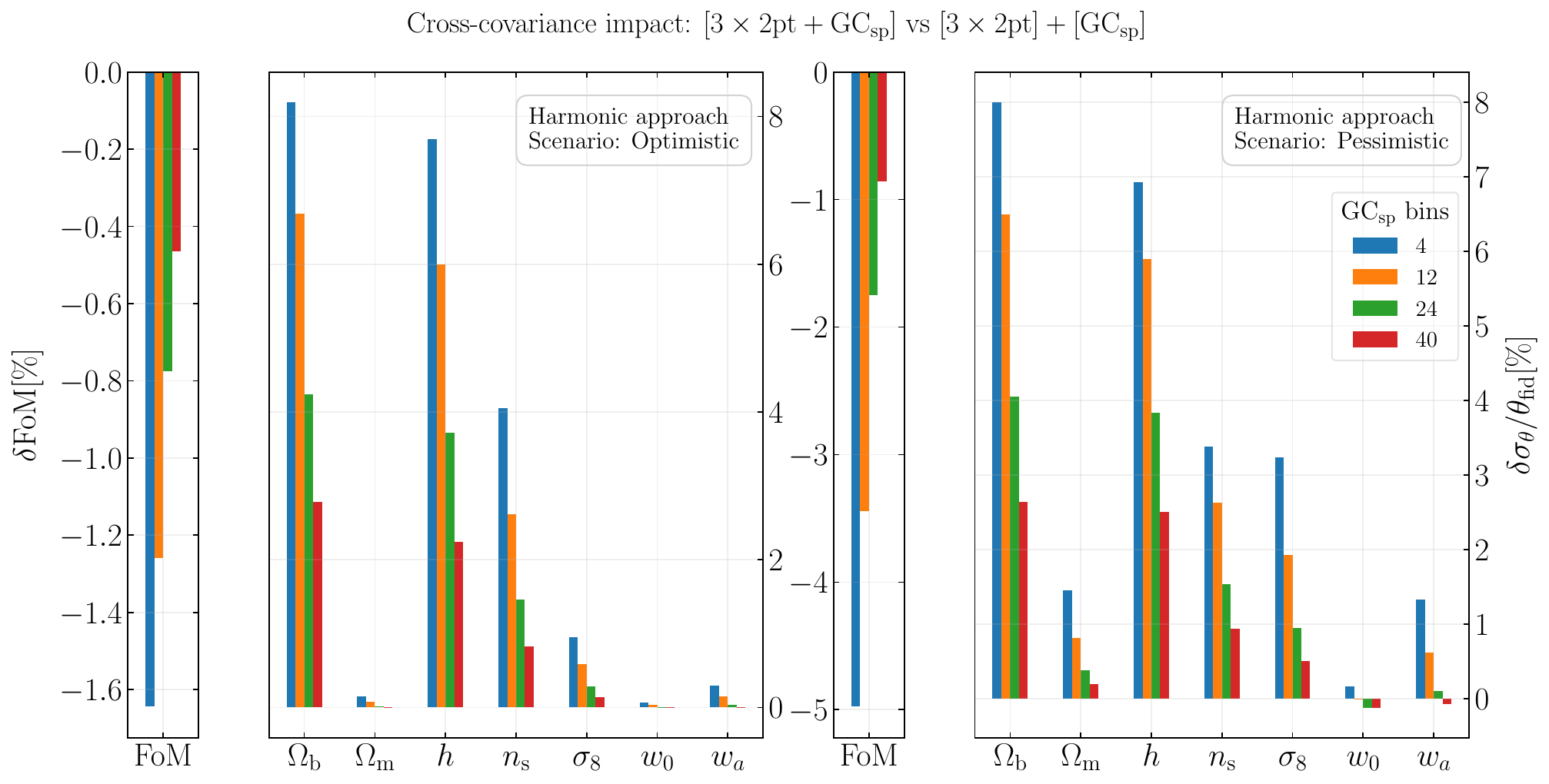}
\caption{Impact on FoM and marginalised 1-$\sigma$ uncertainties, of cross covariances between $\GCsp$ and $\threetwoptShort\, ([\WL+\GCph+\xc{\WL}{\GCph}])$, quantified with percentage differences on the constraints, in the optimistic (left) and pessimistic (right) scenarios. The reference for the percentage are the constraints of $[\threetwoptShort] + [\GCsp]$, where $\GCsp$ is considered as \emph{independent} from the rest. Note the opposite sign of the percentage differences for FoM and uncertainties.}
\label{fig:3x2pt_sp_cov_impact}
\end{figure*}

\subsection{Impact of the $\GCsp$ cross-covariances on parameter constraints}

The impact on the constraints of the cross-covariance between $\GCsp$ and \threetwoptShort\ is shown in the two panels of \cref{fig:3x2pt_sp_cov_impact}. The plot compares the constraints from the $[\threetwoptShort+\GCsp]$ and $[\threetwoptShort]+[\GCsp]$ combinations: in the former the $\GCsp$-\threetwoptShort\ cross-covariance is taken into account, while it is not in the latter. The impact of the cross-covariance is almost the same in all the scenarios and decreases as the number of $\GCsp$ bins increases, confirming the same trend observed in the two pairwise combinations $\GCph$-$\GCsp$ and $\WL$-$\GCsp$. The percentage variations on the constraints are always below $10\%$ ($5\%$) with $4$ ($40$) spectroscopic bins. The covariance almost always worsens the constraints with respect to considering $\GCsp$ and \threetwoptShort\ as independent. The parameters whose uncertainties are affected the most by the $\GCsp$-\threetwoptShort\ cross-covariance are $\Omb$ and $h$, with variations of $\sim 8\%$ with $4$ bins. When $40$ bins are used for $\GCsp$ the variation reduces to $\sim 2\%$ for both parameters.

One of the most evident differences between the optimistic and the pessimistic scenario is the impact on the FoM, which is slightly higher in the pessimistic than in the optimistic setting. However the percentage difference is always below $5\%$, the worst case being the pessimistic scenario with $4$ bins, where it is $\sim 4\%$. The percentage variation on the $\Omm$ uncertainty is at the sub-percent level in the optimistic scenario, while it is about at the percent level in the pessimistic scenario. The uncertainty on $\sige$ is always smaller than $1\%$ in the optimistic scenario, while it ranges from $4\%$ to $1\%$ in the pessimistic scenario.

\begin{figure*}[t]
\centering
\includegraphics[width=0.95\textwidth]{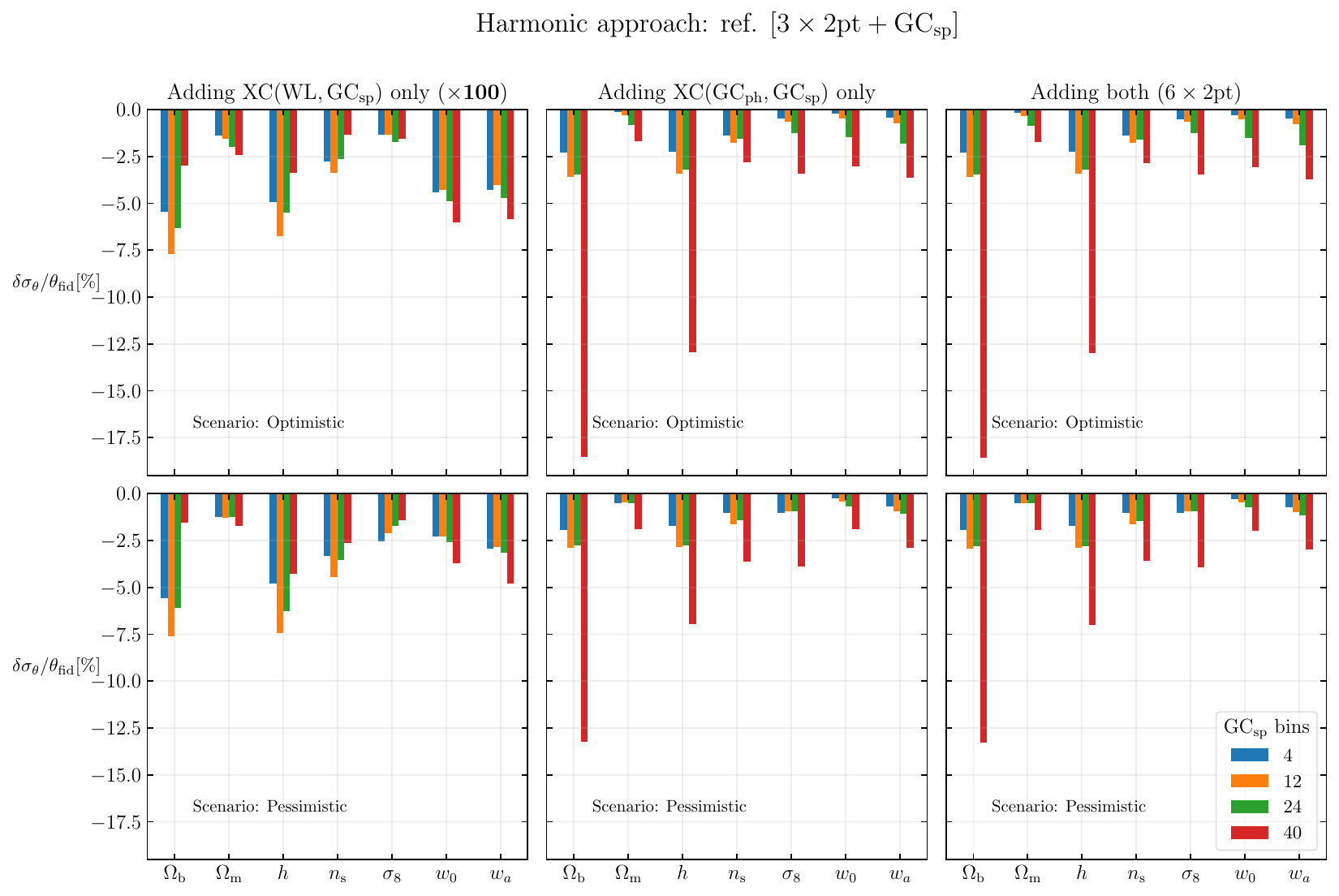}
\caption{Impact of the $\xc{\GCph}{\GCsp}$ and $\xc{\WL}{\GCsp}$ cross-correlations on marginalised $\onesigma$ uncertainties, with respect to the constraints given by the $[\threetwoptShort+\GCsp]$ Fisher matrix, in which the cross-covariance between \threetwoptShort\ statistics and $\GCsp$ is taken into account. The top panels refer to the optimistic scenario, while the bottom panels refer to the pessimistic one. The percentage differences related to $\xc{\WL}{\GCsp}$ have been multiplied by $100$ to make them visible when using a single scale on the $y$ axis.}
\label{fig:6x2pt_harmonic_xcs_impact_ref_3x2pt_gcsp_cov}
\end{figure*}

\subsection{Impact of the $\rm \mathbf{XC}$ signals on parameter constraints}
\begin{figure*}[t]
\centering
\includegraphics[width=0.95\textwidth]{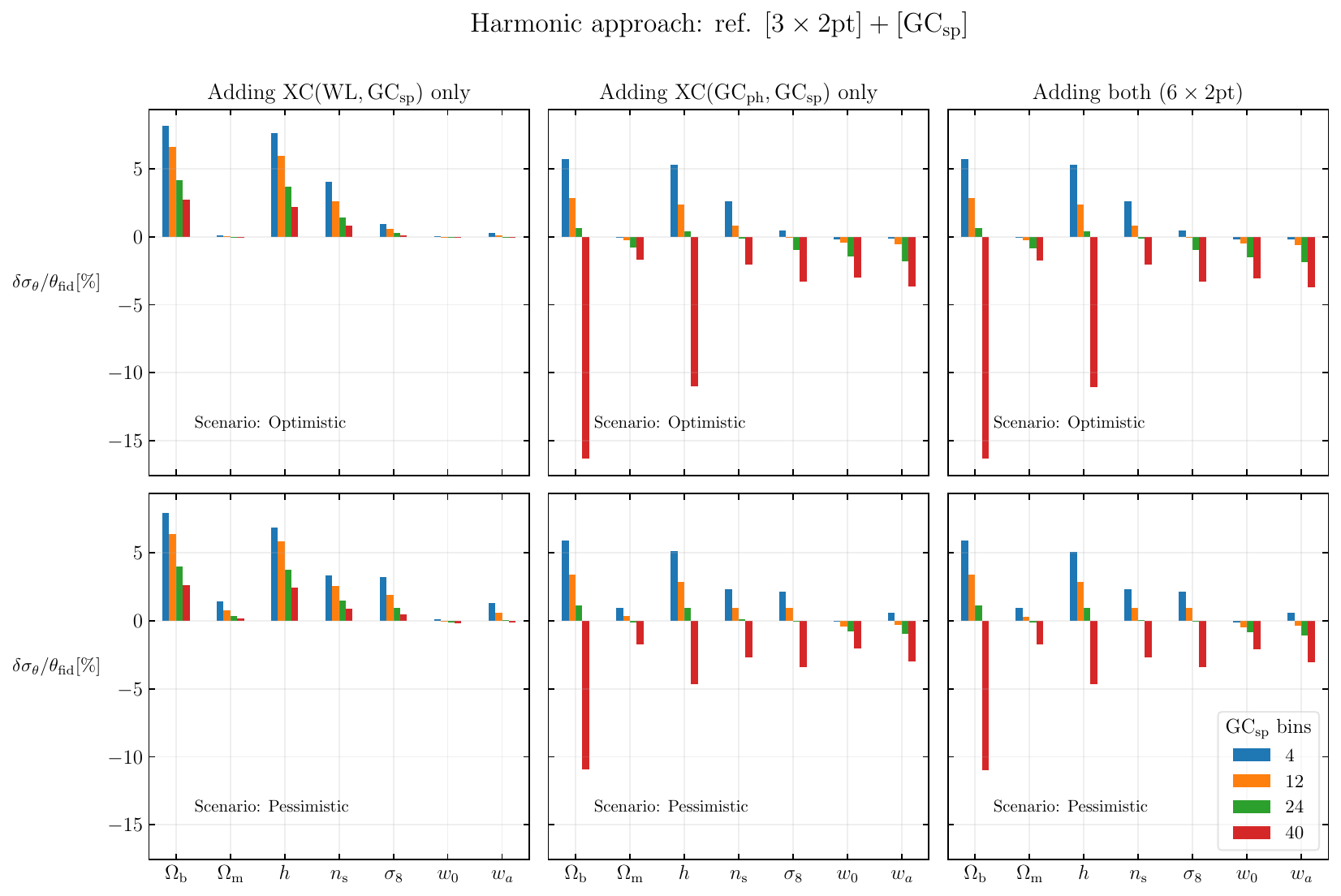}
\caption{Impact of the $\xc{\GCph}{\GCsp}$ and $\xc{\WL}{\GCsp}$ cross-correlations on marginalised $\onesigma$ uncertainties, with respect to the constraints given by $[\threetwoptShort]+[\GCsp]$, \threetwoptShort\ and $\GCsp$ are combined as independent. The top panels refer to the optimistic scenario, while the bottom panels refer to the pessimistic one.}
\label{fig:6x2pt_harmonic_xcs_impact_ref_3x2pt_gcsp_nocov}
\end{figure*}

\begin{figure*}[t]
\centering
\includegraphics[width=0.95\textwidth]{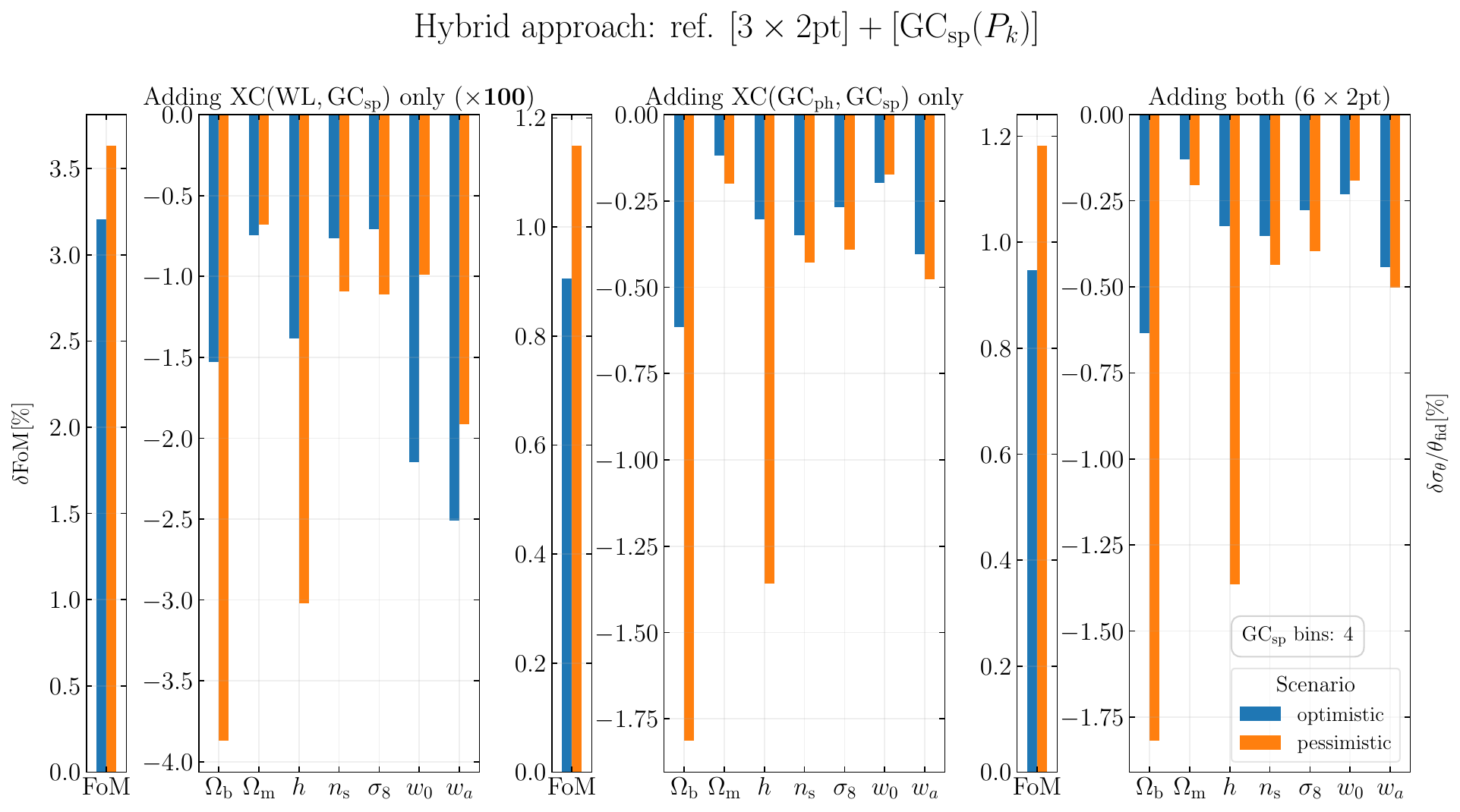}
\caption{Impact on FoM and marginalised $\onesigma$ errors, of the $\xc{\GCph}{\GCsp}$ and $\xc{\WL}{\GCsp}$ cross-correlations in the hybrid approach. The reference for the percentage differences is $[\WL+\GCph+\xc{\WL}{\GCph}] + [\GCsp(P_k)]$. In any case the cross-correlations are always considered to be \emph{covariant} with the \threetwoptShort\ statistics observables. The percentage differences in left panel have been multiplied by $10$, to make them visible to the naked eye. Note the opposite sign of the percentage differences for FoM and uncertainties.}
\label{fig:6x2pt_hybrid_xcs_impact}
\end{figure*}

Here we discuss the impact on the constraints of the $\xc{\GCph}{\GCsp}$ and $\xc{\WL}{\GCsp}$ cross-correlations. The results in the harmonic approach is shown in \cref{fig:6x2pt_harmonic_xcs_impact_ref_3x2pt_gcsp_nocov,fig:6x2pt_harmonic_xcs_impact_ref_3x2pt_gcsp_cov}, while the hybrid approach result is displayed in \cref{fig:6x2pt_hybrid_xcs_impact}. The figures show the effect of the inclusion of the $\xc{\WL}{\GCsp}$ cross-correlation, the impact of the $\xc{\GCph}{\GCsp}$ cross-correlation, and the impact of both cross-correlations.

\paragraph{\textit{Harmonic approach}}
For the harmonic approach, two Fisher matrices have been used as reference for the percentage differences: $[\threetwoptShort] + [\GCsp]$, where the \threetwoptShort\ statistics and $\GCsp$ are combined as independent, and $[\threetwoptShort+\GCsp]$, where the cross-covariance between the two is accounted (see \cref{sec:fisher_fullcl}). As explained in the previous section, these two combinations do not produce the same constraints (see \cref{fig:3x2pt_sp_cov_impact}), as the independent combination yields slightly better constraints.

We consider the two different Fisher matrices above since, when the cross-correlations are added in the harmonic approach, the cross-covariance between $\GCsp$ and the \threetwoptShort\ statistics is always accounted for. Thus, on the one hand, we focus on the improvement due to the inclusion of the $\rm XC$ signals alone, and this is done when the Fisher matrix $[\threetwoptShort+\GCsp+\xc{\WL}{\GCsp}]$ is compared to the reference $[\threetwoptShort+\GCsp]$, and the same for $[\threetwoptShort+\GCsp+\xc{\WL}{\GCsp}]$ and $[\threetwoptShort+\GCsp+\xc{\WL}{\GCsp}+\xc{\GCph}{\GCsp}]$ (see \cref{fig:6x2pt_harmonic_xcs_impact_ref_3x2pt_gcsp_cov}). In this case, the percentage differences are representative of the net effect of the cross-correlation signals on the constraints, which is expected to be always positive.

On the other hand, when the Fisher matrices comprising the $\rm XC$'s information are compared to the independent combination $[\threetwoptShort] + [\GCsp]$, we focus on the total effect, which not only contains the gain from the inclusion of the $\rm XC$ signals, but also the penalty from the cross-covariance between $\GCsp$ and the \threetwoptShort\ statistics (see \cref{fig:6x2pt_harmonic_xcs_impact_ref_3x2pt_gcsp_nocov}). This comparison is useful in order to evaluate the overall impact of the cross angular power spectra.

\paragraph{\textit{Harmonic approach -- adding $\xc{\WL}{\GCsp}$}}
The net effect of the $\xc{\WL}{\GCsp}$ inclusion is shown in the left panels of \cref{fig:6x2pt_harmonic_xcs_impact_ref_3x2pt_gcsp_cov}, where the Fisher matrix $[\threetwoptShort+\GCsp+\xc{\WL}{\GCsp}]$ is compared to the reference $[\threetwoptShort+\GCsp]$. The variation on the constraints due to the addition of $\xc{\WL}{\GCsp}$ is always about $0.01\%$-$0.05\%$, with no significant differences between the optimistic and pessimistic scenarios. Therefore, it seems that this cross-correlation does not give any contribution to the constraints, i.e. it looks like the computation of the total $[\threetwoptShort+\GCsp+\xc{\WL}{\GCsp}]$ Fisher matrix is not useful to improve the \Euclid performance.
This might seem in contrast to what happens in the pairwise combination of $\WL$ and $\GCsp$, where the $\xc{\WL}{\GCsp}$ cross-correlation signal has a significant impact on the constraints. However, the latter case does not include the $\xc{\WL}{\GCph}$ signal. Instead, in the case now under discussion of $[\threetwoptShort+\GCsp+\xc{\WL}{\GCsp}]$, the reference for the net effect here considered is $[\threetwoptShort+\GCsp]$, which contains the contribution of the $\xc{\WL}{\GCph}$ cross-correlation, proven to be dominant.

From the above reasoning, it is also possible to infer the reason why the constraints from the $[\threetwoptShort+\GCsp+\xc{\WL}{\GCsp}]$ Fisher matrix are worse by $6\%$ at most (see \cref{fig:6x2pt_harmonic_xcs_impact_ref_3x2pt_gcsp_nocov}) than the ones from the independent combination $[\threetwoptShort]+[\GCsp]$: the impact of the cross-correlation signal is so small that its possible improvements are completely dominated by the cross-covariance between $\GCsp$ and the \threetwoptShort\ statistics (see \cref{fig:3x2pt_sp_cov_impact}), which is present when adding the cross signal, but not kept into account in the reference. Therefore, the total effect of the inclusion of $\xc{\WL}{\GCsp}$ in the combination of $\GCsp$ with the \threetwoptShort\ statistic is to worsen the parameter constraints. However, while it has been just shown that the $\xc{\WL}{\GCsp}$ signal can be safely neglected, the cross-covariance between $\GCsp$ and the \threetwoptShort\ statistic needs to be taken with caution.  

\paragraph{\textit{Harmonic approach -- adding $\xc{\GCph}{\GCsp}$}}
The middle panel of \cref{fig:6x2pt_harmonic_xcs_impact_ref_3x2pt_gcsp_cov} shows the \emph{positive} net effect of the $\xc{\GCph}{\GCsp}$ inclusion, which increases with the number of spectroscopic bins. This confirms the same behaviour observed in the pairwise combination of $\GCph$ and $\GCsp$. The gain on the FoM relative to the Fisher matrix $[\threetwoptShort+\GCsp]$ is about $1\%$ for $4$ bins, and increases up to $8$-$10\%$ for $40$ bins, with practically no differences between the optimistic and the pessimistic scenario. \Cref{fig:6x2pt_harmonic_xcs_impact_ref_3x2pt_gcsp_nocov} shows that when the independent combination $[\threetwoptShort]+[\GCsp]$ is used as reference instead, the FoM variation due to the $\xc{\GCph}{\GCsp}$ inclusion is $-2\%$ ($+7\%$) for $4$ ($40$) $\GCsp$ bins. The small worsening at $4$ bins is due to the fact that the positive contribution of the cross-correlation is cancelled by the negative contribution of the $\GCsp$-\threetwoptShort\ cross-covariance (see \cref{fig:3x2pt_sp_cov_impact}). In fact, this cross-covariance is taken into account in the Fisher matrix $[\threetwoptShort+\GCsp+\xc{\GCph}{\GCsp}]$, while it is not in the $[\threetwoptShort]+[\GCsp]$, which is used as the reference in this last case.

Analogously, the marginalised $\onesigma$ uncertainties on the cosmological parameters exhibit a similar behaviour, with no significant differences between the optimistic and the pessimistic scenarios. When using $4$ spectroscopic bins the inclusion of the $\xc{\GCph}{\GCsp}$ cross-correlation produces a small improvement when the Fisher matrix $[\threetwoptShort+\GCsp]$ is used as reference. For a small number of bins, this positive contribution is in general compensated by cross-covariance effects when the percentage differences are referred to the $[\threetwoptShort]+[\GCsp]$ Fisher matrix. When using $40$ spectroscopic bins the cross-correlation dominates and the cross-covariance effects become negligible, and the percentage differences become always positive, independently of the reference that is used. The parameters whose uncertainties decrease the most are $\Omb$ and $h$, gaining $15\%$ and $10\%$ respectively in the optimistic scenario, $10\%$ and $5\%$ in the pessimistic.

Therefore, the total effect of the inclusion of $\xc{\GCph}{\GCsp}$ in the combination of $\GCsp$ with the \threetwoptShort\ statistic depends on the chosen binning set, and may be dominant with respect to $\GCsp$-\threetwoptShort\ cross-covariance effects for a large number of bins.

\paragraph{\textit{Harmonic approach -- the \sixtwoptShort\ statistics}}
In the harmonic approach of \cref{eq:6x2pt_harmonic_defs_from_3x2pt}, the constraints produced by the \sixtwoptShort\ analysis  are equivalent to the ones given by including the $\xc{\GCph}{\GCsp}$ alone. The percentage differences between the constraints from the \sixtwoptShort\ Fisher matrix and the $[\threetwoptShort+\GCsp]$ Fisher matrix are reported in  \cref{fig:6x2pt_harmonic_xcs_impact_ref_3x2pt_gcsp_cov}. These are indistinguishable from the ones which refer to the impact of the $\xc{\GCph}{\GCsp}$ cross-correlation only with respect to $[\threetwoptShort+\GCsp]$. This is expected, since in the above paragraphs it has been shown that $\xc{\WL}{\GCsp}$ provides a negligible contribution with respect to $\xc{\GCph}{\GCsp}$.

\paragraph{\textit{Hybrid approach}}
The contribution of the $\xc{\WL}{\GCsp}$ and $\xc{\GCph}{\GCsp}$ signals in the hybrid approach is reported in \cref{fig:6x2pt_hybrid_xcs_impact}. In this case the reference used is always the independent combination $[\threetwoptShort]+[\GCsp(P_k)]$, since in the hybrid approach the cross-covariance between $\GCsp(P_k)$ and the \threetwoptShort\ is neglected. 

\paragraph{\textit{Hybrid approach -- adding $\xc{\WL}{\GCsp}$}}
In the hybrid approach the impact of the $\xc{\WL}{\GCsp}$ cross-correlation on the constraints from $[\threetwoptShort+\xc{\WL}{\GCsp}]+[\GCsp(P_k)]$, relative to $[\threetwoptShort]+[\GCsp(P_k)]$, is negligible, being always less than $0.05\%$, both in the optimistic and in the pessimistic scenarios. This result is similar to what is found with the harmonic approach, in which we consider the comparison of $[\threetwoptShort+\GCsp+\xc{\WL}{\GCsp}]$ with respect to $[\threetwoptShort+\GCsp]$, to isolate the impact of $\xc{\WL}{\GCsp}$. This is expected, since there are no differences in the \threetwoptShort\ between the two approaches, and, in particular, the $\xc{\WL}{\GCph}$ cross-correlation is computed in the same way in the two cases.

\paragraph{\textit{Hybrid approach -- adding $\xc{\GCph}{\GCsp}$}}
In the hybrid approach the impact of the $\xc{\GCph}{\GCsp}$ cross-correlation on the constraints from $[\threetwoptShort+\xc{\GCph}{\GCsp}]+[\GCsp(P_k)]$, relative to $[\threetwoptShort]+[\GCsp(P_k)]$, is slightly smaller than in the harmonic one with $4$ spectroscopic bins. \Cref{fig:6x2pt_hybrid_xcs_impact} shows that the absolute percentage differences on all constraints is always below $2\%$. The gain on the FoM is $\sim 1\%$ both in the optimistic and the pessimistic scenario. The parameters whose uncertainties are affected the most are $\Omb$ and $h$, with a gain of $1.5\%$ at most in the pessimistic scenario, and less than $0.5\%$ in the optimistic scenario.

\paragraph{\textit{Hybrid approach -- the \sixtwoptShort\ statistics}}
The constraints given by the hybrid \sixtwoptShort\ statistics are similar to the ones given by the inclusion of the $\xc{\GCph}{\GCsp}$ cross-correlation only. This is manifest in \cref{fig:6x2pt_hybrid_xcs_impact}. Nonetheless, in the hybrid approach the inclusion of the $\xc{\GCph}{\GCsp}$ cross-correlation has a negligible impact on the constraints, as discussed in the above paragraph. Therefore, the Fisher matrix of the hybrid \sixtwoptShort\ statistics produces constraints that are almost equivalent to the $[\threetwoptShort]+[\GCsp(P_k)]$ Fisher matrix.

\begin{figure*}[t]
\centering
\includegraphics[width=0.95\textwidth]{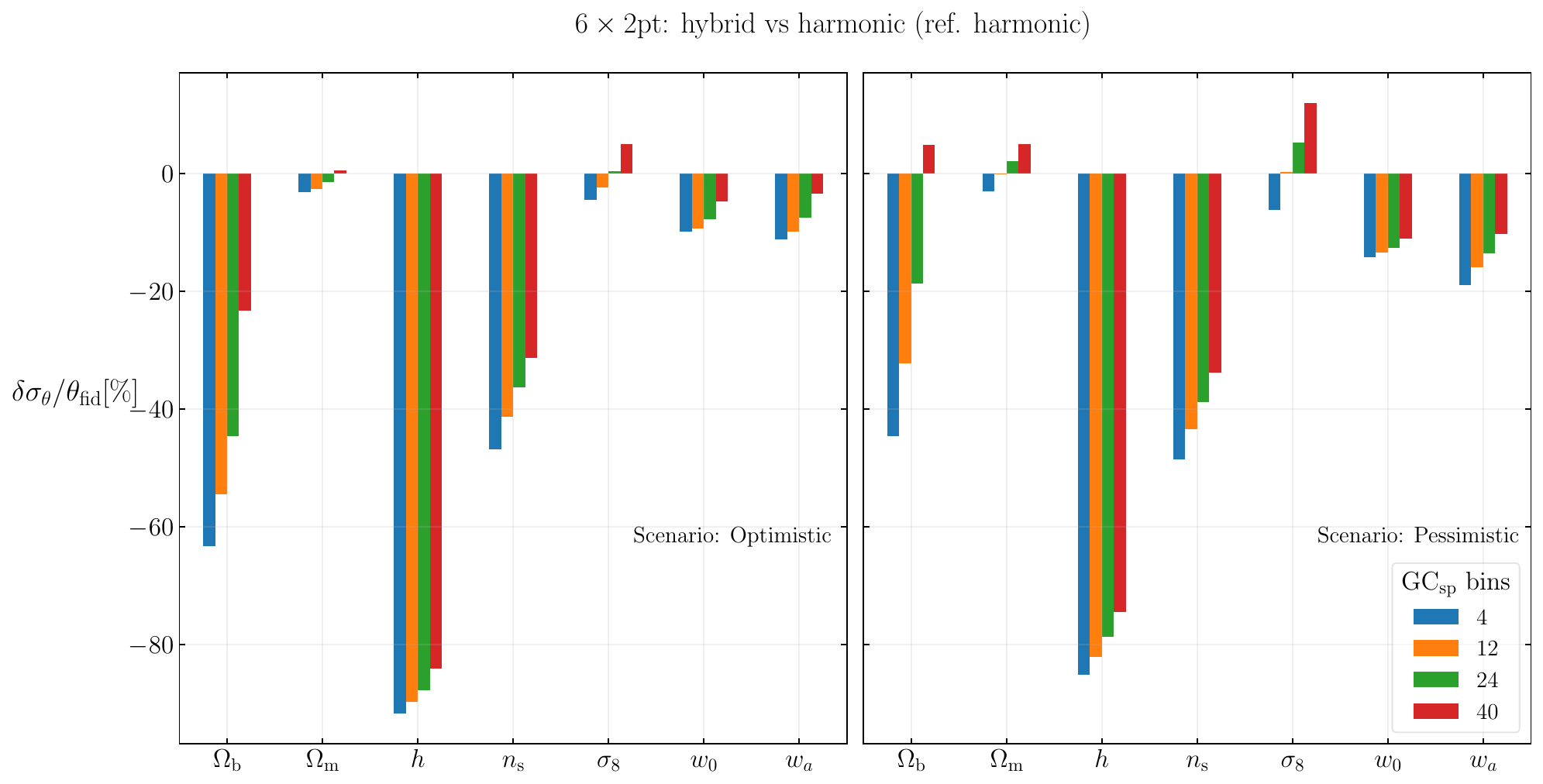}
\caption{Comparison between marginalised $\onesigma$ uncertainties of the hybrid \sixtwoptShort\ and the harmonic \sixtwoptShort\ statistics, quantified using percentage differences taking the latter as reference.}
\label{fig:6x2pt_hybrid_vs_harmonic}
\end{figure*}

\subsection{The \sixtwoptShort\ statistics: hybrid approach vs harmonic approach}
For the \sixtwoptShort\ statistics the hybrid approach performs better than the harmonic one, especially when a small number of spectroscopic bins is used for the latter. The comparison of the two approaches is reported in \cref{fig:6x2pt_hybrid_vs_harmonic}. In this case, the reference adopted is the Fisher matrix of the harmonic approach, \cref{eq:6x2pt_harmonic_defs_from_3x2pt}. The hybrid approach produces a FoM that is $20\%$ ($40\%$) larger than the harmonic one in the optimistic (pessimistic) scenario, when $4$ spectroscopic bins are used for the latter. When using $40$ bins, the harmonic approach performs instead slightly better ($\sim 6\%$) than the hybrid one in the optimistic scenario, while it is equivalent to it in the pessimistic scenario. 

Regarding the marginalised $\onesigma$ uncertainties on the dark energy parameters, $\wz$-$\wa$, the hybrid approach always provides better constraints than the harmonic one, regardless the number of spectroscopic bins used for the latter. However, the harmonic approach with $40$ bins produces a slightly higher FoM than the hybrid one in the optimistic scenario.

\begin{table*}[t]
\centering
\caption{$\FoM$ of the \sixtwoptShort\ statistics in the harmonic and hybrid approaches. The $\Delta\FoM$ columns quantify the differences with respect to the $[\threetwoptShort]+[\GCsp(P_k)]$ combination, which is used as reference to assess the impact on the $\FoM$ of the $\xc{\GCph}{\GCsp}$ and $\xc{\WL}{\GCsp}$ cross-correlations.}
\begin{resultstable}{*{1}{l}*{4}{c}}
\multicolumn{5}{c}{\sixtwoptShort\ FoM forecasts} \\
$\GCsp$\, \text{bins} & \text{Fisher matrix} & $\FoM$ & $\Delta\FoM$ & $\Delta\FoM$ (\%) \\
\midrule
\multicolumn{5}{c}{Optimistic scenario} \\
\midrule
\multirow{3}{*}{4} 
& [\threetwoptShort] + [$\GCsp(P_k)$] & 1216.16  & \text{--} & \text{--} \\
& \sixtwoptShort\ \, $\mathrm{hybrid}$ &  1227.69 &   11.53 &  $\positive{+0.95\%}$ \\
& \sixtwoptShort\ \, $\mathrm{harmonic}$ &  1018.43 & $-197.73$ & $\negative{-16.26\%}$ \\
\midrule
\multirow{1}{*}{12} 
& \sixtwoptShort\ \, $\mathrm{harmonic}$ &  1073.31 & $-142.85$ & $\negative{-11.75\%}$ \\
\midrule
\multirow{1}{*}{24} 
& \sixtwoptShort\ \, $\mathrm{harmonic}$ &  1151.13 &  $-65.03$ &  $\negative{-5.35\%}$ \\
\midrule
\multirow{1}{*}{40} 
& \sixtwoptShort\ \, $\mathrm{harmonic}$ &  1296.44 &   80.28 &  $\positive{+6.60\%}$ \\
\midrule
\multicolumn{5}{c}{Pessimistic scenario} \\
\midrule
\multirow{3}{*}{4} 
& [\threetwoptShort] + [$\GCsp(P_k)$]    & 549.37 & \text{--} & \text{--} \\
& \sixtwoptShort\ \, $\mathrm{hybrid}$ &  555.87 &    6.50 &  $\positive{+1.18\%}$ \\
& \sixtwoptShort\ \, $\mathrm{harmonic}$ &  379.85 & $-169.52$ & $\negative{-30.86\%}$ \\
\midrule
\multirow{1}{*}{12} 
& \sixtwoptShort\ \, $\mathrm{harmonic}$ &  434.00 & $-115.37$ & $\negative{-21.00\%}$ \\
\midrule
\multirow{1}{*}{24} 
& \sixtwoptShort\ \, $\mathrm{harmonic}$ &  486.40 &  $-62.97$ & $\negative{-11.46\%}$ \\
\midrule
\multirow{1}{*}{40} 
& \sixtwoptShort\ \, $\mathrm{harmonic}$ &  550.82 &    1.45 &  $\positive{+0.26\%}$ \\
\end{resultstable}
\label{tab:6x2pt_fom_table}
\end{table*}

\Cref{fig:6x2pt_hybrid_vs_harmonic} shows that the hybrid approach performs drastically better in constraining $h$, producing a $\onesigma$ uncertainty on it which is always more than $70\%$ smaller than the one in the harmonic approach. The hybrid approach gives better uncertainties than the harmonic one for $\ns$ and $\Omb$ too. For $\ns$ the uncertainty of the hybrid approach is always smaller than the one of the harmonic approach by $30$-$40\%$. For $\Omb$ the hybrid approach gives a $60\%$ smaller uncertainty than the harmonic approach with $4$ bins, while the difference is about $20\%$ with $40$ bins. Concerning the uncertainty on $\Omm$ the two approaches produce results that are always comparable within $5\%$. Finally, on $\sige$ the harmonic approach with $12$ bins performs slightly better than the hybrid one. In the optimistic scenario the uncertainties on $\sige$ are always comparable, while in the pessimistic case the harmonic approach produces a $10\%$ smaller uncertainty when using $40$ tomographic bins.

\begin{figure*}[!ht]
\centering
\begin{tabular}{c}
\includegraphics[width=0.95\textwidth]{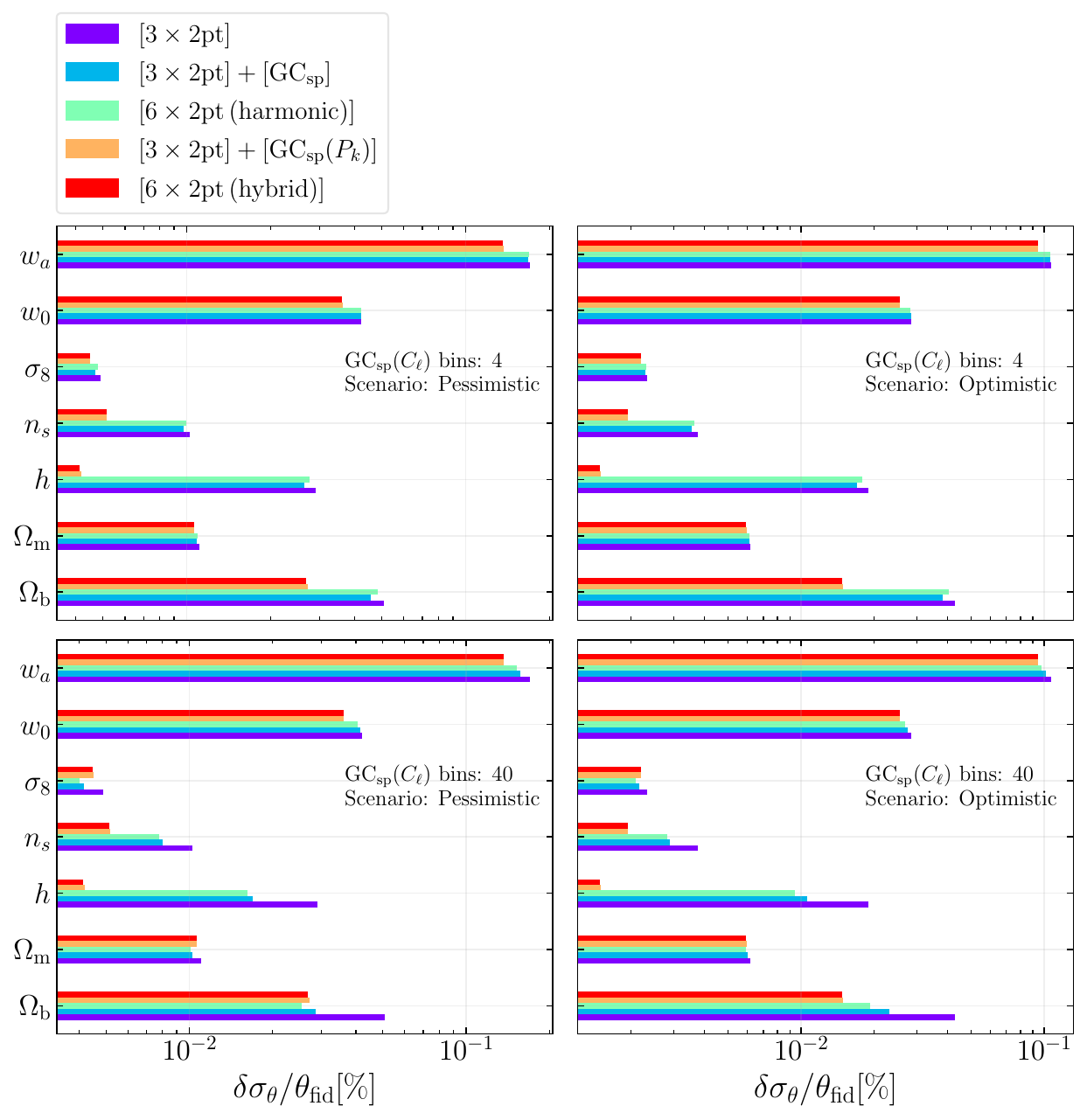}
\end{tabular}
\caption{Comparison between harmonic and hybrid approaches for the case of \sixtwoptShort\ statistics. The top panels refer to the optimistic scenario, while the bottom panels refer to the pessimistic one.}
\label{fig:6x2pt_fullcl_vs_2dx3d_error_barplot}
\end{figure*}

\begin{figure*}[!ht]
\centering
\begin{tabular}{cc}
\includegraphics[width=0.4500\textwidth]{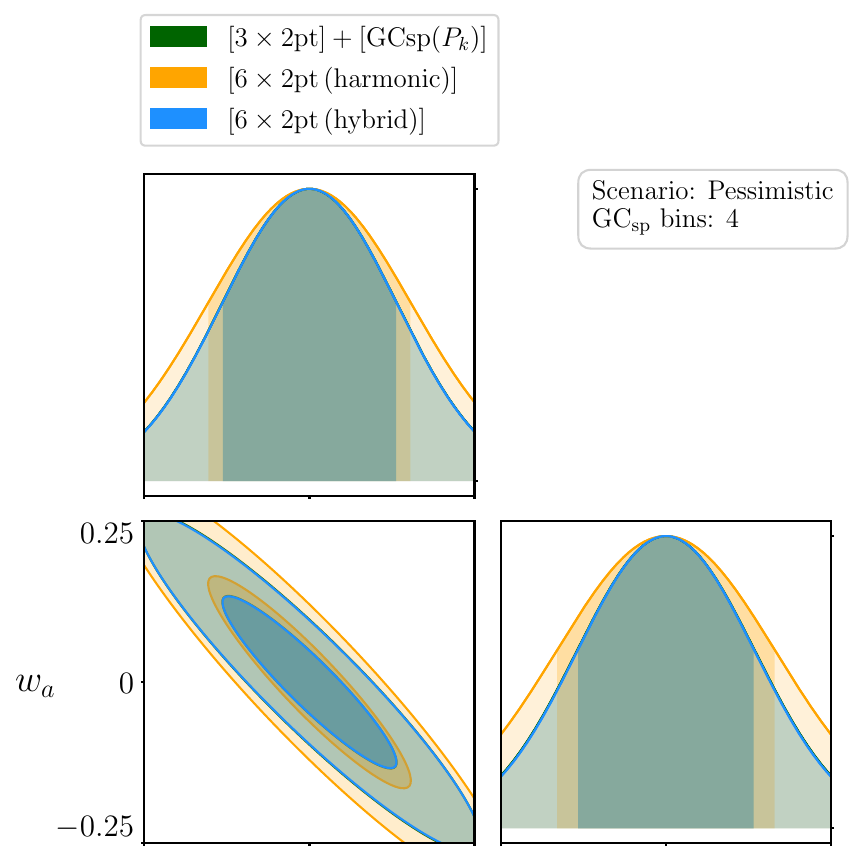} &
\includegraphics[width=0.4220\textwidth]{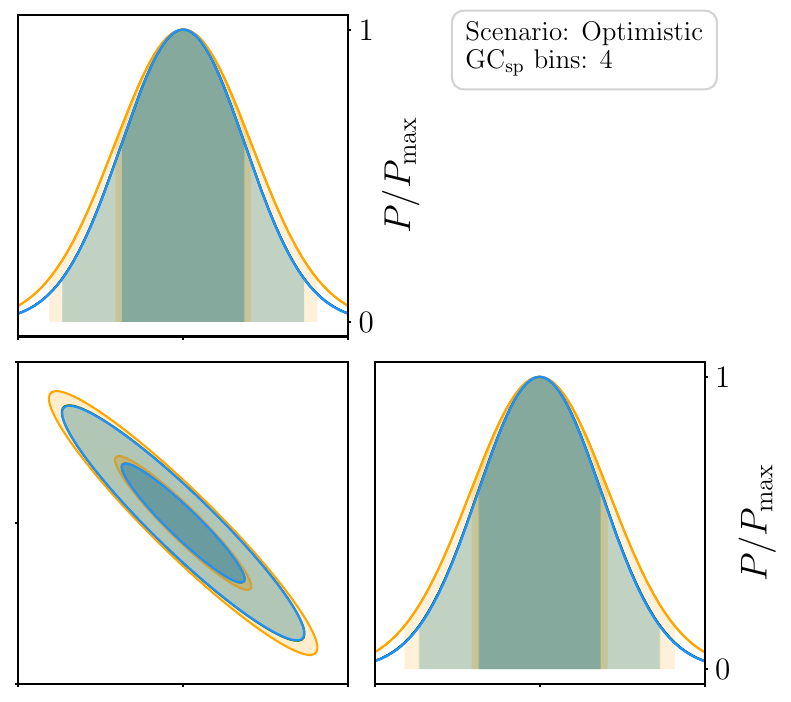} \\
\includegraphics[width=0.4500\textwidth]{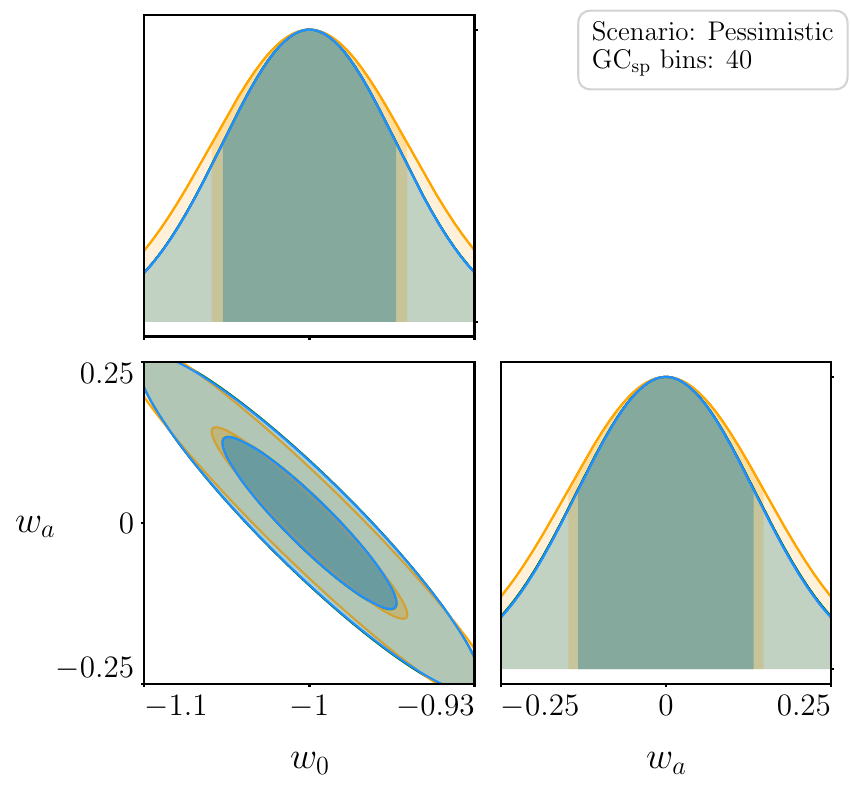} &
\includegraphics[width=0.4220\textwidth]{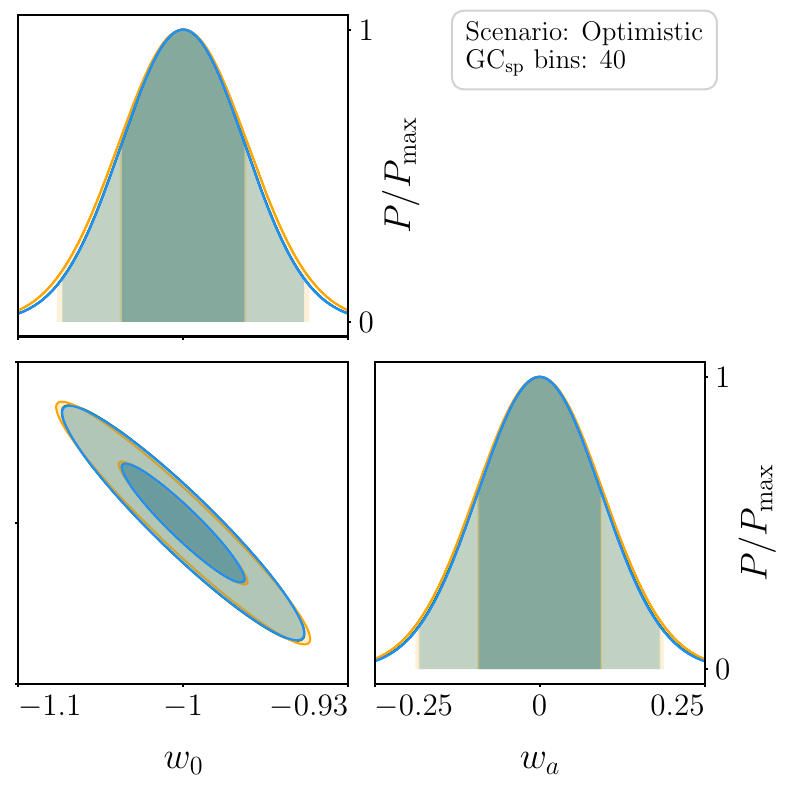} \\
\end{tabular}
\caption{Contour plots showing the comparison between the two approaches in the case of the \sixtwoptShort\ statistics.}
\label{fig:contour_w0wa_6x2pt_fullcl_vs_2dx3d}
\end{figure*}

\section{Main results and conclusions}
In this work we presented the results of the first \Euclid forecasts that include all the pairwise correlations between the main probes: weak lensing ($\WL$), photometric galaxy clustering ($\GCph$), and spectroscopic galaxy clustering ($\GCsp$). We have studied these correlations from two points of view. First, off-diagonal terms of the covariance matrix, the \emph{cross-covariances}, that account for the statistical correlation between two different probes. Second, as additional observables, the \emph{cross-correlation signals} which can be sensitive to cosmological parameters.

This work is a natural extension of the \Euclid IST forecast \citetalias{2020A&A...642A.191E}, considering $\WL$ and $\GCph$ and their correlation in the 2D harmonic domain. The $\GCsp$ was instead studied in the 3D Fourier domain, and to a first approximation was assumed to be independent from the other two probes. Here we extend the harmonic formalism also to $\GCsp$, with the aim of including the $\xc{\GCph}{\GCsp}$ and $\xc{\WL}{\GCsp}$ correlations in the analysis as well. Differently, a work\footnote{Dournac et al, in preparation.} complementary to this one introduces a new data vector, essentially the ratio of the correlation of the two samples. Since this data set is free from sampling variance, it aims to achieve a
significant improvement in the final constraints on the cosmological parameters.

We have considered two different approaches to include the cross-correlations in the analysis: the \emph{harmonic} approach and the \emph{hybrid} approach.

In the harmonic approach all the observables -- the two-point correlation functions -- are treated in the harmonic domain, i.e. using the $C(\ell)$'s formalism. This approach allows us to naturally include all cross-covariances between the observables, computed via \cref{eq:4thorder_cov}. Nonetheless, it has the disadvantage of significantly lowering the constraining power of $\GCsp$, since the integral along the line of sight prevents to fully exploit the accurate radial information provided by the spectroscopic clustering. In order to recover such information, we tried to refine the tomographic binning of the projected $\GCsp$, from $4$ bins (the baseline setting) up to a maximum of $40$ bins, but in any case we neglected RSD in the angular $\GCsp$ power spectra. 

In the hybrid approach, all the observables are studied in the harmonic domain -- including $\xc{\WL}{\GCsp}$ and $\xc{\GCph}{\GCsp}$ -- except for the $\GCsp$ auto-correlation function. We have considered it as an independent observable, adding the $\GCsp$ Fisher matrix that had been computed in \citetalias{2020A&A...642A.191E}, using the 3D Fourier power spectrum as observable. The main advantage of this approach is that it fully exploits the potential of $\GCsp$, keeping the information from radial BAO and RSD available thanks to accurate spectroscopic redshift measurements. At the same time the $\xc{\GCph}{\GCsp}$ and $\xc{\WL}{\GCsp}$ cross-correlations are correctly included in the analysis as harmonic two-point functions, i.e. $C(\ell)$'s. However, this approach comes with a drawback.
It is not obvious how to compute the cross-covariance terms between the 3D $\GCsp$ auto-correlation and the 2D \threetwoptShort observables. Therefore, according to the conclusions from the analysis in the harmonic domain (which correctly accounts for the projected part of such cross-covariances), indicating that the impact of this ``projected'' cross-covariance on the \Euclid\ performance is negligible, and assuming that it is a good approximation of the full 3D-2D covariance (especially because the \threetwoptShort\ statistics only depends on transverse modes), in the hybrid approach we neglect the cross-covariances between 3D and 2D probes.
\textcolor{black}{However, we would like to point out that the methodology presented in this paper is not meant to be used with forthcoming Euclid data, but as a tool to forecast the impact of cross-correlations and covariances among the Euclid photometric and spectroscopic probes.}

We summarise the results of our analysis in the three cases considered: the combination of $\GCph$ and $\GCsp$, the combination of $\WL$ and $\GCsp$, and the so-called \sixtwoptShort, i.e. the full combination of $\WL, \GCph, \GCsp$ altogether. In the latter case -- which is the most interesting one for the \Euclid data analysis -- we present the results in an optimistic and pessimistic scenarios. Instead, for the two pairwise combinations $\GCph$-$\GCsp$ and $\WL$-$\GCsp$, we report the results for the optimistic scenario alone. 

\paragraph{\textit{Combination of $\GCph$ and $\GCsp$}} 
In the harmonic approach, the full analysis, $[\GCph+\GCsp+\xc{\GCph}{\GCsp}]$, i.e. including the cross-covariance and cross-spectrum, provides a FoM of 69 in the baseline setting of 4 spectroscopic bins. For the combination of $\GCph$ and $\GCsp$, the cross-covariance between the two can be considered practically negligible, as it does not change the uncertainties on the cosmological parameters by more than $6\%$ and the FoM (computed considering only the \threetwoptShort\ statistics associated with GC) by more than $3\%$, with respect to the reference combination in the harmonic domain, $[\GCph]+[\GCsp]$, of the two probes taken as independent. In the harmonic approach, the $\xc{\GCph}{\GCsp}$ cross-correlation signal starts to be significant only when using $40$ bins, yielding a maximum FoM of $153$, which is $38\%$ higher than for the $[\GCph]+[\GCsp]$ combination. Regarding the uncertainties on $\wz$ and $\wa$, the improvement given by the $\xc{\GCph}{\GCsp}$ cross-correlation is 5 and 6\% respectively in the baseline $4$ bins setting, while it is $23\%$ and $25\%$ when using $40$ bins. The gain on the uncertainties on the other parameters is less than $5\%$ with $4$ bins and ranges from $10\%$ to $20\%$ with $40$ bins.

In the hybrid approach, the full analysis, $[\GCph+\xc{\GCph}{\GCsp}] + [\GCsp(P_k)]$, including the cross-spectrum but neglecting the cross-covariance, gives instead a FoM of $234$, which is much higher than in the harmonic case. However, in the hybrid case the impact of the $\xc{\GCph}{\GCsp}$ signal is negligible, since combining the probes as independent, $[\GCph]+[\GCsp(P_k)]$, provides a FoM of $230$, which is only $1.7\%$ lower than in the full case. The gain on the marginalised uncertainties is even smaller, being always less than $1\%$. Therefore, it is possible to conclude that the $\xc{\GCph}{\GCsp}$ cross-correlation can be neglected in the combination of $\GCph$ with $\GCsp$.

\paragraph{\textit{Combination of $\WL$ and $\GCsp$}}
In the harmonic approach, the full analysis, $[\WL+\GCsp+\xc{\WL}{\GCsp}]$, i.e. including the cross-covariance and cross-spectrum, provides a FoM of 103 in the baseline setting of 4 spectroscopic bins. For the combination of $\WL$ and $\GCsp$, the cross-covariance is even more negligible than for the GC case above, impacting the constraints always by less than $1\%$ with respect to the independent combination, $[\WL] + [\GCsp]$, computed in the harmonic domain. In this domain, the $\xc{\WL}{\GCsp}$ cross-correlation signal improves the constraints almost independently of the number of tomographic bins used for $\GCsp$. The percentage gain on the corresponding FoM, produced by the $\rm XC$ inclusion, is $38\%$ and $33\%$ with $4$ and $40$ bins, respectively. This seemingly counterintuitive trend has been explained in \cref{sec:results_wlxsp}: it is due to the fact that the performance of the harmonic $\GCsp$ auto-spectrum alone improves faster than for $\xc{\WL}{\GCsp}$ when refining the spectroscopic binning. The value of the FoM provided by the full combination, $[\WL+\GCsp+\xc{\WL}{\GCsp}]$, increases to $188$ with $40$ bins, i.e. by $\sim 45\%$ with respect to the baseline $4$-bin setting.
The $\xc{\WL}{\GCsp}$ signal also improves the uncertainties on $\Omm$ and $\sige$, by $12\%$--$18\%$ for the former and $11\%$--$16\%$ for the latter. The error decrease for $\Omb$ and $\ns$ ranges instead from $5\%$ to $8\%$ for the former and from $5\%$ to $2\%$ on the latter. Finally, the uncertainty on $h$ is the only exception to this trend, with an improvement only by $1\%$--$4\%$.

In the hybrid approach, the full analysis, $[\WL+\xc{\WL}{\GCsp}] + [\GCsp(P_k)]$ provides a FoM of $183$, which is comparable to the harmonic case for the 40-bin setting, but is $\sim 56\%$ larger than the 4-bin case. In the hybrid approach, the impact of the $\xc{\WL}{\GCsp}$ signal is not negligible, since the FoM is $15\%$ larger than for the independent combination, $[\WL]+[\GCsp(P_k)]$, of the two probes.
The improvements on the marginalised uncertainties are all in the range between $4\%$ and $8\%$, with no significant differences between the parameters. The uncertainty on $\Omm$ improves the most, by about $8\%$, while the smallest improvement is on $\ns$, with an error decreasing by slightly less than $5\%$.

\paragraph{\textit{Combination of $\WL, \GCph, \GCsp$ and the \sixtwoptShort\ statistics}}
In the harmonic approach, the full \sixtwoptShort\ analysis, \cref{eq:6x2pt_harmonic_defs_from_3x2pt}, i.e. including all cross-covariances and cross-spectra, provides a \Euclid\ FoM of 1018 in the baseline 4-bin setting.
The overall impact of the cross-covariances between the angular $\GCsp$ and the \threetwoptShort\ statistics, with respect to the independent combination $[\threetwoptShort]+[\GCsp]$, is slightly higher than for the pairwise combinations $\GCph$-$\GCsp$ and $\WL$-$\GCsp$, reported above. This is presumably due to the cumulative effect of three independent off-diagonal covariance blocks. 
However, the increase on parameter uncertainties is never larger than $8\%$, as \cref{fig:3x2pt_sp_cov_impact} shows.

Concerning the impact of the cross-correlation signals between $\GCsp$ and 2D probes, this is always negligible both in the harmonic and hybrid approaches. 
Here below we summarise the different contributions.

\begin{itemize}
\item $\xc{\WL}{\GCsp}$ case:
    \begin{itemize}[label=$\bullet$]
      \item In the harmonic approach, the $\xc{\WL}{\GCsp}$ is definitely negligible both in the optimistic and pessimistic scenarios: it always improves parameter constraints by less than $0.05\%$, with respect to the independent combination, $[\threetwoptShort]+[\GCsp]$, computed in the same approach. This is because the contribution brought by the $\xc{\WL}{\GCsp}$ is dominated by the $\xc{\WL}{\GCph}$, which is already present in the Fisher matrix used as reference.
      \item In the hybrid approach, similarly the $\xc{\WL}{\GCsp}$ cross-correlation improves both the FoM and the uncertainties by always less than $0.05\%$ with respect to the independent $[\threetwoptShort]+[\GCsp(P_k)]$ combination. 
    \end{itemize}

    \item $\xc{\GCph}{\GCsp}$ case:
    \begin{itemize}[label=$\bullet$]
        \item In the harmonic approach, the performance improvement due to $\xc{\GCph}{\GCsp}$ is always below $5\%$ when compared to $[\threetwoptShort+\GCsp]$, while it is dominated by the covariance effect when compared to $[\threetwoptShort]+[\GCsp]$. The only exception is when using $40$ bins for $\GCsp$: in this case the $\FoM$ improves by about $10\%$, while the marginalised uncertainties on $\Omb$ and $h$ decrease by $\sim 15\%$. The uncertainties on the other parameters decrease by always less than $5\%$ instead.
        \item In the hybrid approach, the $\xc{\GCph}{\GCsp}$ improvement is always below $2\%$, for both FoM and uncertainties with respect to the independent combination $[\threetwoptShort]+[\GCsp(P_k)]$, since, in this case, only the baseline setting of $4$ spectroscopic bins can be used for consistency with the official 3D spectroscopic galaxy clustering approach.
    \end{itemize}
\end{itemize}
In general, the effect of the $\xc{\GCph}{\GCsp}$ is larger than for $\xc{\WL}{\GCsp}$, both in the optimistic and pessimistic scenarios and both in the harmonic and in the hybrid approaches. For this reason the \sixtwoptShort\  statistics is essentially equivalent to adding $\xc{\GCph}{\GCsp}$ only. Nonetheless, the improvement on the constraints produced by the $\xc{\GCph}{\GCsp}$ cross-correlation is almost always smaller than $10\%$. 

Finally, looking at the absolute performance, the values of the FoM are reported in \cref{tab:6x2pt_fom_table}, and summarised as follows:
\begin{enumerate}
    \item the independent combination in flat space, $[\threetwoptShort]+[\GCsp(P_k)]$, from \citetalias{2020A&A...642A.191E} is taken as reference, and it gives a FoM of $1216$ ($549$) in the optimistic (pessimistic) scenario;
    
    \item in the harmonic domain, the angular \sixtwoptShort\ statistics with $40$ spectroscopic bins provides a total FoM of $1296$ ($550$) in the optimistic (pessimistic) scenario, which is only $6.6\%$ ($0.26\%$) better than 1. When using the standard $4$ spectroscopic bins it gives a FoM of $1018$ ($380$) in the optimistic (pessimistic) scenario, which instead is $15\%$ ($31\%$) worse than 1. 
    
    \item The hybrid \sixtwoptShort\ statistics is essentially equivalent to the $[\threetwoptShort]+[\GCsp(P_k)]$ case, with a FoM of $1228$ ($556$) in the optimistic (pessimistic) scenario, i.e. only $1.18\%$ ($0.95\%$) better than 1. 
\end{enumerate}

Therefore, we can affirm that the $\xc{\GCph}{\GCsp}$ and $\xc{\WL}{\GCsp}$ cross-correlations have negligible impact on the \Euclid\ performance when added to the combination of $\GCsp$ and the \threetwoptShort\ statistics, both in the harmonic and hybrid approaches.\\

In conclusion, either the cross-covariances (here computed only in the harmonic approach) or the cross-correlations (computed both in the harmonic and hybrid approaches) between the two \Euclid\ main probes, i.e. $\GCsp$ and the so-called \threetwoptShort\ statistics, have a negligible impact on the cosmological parameter constraints and, therefore, on the \Euclid\ performance. \textcolor{black}{Regarding the cross-covariance impact, this issue was addressed also in \cite{Taylor:2022rgy}, but following a different approach: they derived an analytical expression for the Gaussian cross-covariance between the $\threetwoptShort$ statistics and the $\GCsp$ multipoles. Also this approach leads to a negligible impact of the cross-covariance, hence corroborating our findings.}

In the case of the hybrid approach, we attribute this result to the effect of the $\xc{\WL}{\GCph}$ cross-correlation which is dominant with respect to the other cross-correlations, and to the higher performance of the full anisotropic 3D $\GCsp$ probe with respect to the projected one.

In the case of the 2D harmonic approach, we attribute this result to two main limitations of the 2D projected $\GCsp$ approximation: the high shot noise and the limited redshift range of the sample, with respect to the \threetwoptShort\ statistics, together with the suppression of radial information, as RSD. We have found that, under two conditions, $\GCsp$ in harmonic space becomes equivalent to $\GCph$ in terms of constraining power, as it can be seen from \cref{fig:gcph_vs_gcsp_harmonic}. The first condition is to reduce the shot noise of $\GCsp$ to the same level of $\GCph$. The second is to restrict the tomographic bins of $\GCph$ to the $4$ photometric bins contained in the $\GCsp$ redshift range. Under these same conditions, $\xc{\WL}{\GCsp}$ and $\xc{\WL}{\GCph}$ equally contribute to the \Euclid\ performance, as we show in \cref{fig:xcwlph_vs_xcwlsp}. Nonetheless, these conditions are not realistic.

\textcolor{black}{Finally, we would like to point out to the reader that in our work we focused on the impact of the Euclid photometric and spectroscopic probes on the cosmological parameters' determination, neglecting the implications for systematics. For instance, as studied in \cite{Newman:2008mb}, it is possible to use the photometric and spectroscopic cross-correlations to calibrate the photometric galaxy density. On the same side, with a $\sixtwoptShort$ pt analysis it is possible to perform the so-called “shear-ratio test”~\citep{Jain:2003tba}; although there is not a great deal of cosmological information encoded in these data, they can be used to calibrate the shears and redshifts of the photometric sources~\citep{Johnson:2016ucr}.}

Future extensions of this work will overcome some approximations that have been made.
First, we have computed the $C(\ell)$'s making use of the Limber approximation. It has been shown in \cite{Fang:2019xat} that this may result in a biased analysis. Using the exact expression for computing the angular power spectra would help to prevent this issue. Second, we have computed the $C(\ell)$'s covariance with \cref{eq:4thorder_cov} as in \citetalias{2020A&A...642A.191E}, and this formula only accounts for the Gaussian contributions. Comparison with covariances estimated from $N$-body simulations showed that the inclusion of non-Gaussian effects may be necessary in order to reach a better agreement with simulations \citep{DES:2017tss}. In order to obtain a more realistic signal-to-noise ratio we have performed some forecasts including the SSC contribution as computed in \cite{Lacasa:2018hqp}: the results of this study are not modified, as we found both the cross-covariance and the cross-correlation between the photometric and spectroscopic probes to be negligible even in this scenario. Third, in what we have called the hybrid approach we have neglected the covariances between the Fourier $\GCsp$ auto-correlation and the \threetwoptShort{} probe. Our calculations in harmonic space suggest that these terms may be negligible. Nonetheless, providing analytical modelling of these terms when $\GCsp$ is studied in 3D Fourier space would surely help to confirm our findings.

\begin{acknowledgements}
MB acknowledges financial support from the ASI agreement n. I/023/12/0 "Euclid attivitá relativa alla fase B2/C".
SC acknowledges support from the Italian Ministry of University and Research, PRIN 2022 `EXSKALIBUR -- Euclid-Cross-SKA: Likelihood Inference Building for Universe Research', from the Italian Ministry of Foreign Affairs and International
Cooperation (grant no.\ ZA23GR03), and from the European Union -- Next Generation EU.
\AckEC
\end{acknowledgements}

\bibliographystyle{aa} 
\bibliography{biblio.bib}

\begin{thebibliography}{39}
\expandafter\ifx\csname natexlab\endcsname\relax\def\natexlab#1{#1}\fi

\bibitem[{{Asorey} {et~al.}(2012){Asorey}, {Crocce}, {Gazta{\~n}aga}, \&
  {Lewis}}]{Asorey2012}
{Asorey}, J., {Crocce}, M., {Gazta{\~n}aga}, E., \& {Lewis}, A. 2012, \mnras,
  427, 1891

\bibitem[{Bacon {et~al.}(2000)Bacon, Refregier, \& Ellis}]{Bacon:2000sy}
Bacon, D.~J., Refregier, A.~R., \& Ellis, R.~S. 2000, \mnras, 318, 625

\bibitem[{Baldauf {et~al.}(2013)Baldauf, Seljak, Smith, Hamaus, \&
  Desjacques}]{Baldauf:2013hka}
Baldauf, T., Seljak, U., Smith, R.~E., Hamaus, N., \& Desjacques, V. 2013,
  Phys. Rev. D, 88, 083507

\bibitem[{{Bird} {et~al.}(2012){Bird}, {Viel}, \& {Haehnelt}}]{bird2012}
{Bird}, S., {Viel}, M., \& {Haehnelt}, M.~G. 2012, \mnras, 420, 2551

\bibitem[{Bridle \& King(2007)}]{Bridle:2007ft}
Bridle, S. \& King, L. 2007, New J. Phys., 9, 444

\bibitem[{Camera {et~al.}(2018)Camera, Fonseca, Maartens, \&
  Santos}]{Camera:2018jys}
Camera, S., Fonseca, J., Maartens, R., \& Santos, M.~G. 2018, \mnras, 481, 1251

\bibitem[{Chevallier \& Polarski(2001)}]{chevallier2001accelerating}
Chevallier, M. \& Polarski, D. 2001, Int. J. Mod. Phys., D, 10, 213

\bibitem[{{Cropper} {et~al.}(2016){Cropper}, {Pottinger}, {Niemi}, {Azzollini},
  {Denniston}, {Szafraniec}, {Awan}, {Mellier}, {Berthe}, {Martignac}, {Cara},
  {Di Giorgio}, {Sciortino}, {Bozzo}, {Genolet}, {Cole}, {Philippon}, {Hailey},
  {Hunt}, {Swindells}, {Holland}, {Gow}, {Murray}, {Hall}, {Skottfelt},
  {Amiaux}, {Laureijs}, {Racca}, {Salvignol}, {Short}, {Lorenzo Alvarez},
  {Kitching}, {Hoekstra}, {Massey}, \& {Israel}}]{cropper2016}
{Cropper}, M., {Pottinger}, S., {Niemi}, S., {et~al.} 2016, in Society of
  Photo-Optical Instrumentation Engineers (SPIE) Conference Series, Vol. 9904,
  Space Telescopes and Instrumentation 2016: Optical, Infrared, and Millimeter
  Wave, ed. H.~A. {MacEwen}, G.~G. {Fazio}, M.~{Lystrup}, N.~{Batalha},
  N.~{Siegler}, \& E.~C. {Tong}, 99040Q

\bibitem[{Eriksen \& Gaztanaga(2015)}]{Eriksen:2014wda}
Eriksen, M. \& Gaztanaga, E. 2015, \mnras, 452, 2149

\bibitem[{{Euclid Collaboration: Blanchard}
  {et~al.}(2020)}]{2020A&A...642A.191E}
{Euclid Collaboration: Blanchard}, A. {et~al.} 2020, \aap, 642, A191

\bibitem[{{Euclid Collaboration: Cropper} {et~al.}(2024)}]{Euclid:2024sqd}
{Euclid Collaboration: Cropper}, M. {et~al.} 2024, arXiv:2405.13492

\bibitem[{{Euclid Collaboration: Jahnke} {et~al.}(2024)}]{Euclid:2024yvv}
{Euclid Collaboration: Jahnke}, K. {et~al.} 2024, arXiv:2405.13493

\bibitem[{{Euclid Collaboration: Mellier} {et~al.}(2024)}]{Euclid:2024yrr}
{Euclid Collaboration: Mellier}, Y. {et~al.} 2024, arXiv:2405.13491

\bibitem[{Fang {et~al.}(2020)Fang, Krause, Eifler, \& MacCrann}]{Fang:2019xat}
Fang, X., Krause, E., Eifler, T., \& MacCrann, N. 2020, JCAP, 05, 010

\bibitem[{Grasshorn~Gebhardt \& Jeong(2020)}]{GrasshornGebhardt:2020wsw}
Grasshorn~Gebhardt, H.~S. \& Jeong, D. 2020, Phys. Rev. D, 102, 083521

\bibitem[{Gupta A.~K.(2000)}]{GuptaNagar}
Gupta A.~K., N. D.~K. 2000, {Matrix Variate Distributions} (Chapman {\&} Hall)

\bibitem[{Jain \& Taylor(2003)}]{Jain:2003tba}
Jain, B. \& Taylor, A. 2003, Phys. Rev. Lett., 91, 141302

\bibitem[{Joachimi {et~al.}(2015)Joachimi, Cacciato, Kitching,
  {et~al.}}]{Joachimi:2015mma}
Joachimi, B., Cacciato, M., Kitching, T.~D., {et~al.} 2015, Space Sci. Rev.,
  193, 1

\bibitem[{Joachimi {et~al.}(2021)}]{Joachimi:2020abi}
Joachimi, B. {et~al.} 2021, A\&A, 646, A129

\bibitem[{Johnson {et~al.}(2017)}]{Johnson:2016ucr}
Johnson, A. {et~al.} 2017, \mnras, 465, 4118

\bibitem[{Joudaki {et~al.}(2018)}]{Joudaki:2017zdt}
Joudaki, S. {et~al.} 2018, \mnras, 474, 4894

\bibitem[{{Kaiser}(1992)}]{1992ApJ...388..272K}
{Kaiser}, N. 1992, \apj, 388, 272

\bibitem[{Kaiser {et~al.}(2000)Kaiser, Wilson, \& Luppino}]{Kaiser:2000if}
Kaiser, N., Wilson, G., \& Luppino, G.~A. 2000, arXiv:0003338

\bibitem[{Kiessling {et~al.}(2015)Kiessling, Cacciato, Joachimi,
  {et~al.}}]{Kiessling:2015sma}
Kiessling, A., Cacciato, M., Joachimi, B., {et~al.} 2015, Space Sci. Rev., 193,
  67, [Erratum: Space Sci.Rev. 193, 137 (2015)]

\bibitem[{Kirk {et~al.}(2015)Kirk, Brown, Hoekstra, {et~al.}}]{Kirk:2015nma}
Kirk, D., Brown, M.~L., Hoekstra, H., {et~al.} 2015, Space Sci. Rev., 193, 139

\bibitem[{Kitching {et~al.}(2017)Kitching, Alsing, Heavens, Jimenez, McEwen, \&
  Verde}]{Kitching:2016zkn}
Kitching, T.~D., Alsing, J., Heavens, A.~F., {et~al.} 2017, \mnras, 469, 2737

\bibitem[{Krause {et~al.}(2017)Krause, Eifler, Zuntz, {et~al.}}]{DES:2017tss}
Krause, E., Eifler, T., Zuntz, J., {et~al.} 2017, arXiv:1706.09359

\bibitem[{Lacasa \& Grain(2019)}]{Lacasa:2018hqp}
Lacasa, F. \& Grain, J. 2019, A\&A, 624, A61

\bibitem[{{Laureijs} {et~al.}(2011){Laureijs}, {Amiaux}, {Arduini},
  {Augu{\`e}res}, {Brinchmann}, {Cole}, {Cropper}, {Dabin}, {Duvet}, {Ealet},
  {Garilli}, {Gondoin}, {Guzzo}, {Hoar}, {Hoekstra}, {Holmes}, {Kitching},
  {Maciaszek}, {Mellier}, {Pasian}, {Percival}, {Rhodes}, {Saavedra Criado},
  {Sauvage}, {Scaramella}, {Valenziano}, {Warren}, {Bender}, {Castander},
  {Cimatti}, {Le F{\`e}vre}, {Kurki-Suonio}, {Levi}, {Lilje}, {Meylan},
  {Nichol}, {Pedersen}, {Popa}, {Rebolo Lopez}, {Rix}, {Rottgering},
  {Zeilinger}, {Grupp}, {Hudelot}, {Massey}, {Meneghetti}, {Miller}, {Paltani},
  {Paulin-Henriksson}, {Pires}, {Saxton}, {Schrabback}, {Seidel}, {Walsh},
  {Aghanim}, {Amendola}, {Bartlett}, {Baccigalupi}, {Beaulieu}, {Benabed},
  {Cuby}, {Elbaz}, {Fosalba}, {Gavazzi}, {Helmi}, {Hook}, {Irwin}, {Kneib},
  {Kunz}, {Mannucci}, {Moscardini}, {Tao}, {Teyssier}, {Weller}, {Zamorani},
  {Zapatero Osorio}, {Boulade}, {Foumond}, {Di Giorgio}, {Guttridge}, {James},
  {Kemp}, {Martignac}, {Spencer}, {Walton}, {Bl{\"u}mchen}, {Bonoli},
  {Bortoletto}, {Cerna}, {Corcione}, {Fabron}, {Jahnke}, {Ligori}, {Madrid},
  {Martin}, {Morgante}, {Pamplona}, {Prieto}, {Riva}, {Toledo}, {Trifoglio},
  {Zerbi}, {Abdalla}, {Douspis}, {Grenet}, {Borgani}, {Bouwens}, {Courbin},
  {Delouis}, {Dubath}, {Fontana}, {Frailis}, {Grazian}, {Koppenh{\"o}fer},
  {Mansutti}, {Melchior}, {Mignoli}, {Mohr}, {Neissner}, {Noddle}, {Poncet},
  {Scodeggio}, {Serrano}, {Shane}, {Starck}, {Surace}, {Taylor},
  {Verdoes-Kleijn}, {Vuerli}, {Williams}, {Zacchei}, {Altieri}, {Escudero
  Sanz}, {Kohley}, {Oosterbroek}, {Astier}, {Bacon}, {Bardelli}, {Baugh},
  {Bellagamba}, {Benoist}, {Bianchi}, {Biviano}, {Branchini}, {Carbone},
  {Cardone}, {Clements}, {Colombi}, {Conselice}, {Cresci}, {Deacon}, {Dunlop},
  {Fedeli}, {Fontanot}, {Franzetti}, {Giocoli}, {Garcia-Bellido}, {Gow},
  {Heavens}, {Hewett}, {Heymans}, {Holland}, {Huang}, {Ilbert}, {Joachimi},
  {Jennins}, {Kerins}, {Kiessling}, {Kirk}, {Kotak}, {Krause}, {Lahav}, {van
  Leeuwen}, {Lesgourgues}, {Lombardi}, {Magliocchetti}, {Maguire}, {Majerotto},
  {Maoli}, {Marulli}, {Maurogordato}, {McCracken}, {McLure}, {Melchiorri},
  {Merson}, {Moresco}, {Nonino}, {Norberg}, {Peacock}, {Pello}, {Penny},
  {Pettorino}, {Di Porto}, {Pozzetti}, {Quercellini}, {Radovich}, {Rassat},
  {Roche}, {Ronayette}, {Rossetti}, {Sartoris}, {Schneider}, {Semboloni},
  {Serjeant}, {Simpson}, {Skordis}, {Smadja}, {Smartt}, {Spano}, {Spiro},
  {Sullivan}, {Tilquin}, {Trotta}, {Verde}, {Wang}, {Williger}, {Zhao},
  {Zoubian}, \& {Zucca}}]{laureijs2011euclid}
{Laureijs}, R., {Amiaux}, J., {Arduini}, S., {et~al.} 2011, arXiv:1110.3193

\bibitem[{Lemos {et~al.}(2017)Lemos, Challinor, \& Efstathiou}]{Lemos:2017arq}
Lemos, P., Challinor, A., \& Efstathiou, G. 2017, JCAP, 05, 014

\bibitem[{Linder(2002)}]{Linder2002}
Linder, E.~V. 2002, Phys. Rev. Lett., 90, 4

\bibitem[{Loureiro {et~al.}(2019)}]{Loureiro:2018qva}
Loureiro, A. {et~al.} 2019, \mnras, 485, 326

\bibitem[{{Maciaszek} {et~al.}(2022){Maciaszek}, {Ealet}, {Gillard}, {Jahnke},
  {Barbier}, {Prieto}, {Bon}, {Bonnefoi}, {Caillat}, {Carle}, {Costille},
  {Ducret}, {Fabron}, {Foulon}, {Gimenez}, {Grassi}, {Jaquet}, {Le Mignant},
  {Martin}, {Pamplona}, {Sanchez}, {Cl{\'e}mens}, {Caillat}, {Niclas},
  {Secroun}, {Kubik}, {Ferriol}, {Berthe}, {Barri{\`e}re}, {Fontignie},
  {Valenziano}, {Auricchio}, {Battaglia}, {De Rosa}, {Farinelli}, {Franceschi},
  {Medinaceli}, {Morgante}, {Sortino}, {Trifoglio}, {Corcione}, {Capobianco},
  {Ligori}, {Dusini}, {Borsato}, {Dal Corso}, {Laudisio}, {Sirignano},
  {Stanco}, {Ventura}, {Patrizii}, {Chiarusi}, {Fornari}, {Giacomini},
  {Margiotta}, {Mauri}, {Pasqualini}, {Sirri}, {Spurio}, {Tenti}, {Travaglini},
  {Bonoli}, {Bortoletto}, {Balestra}, {Dalessandro}, {Grupp}, {Penka},
  {Steinwagner}, {Hormuth}, {Schirmer}, {Seidel}, {Padilla}, {Casas}, {Lloro},
  {Toledo-Moreo}, {Gomez}, {Colodro-Conde}, {Liz{\'a}n}, {Diaz}, {Lilje},
  {Andersen}, {Andersen}, {S{\o}rensen}, {Hornstrup}, {Jessen}, {Thizy},
  {Holmes}, {Pniel}, {Jhabvala}, {Pravdo}, {Seiffert}, {Waczynski}, {Laureij},
  {Racca}, {Salvignol}, {Boenke}, {Strada}, \& {Mellier}}]{Maciaszek}
{Maciaszek}, T., {Ealet}, A., {Gillard}, W., {et~al.} 2022, in Society of
  Photo-Optical Instrumentation Engineers (SPIE) Conference Series, Vol. 12180,
  Space Telescopes and Instrumentation 2022: Optical, Infrared, and Millimeter
  Wave, ed. L.~E. {Coyle}, S.~{Matsuura}, \& M.~D. {Perrin}, 121801K

\bibitem[{Newman(2008)}]{Newman:2008mb}
Newman, J.~A. 2008, \apj, 684, 88

\bibitem[{Pozzetti {et~al.}(2016)Pozzetti, Hirata, Geach, Cimatti, Baugh,
  Cucciati, Merson, Norberg, \& Shi}]{Pozzetti:2016cch}
Pozzetti, L., Hirata, C., Geach, J., {et~al.} 2016, A\&A, 590, A3

\bibitem[{Stebbins(1996)}]{Stebbins:1996wx}
Stebbins, A. 1996, arXiv:9609149

\bibitem[{Takahashi {et~al.}(2012)Takahashi, Sato, Nishimichi, Taruya, \&
  Oguri}]{Takahashi:2012em}
Takahashi, R., Sato, M., Nishimichi, T., Taruya, A., \& Oguri, M. 2012, \apj,
  761, 152

\bibitem[{Taylor \& Markovi\v{c}(2022)}]{Taylor:2022rgy}
Taylor, P.~L. \& Markovi\v{c}, K. 2022, Phys. Rev. D, 106, 063536

\bibitem[{Taylor {et~al.}(2022)Taylor, Markovi\v{c}, Pourtsidou, \&
  Huff}]{Taylor:2021bhg}
Taylor, P.~L., Markovi\v{c}, K., Pourtsidou, A., \& Huff, E. 2022, Phys. Rev.
  D, 105, 084007

\end{thebibliography}

\appendix
\section{Details of the cross-covariance}
\label{sec:appendix_cross}
There are three possible angular power spectra that can be constructed from two probes $\rm A$ and $\rm B$:
\begin{itemize}
\item auto power spectrum of $\rm A$, $C^{\rm AA}(\ell)$;
\item auto power spectrum of $\rm B$, $C^{\rm BB}(\ell)$;
\item cross power spectrum between $\rm A$ and $\rm B$, $C^{\rm AB}(\ell)$.
\end{itemize}
As a concrete example, consider $\mathrm{A}=\GCph$ and $\mathrm{B}=\WL$.

\paragraph{\textit{Combining $\GCph$ and $\WL$ as independent probes}}
When assuming $\GCph$ and $\WL$ to be independent, the resulting Fisher matrix will be given by the sum of the Fishers of the two single probes. According to the conventions explained in \cref{sec:naming}, the resulting Fisher is denoted as $[\GCph] + [\WL]$. This case is equivalent to building a data-vector including the two auto-correlations
\begin{equation}\label{eq:datavector_ph+wl}
\mathbf{C}(\ell) = 
\left\lbrace \bcl{\phph}{\ell}, \bcl{\wlwl}{\ell} \right\rbrace \;.
\end{equation}
and setting to zero the off-diagonal blocks of the associated covariance matrix
\begin{multline}\label{eq:cov_ph+wl_indep}
\cov{\mathbf{C}(\ell)}{\mathbf{C}(\ell)} =\\ 
\begin{pmatrix}
\cov{\bcl{\phph}{\ell}}{\bcl{\phph}{\ell}} & 0 \\[1ex]
0 & \cov{\bcl{\wlwl}{\ell}}{\bcl{\wlwl}{\ell}} 
\end{pmatrix}\;.
\end{multline}
This is equivalent to neglecting the cross-covariance between the auto-correlations of the two single probes, which is given by the block
\[
\cov{\bcl{\phph}{\ell}}{\bcl{\wlwl}{\ell}}\;.
\]
Since the matrix in \cref{eq:cov_ph+wl_indep} is block diagonal, its inverse is of the same form:
\begin{align}
\begin{pmatrix}
A & 0 \\[1ex]
0 & B
\end{pmatrix}^{-1} 
= 
\begin{pmatrix}
A^{-1} & 0 \\[1ex]
0 & B^{-1} 
\end{pmatrix}\;.
\end{align}
Therefore the matrix product entering the Fisher matrix element is simply given by the bilinear form between $(\bcl{\phph}{\ell}_{,\alpha},\bcl{\wlwl}{\ell}_{,\alpha})^{\sf T}$ and a block-diagonal covariance matrix, whose diagonal blocks are $\cov{\bcl{\phph}{\ell}}{\bcl{\phph}{\ell}}^{-1}$ and $\cov{\bcl{\wlwl}{\ell}}{\bcl{\wlwl}{\ell}}^{-1}$. The resulting Fisher matrix element is
\begin{align}
F^{\GCph+\WL}_{\alpha\beta}(\ell) &=
\left[\bcl{\phph}{\ell}_{,\alpha}\right]^T\,
\cov{\bcl{\phph}{\ell}}{\bcl{\phph}{\ell}}^{-1}\,
\bcl{\phph}{\ell}_{,\beta}\nonumber \\
&+\left[\bcl{\wlwl}{\ell}_{,\alpha}\right]^T\,
\cov{\bcl{\wlwl}{\ell}}{\bcl{\wlwl}{\ell}}^{-1}\,
\bcl{\wlwl}{\ell}_{,\beta}\nonumber \\
&= F^{\GCph}_{\alpha\beta}(\ell) + F^{\WL}_{\alpha\beta}(\ell)\;.
\end{align}
which is the simple sum of the fisher elements associated with the single probes.
\paragraph{\textit{Combining $\GCph$ and $\WL$ with cross-covariance}}
The Fisher matrix in this case is denoted as $[\GCph+\WL]$. The data-vector is the same as the previous one (Eq.\ \ref{eq:datavector_ph+wl}), but the off-diagonal blocks of the covariance matrix are taken into account
\begin{multline}\label{eq:cov_ph+wl_with_cov}
\cov{\mathbf{C}(\ell)}{\mathbf{C}(\ell)} = \\
\begin{pmatrix}
\cov{\bcl{\phph}{\ell}}{\bcl{\phph}{\ell}} &
\cov{\bcl{\phph}{\ell}}{\bcl{\wlwl}{\ell}}\\[1ex]
\cov{\bcl{\wlwl}{\ell}}{\bcl{\phph}{\ell}} & \cov{\bcl{\wlwl}{\ell}}{\bcl{\wlwl}{\ell}} 
\end{pmatrix}\;.
\end{multline}
This matrix is not block-diagonal, hence, when inverting it, the blocks will mix with each other. There exist some formulas based on the Schur complement \citep{GuptaNagar} for the inverse of a $2\times 2$ block matrix, but writing it down does not help to enlighten what happens in this case. From an intuitive point of view, the cross-covariance between two observables should worsen the constraints with respect to combining the two probes as independent. This can be understood with the following argument. If two observables exhibit a non-zero cross-covariance, there will be a mutual correlation between the two. In particular, a change in one of the two -- for example induced by a variation of the cosmological parameters -- will statistically induce a corresponding variation in the other. This in turn means that the two observables will share an amount of cosmological information, and therefore the total information coming from their combination will be \emph{less} than the direct sum of the two pieces of information carried individually by the two of them.
%
\paragraph{\textit{Combining $\GCph$, $\WL$ and their cross-correlation}}
Here both the covariance and the cross-correlation between $\GCph$ and $\WL$ are taken into account. The resulting Fisher matrix is denoted as $[\GCph+\WL+\xc{\GCph}{\WL}]$, and the data-vector includes accordingly the maximal set of the available $C(\ell)$'s
\begin{equation}\label{eq:datavector_ph+sp+xc}
\mathbf{C}(\ell) =
\left\lbrace 
\bcl{\phph}{\ell}, 
\bcl{\phwl}{\ell},
\bcl{\wlwl}{\ell}
\right\rbrace\;,
\end{equation}
and the covariance is the full $3\times 3$ block matrix associated with this data-vector has the following entries
\begin{align}
\label{eq:cov_ph+sp+xc}
\cov{\mathbf{C}(\ell)}{\mathbf{C}(\ell)}_{11} &= 
\cov{\bcl{\phph}{\ell}}{\bcl{\phph}{\ell}} \nonumber\\
\cov{\mathbf{C}(\ell)}{\mathbf{C}(\ell)}_{12} &= \cov{\bcl{\phph}{\ell}}{\bcl{\phwl}{\ell}} \nonumber\\
\cov{\mathbf{C}(\ell)}{\mathbf{C}(\ell)}_{12} &= \cov{\bcl{\phph}{\ell}}{\bcl{\wlwl}{\ell}} \nonumber\\
\cov{\mathbf{C}(\ell)}{\mathbf{C}(\ell)}_{22} &= \cov{\bcl{\phwl}{\ell}}{\bcl{\phwl}{\ell}} \nonumber\\
\cov{\mathbf{C}(\ell)}{\mathbf{C}(\ell)}_{13} &= \cov{\bcl{\phwl}{\ell}}{\bcl{\wlwl}{\ell}} \nonumber\\
\cov{\mathbf{C}(\ell)}{\mathbf{C}(\ell)}_{33} &= \cov{\bcl{\wlwl}{\ell}}{\bcl{\wlwl}{\ell}}\;.
\end{align}
In this case the new information coming from the cross-correlation is added to the data-vector, and this contribution is expected to tighten the resulting constraints with respect to the uncorrelated sum. In particular, the cross-correlation is itself a function of the cosmological parameters, meaning that its value is be sensitive to a variation of the parameters themselves. In this sense it is said that adding the \emph{cross-correlation signal} is expected to provide more cosmological information, therefore improving the combined constraints.
On the other hand, also all the cross-covariances between the $C(\ell)$'s are being considered in this case, and this tends to worsen the constraints, as explained in the previous paragraph. So there are two concurring effects, and in principle it is not obvious which of the two is dominant. The forecasts performed in this work show that the tightest constraints are actually obtained when both the cross-covariance and cross-correlation are included.\\
\subsection{Hybrid approach cross-covariance}
\label{sec:appendix_hybrid_cross}

The hybrid approach is equivalent to
using the data vector
\begin{equation}
\mathbf{D} = \left \lbrace
\bcl{\phsp}{\ell_1, \dots, \ell_N}, \, \bcl{\phsp}{\ell_1, \dots, \ell_N}, \, \GCsp(P_k)
\right \rbrace \;,
\end{equation}
and computing the Fisher matrix with \cref{eq:fisher_most_compact_form} by using a covariance matrix that can be symbolically written as
\begin{equation}
\begin{pmatrix}
\cov{\phph}{\phph} & \cov{\phph}{\phsp} & 0 \\
\cov{\phsp}{\phph} & \cov{\phsp}{\phsp} & 0 \\
0 & 0 & \cov{\spsp}{\spsp} \\
\end{pmatrix} \;.
\end{equation}
The upper left $2\times 2$ sector contains the covariances between the elements of the $\cl{\phph}{}{\ell}$ and $\cl{\phsp}{}{\ell}$ matrices, organised in block-diagonal form for all multipoles as in \cref{eq:block_diag_multipole_covariance}. Analogously, the lower right corner block represents the auto-covariance of the spectroscopic galaxy Fourier power spectrum for all wave-numbers and redshifts considered. The zeroes correspond to the elements containing the unknown covariances between the 2D and 3D power spectra, which are therefore neglected.

%

\end{document}